\documentclass[hidelinks,11pt]{article}
\usepackage{natbib,amsmath,amsthm,amssymb,color,graphicx}
\usepackage{times}
\usepackage{multirow}
\usepackage{hyperref}

\usepackage[usenames,dvipsnames]{xcolor}

\usepackage{algorithm}
\usepackage{booktabs}
\usepackage{subcaption}

\numberwithin{equation}{section}
\theoremstyle{plain}

\theoremstyle{plain}
\theoremstyle{definition}

\newtheorem{alg}{Algorithm}
\newtheorem{defn}{Definition}

\newcommand{\Varek}[2]{\Vare_{#1, #2}}      %
\newcommand{\facloadrowr}[2]{\facload_{#1,#2}} %
\newcommand{\idiovsp}{(\sigma^2_{\lsp}) \new}
\newcommand{\uncon}{u}

\newcommand{\betaIBP}{\gamma}
\newcommand{\bIBP}{b_\nfac}
\newcommand{\alphaIBP}{\alpha}
\newcommand{\aIBP}{a_\nfac}
\newcommand{\Fd}[1]{\mbox{\rm F}\left(#1\right)}
\newcommand{\tauglob}{\kappa}
\newcommand{\tauloc}{\omega}
\newcommand{\taulocv}{\boldsymbol{\omega}}
\newcommand{\taucol}{\theta}
\newcommand{\taucolv}{\boldsymbol{\theta}}
\newcommand{\aalpha}{a^\alpha}
\newcommand{\balpha}{b^\alpha}
\newcommand{\abeta}{a^\betaIBP}
\newcommand{\bbeta}{b^\betaIBP}
\newcommand{\aglob}{a^\tauglob}
\newcommand{\bglob}{b^\tauglob}
\newcommand{\cglob}{c^\tauglob}
\newcommand{\aloc}{a^\tauloc}
\newcommand{\bloc}{b^\tauloc}
\newcommand{\cloc}{c^\tauloc}
\newcommand{\acol}{a^\taucol}
\newcommand{\bcol}{b^\taucol}
\newcommand{\ccol}{c^\taucol}
\newcommand{\rstar}{r^\star}

\newcommand{\leadset}[1]{{\cal L} (#1)}

\newcommand{\Wishart}[2]{ \mathcal{W} _{#1}\left(#2\right)}
\newcommand{\addb}{,b}
\newcommand{\ji}{{j_i}}

\newcommand{\GIG}[1]{\mathcal{GIG} \left(#1\right)}

\newcommand{\rankl}{z}
\newcommand{\SSR}{{\mbox{\rm SSR}}}
\newcommand{\oddpost}{O^{\mbox{\rm \tiny post}}}
\newcommand{\Bodd}{D}
\newcommand{\Count}[1]{\#\{#1\}}
\newcommand{\CountAR}{3579}

\newcommand{\nfacsp}{r_{\footnotesize sp}}
\newcommand{\jzero}{j_{\footnotesize 0}}
\newcommand{\jsp}{j_{\footnotesize \rm sp}}

\newcommand{\lsp}{l_{\footnotesize \rm sp}}
\newcommand{\psplit}{p_{\footnotesize split}}
\newcommand{\pmerge}{p_{\footnotesize merge}}

\newcommand{\Gammainv}[1]{\mathcal{G}^{-1} \left(#1\right)}
\newcommand{\Gammad}[1]{ \mathcal{G}\left(#1\right)}

\newcommand{\Normal}[1]{ \mathcal{N} \left(#1\right)}
\newcommand{\Normult}[2]{ \mathcal{N} _{#1}\left(#2\right)}

\newcommand{\Betafun}[1]{B (#1)}
\newcommand{\Betadis}[1]{\mathcal{B}\left(#1\right)}

\newcommand{\Gamfun}[1]{\Gamma (#1)}

\newcommand{\Uniform}[1]{\mathcal{U}\left[#1\right]}

\newcommand{\rvY}{Y} %
\newcommand{\Real}{\mathbb{R}}

\newcommand{\Sm}{{\mathbf S}}

\newcommand{\hypv}{\boldsymbol{\tau}} %
\newcommand{\pshift}{p_{\mbox{\rm \footnotesize shift}}}

\newcommand{\ym}{{\mathbf y}}             %

\newcommand{\om}{\Omega}                  %
\newcommand{\Vary}{{\mathbf \om}}         %
\newcommand{\dimy}{m}                     %
\newcommand{\load}{\beta}                 %
\newcommand{\facload}{\boldsymbol{\load}} %
\newcommand{\nfac}{k}                     %

\newcommand{\nsp}{\nfacsp}  %
\newcommand{\fac}{f}                      %
\newcommand{\facm}{{\mathbf \fac}}        %
\newcommand{\facmk}[2]{\facm^{#1}_{#2}}
\newcommand{\facmktilde}[2]{\tilde \facm^{#1}_{#2}}
\newcommand{\error}{\epsilon}             %
\newcommand{\errorm}{\boldsymbol{\error}} %
\newcommand{\Vare}{{\mathbf \Sigma}}      %

\newcommand{\Varetrue}{\Vare_0}      %

\newcommand{\Corr}[1]{\mbox{\rm Corr}(#1)}
\newcommand{\idiov}{\sigma^2}             %

\newcommand{\Pm}{{\mathbf P}}             %
\newcommand{\lm}{{\mathbf  l}}             %
\newcommand{\lmr}[2]{\lm_{#1,#2}}             %
\newcommand{\deltav}{\boldsymbol{\delta}} %
\newcommand{\deltavsp}{\deltav_{\loadsp}}
\newcommand{\deltacsp}{\delta^{\loadsp}}
\newcommand{\lmsp}{\lm_{\loadsp}}

\newcommand{\loadsp}{\Xi}
\newcommand{\facloadsp}{\boldsymbol{\loadsp}}

\newcommand{\deltavlam}{\boldsymbol{\delta}^\Lambda} %
\newcommand{\deltacol}[1]{\deltav_{\cdot,#1}} %
\newcommand{\deltacolr}[2]{\deltav^{#1}_{\cdot,#2}} %

\newcommand{\bm}{{\mathbf b}}             %
\newcommand{\bV}{{\mathbf B}}             %
\newcommand{\Xb}{{\mathbf X}}             %
\newcommand{\Psiv}{\boldsymbol{\Psi}}     %
\newcommand{\Fm}{{\mathbf F}}             %
\newcommand{\Pim}{\boldsymbol{\Pi}}       %
\newcommand{\bP}{\Pm}
\newcommand{\jtwo}{\ell}

\newcommand{\traces}{\mbox{\rm tr}}

\newcommand{\trace}[1]{\traces (#1 )}

\newcommand{\Omegav}{{\mathbf P}}     %
\newcommand{\cv}{{\mathbf m}}         %
\newcommand{\Lv}{{\mathbf L}}         %

\newcommand{\xm}{{\mathbf x}}             %
\newcommand{\zm}{{\mathbf z}}             %

\newcommand{\Diag}[1]{\mbox{\rm Diag}\, (#1)} %
\newcommand{\dimmat}[2]{#1\times #2}  %
\newcommand{\bfz}{{\mathbf{0}}}         %
\newcommand{\bfzmat}{{\mathbf{O}}}      %
\newcommand{\identm}{{\mathbf I}}       %
\newcommand{\identy}[1]{{\identm}_{#1}} %
\newcommand{\unit}[1]{{\mathbf{1}}_{#1}} %
\newcommand{\Probsym}{\mbox{\rm Pr}}    %
\newcommand{\Prob}[1]{\Probsym (#1)}    %
\newcommand{\Ew}[1]{\mbox{\rm E}(#1)}   %
\newcommand{\trans}[1]{#1 ^{'}}         %
\newcommand{\Var}{\mbox{\rm V}}         %
\newcommand{\V}[1]{\Var (#1 )}          %

\newcommand{\odd}{O}
\newcommand{\oddpr}{O^{\mbox{\rm \tiny pr}}}
\newcommand{\oddrat}{O^{\mbox{\rm \tiny sp}}}
\newcommand{\oddratl}{R}

\newcommand{\loadtrue}{\Lambda}                 %

\newcommand{\facloadtrue}{\boldsymbol{\loadtrue}} %
\newcommand{\nfactrue}{r}                     %

\newcommand{\new}{^{\mbox{\rm \tiny new}}}

\newcommand{\nullmod}{n}

\newenvironment{Figure}[4]%
{\begin{figure}[t!]
\begin{center}
\scalebox{#4}{\includegraphics{#3}}
\caption{#1}\label{#2}}
{\end{center}\end{figure}}

\newenvironment{Figure2}[5]%
{\begin{figure}[t!]
\begin{center}
\scalebox{#5}{\includegraphics{#3}}
\hspace*{1cm}
\scalebox{#5}{\includegraphics{#4}}
\caption{#1}\label{#2}
}{\end{center}\end{figure}}

\newenvironment{Figure3}[6]%
{\begin{figure}[t!]
\begin{center}
\begin{tabular}{ccc}
\scalebox{#6}{\includegraphics{#3}}
& %
\scalebox{#6}{\includegraphics{#4}}
&
\scalebox{#6}{\includegraphics{#5}}
\end{tabular}
\caption{#1}\label{#2}
}{\end{center}\end{figure}}

\newenvironment{Figurepng}[4]%
{\begin{figure}[t]
\begin{center}
 \scalebox{#4}{\includegraphics{#3.png}}
\caption{#1}\label{#2}
}{\end{center} \end{figure}}

{\begin{figure}[t]
\begin{center}
\scalebox{#5}{\includegraphics{#3.png}}    \scalebox{#5}{\includegraphics{#4.png}}
\caption{#1}\label{#2}
}{\end{center} \end{figure}}

{\begin{figure}[t]
\begin{center}
\scalebox{#7}{\includegraphics{#3.png}}    \scalebox{#7}{\includegraphics{#4.png}} \\
\scalebox{#7}{\includegraphics{#5.png}}    \scalebox{#7}{\includegraphics{#6.png}}
\caption{#1}\label{#2}
}{\end{center} \end{figure}}

{\begin{figure}[t!]
\begin{center}
\scalebox{#6}{\includegraphics{#3.png}}    \scalebox{#6}{\includegraphics{#4.png}} \scalebox{#6}{\includegraphics{#5.png}}
\caption{#1}\label{#2}
}{\end{center} \end{figure}}

\newenvironment{Tabelle}[2]%
{\begin{table}[t!]\begin{center}\caption{#1}\label{#2}}{\end{center}\end{table}}

\begin{document}

\title{Sparse Bayesian factor analysis when the number of factors is unknown}

\author{Sylvia Fr\"uhwirth-Schnatter\footnote{Department of Finance, Accounting, and Statistics, WU Vienna University of Economics and Business, Austria. Email: {\tt sfruehwi@wu.ac.at}} \and
	Darjus Hosszejni\footnote{Department of Finance, Accounting, and Statistics, WU Vienna University of Economics and Business, Austria. Email: {\tt darjus.hosszejni@wu.ac.at}}
	\and Hedibert Freitas Lopes\footnote{School of Mathematical and Statistical Sciences, Arizona State University, Tempe, USA \& Insper Institute of Education and Research, S\~ao Paulo, Brazil. Email: {\tt hedibertfl@insper.edu.br}}
}

\maketitle

\begin{abstract}
There has been increased research interest in the subfield of sparse Bayesian factor analysis with shrinkage priors, which achieve additional sparsity beyond the natural
parsimonity of factor models.
In this spirit, we estimate the number of common factors in the highly implemented sparse latent factor model with spike-and-slab priors on the factor loadings matrix.
Our framework leads to a natural, efficient and simultaneous coupling of model estimation and selection on one hand and model identification and  rank estimation (number of factors) on the other hand.
More precisely, by embedding the unordered generalised lower triangular loadings representation into overfitting sparse factor modelling, we obtain posterior summaries regarding factor loadings, common factors as well as the factor dimension
 via postprocessing draws from our efficient and customised Markov chain Monte Carlo scheme.
\end{abstract}

\vspace{0.5cm}
{\em Keywords:}
Hierarchical model;
identifiability;
point-mass mixture priors;
marginal data augmentation;
reversible jump MCMC;
prior distribution;
sparsity;
Heywood problem;
rotational invariance;
ancillarity-sufficiency interweaving strategy;
fractional priors
\vspace{0.5cm}

\centerline{JEL classification: C11, C38, C63}

\section{Introduction}

Factor analysis aims at identifying common variation in multivariate observations and relating
it to hidden causes, the so-called common factors, see %
\citet{thu:mul} and,  more recently,  \citet{and:int}. The common setup consists
of a sample $\ym=\{ \ym_1, \ldots, \ym_T\}$ of  $T$  multivariate observations $\ym_t=\trans{(y_{1t}, \ldots,  y_{\dimy t})}$ of dimension $\dimy$.
For a given factor dimension $\nfactrue$,  the basic factor model is defined as a latent
variable model, involving the   common factors $\facm_t=\trans{(\fac_{1t} \cdots \fac_{\nfactrue t})}$:
\begin{eqnarray}  \label{fac1}
 \facm_t  \sim  \Normult{\nfactrue}{\bfz,\identy{\nfactrue}}, \quad \ym_t =  \facloadtrue \facm_t + \errorm_t, \quad \errorm_t \sim \Normult{\dimy}{\bfz,\Varetrue} ,  \quad  \Varetrue=\Diag{\idiov_1,\ldots,\idiov_{\dimy}},
\end{eqnarray}
where $\facloadtrue$ is the $\dimmat{\dimy}{\nfactrue}$ matrix of
factor loadings $\loadtrue_{ij}$ and $\Varetrue$ is the covariance matrix of the idiosyncratic errors   $\errorm_t$.
Model (\ref{fac1}) implies that conditional on   $\facm_t$   the $\dimy$ elements of $\ym_t$ are independent and all dependence among these variables is explained through the common factors. %
 Assuming independence of $\facm_t$ and $\errorm_t$
implies that, marginally,   $\ym_t$ arises  from a multivariate normal distribution, $\ym_t  \sim \Normult{\dimy}{\bfz,\Vary}$, with zero mean and a covariance matrix  $\Vary$
with the following constrained  structure:
 \begin{eqnarray}
\Vary=     \facloadtrue  \trans{\facloadtrue } + \Varetrue .  \label{fac4}
\end{eqnarray}
Since $\nfactrue$ typically  is (much) smaller than $\dimy$, factor models yield a parsimonious
representation of $\Vary$  with (at most) $\dimy (\nfactrue +1)$  instead of the $\dimy(\dimy+1)/2$ parameters of an unconstrained covariance matrix. Hence, factor models proved to be very useful for  covariance estimation,
especially if $\dimy$ is large; see \citet{fan-etal:hig_je}, \citet{for-etal:ope}, \citet{bha-dun:spa} and \citet{kas:spa}, among others. %

From the very beginning, the goal
 of factor analysis has been to extract the hidden factors 
 to understand the driving forces behind the observed covariance.
 Using (\ref{fac4}), this requires a decomposition of the covariance matrix  $\Vary$ into the cross-covariance matrix $\facloadtrue  \trans{\facloadtrue }$
and the covariance matrix $ \Varetrue $  of the uncorrelated idiosyncratic errors. With the only source of information
being the observed covariance of the data, this is more challenging than estimating  $\Vary$ itself.
A huge literature, going  back to \citet{koo-rei:ide}, \citet{rei:ide} and \citet{and-rub:sta} has addressed this problem of identification
 which can be resolved by imposing additional structure on the factor model, see
 \citet{neu:ide}, \citet{gew-zho:mea}, \citet{bai-ng:pri}, \citet{cha-etal:inv}, and \citet{wil:ide}, among many others.
Recently, a new  identification strategy  based on unordered generalized lower triangular (UGLT) structures
\citep{fru-lop:spa,fru-etal:whe} was introduced that allows to address not only the commonly known rotational invariance problem, but also the problem of variance identification \citep{and-rub:sta} of which the literature is still less aware.

The main goal of the present paper is to foster mathematically rigorous identification in Bayesian factor analysis.
The recent years  have seen many contributions in the field of sparse Bayesian factor analysis  which
achieves additional sparsity beyond the  natural parsimonity of  factor models through the choice of shrinkage priors.
One strand of literature considers continuous shrinkage priors on the factor loadings, often in combination with a prior that allows infinitely many columns in the loading matrix
of which only a finite number is non-zero such as the Indian buffet process prior \citep{gri-gha:inf} or the multiplicative Gamma process \citep{bha-dun:spa}, see \citet{zha-etal:bay_gro}, \citet{roc-geo:fas}, \citet{kas:spa}, and \citet{leg-etal:bay} for recent contributions.
Alternatively, following the pioneering paper by   \citet{wes:bay_fac},  many authors considered  point mass mixture
   priors on the factor loadings in   basic factor models  \citep{car-etal:hig,fru-lop:par},  in dedicated factor models with correlated
  factors \citep{con-etal:bay_exp}  and
   dynamic factor models \citep{kau-sch:bay}. %

 Sparse Bayesian factor analysis with point mass mixture (also called spike-and-slab)
   priors  allow   factor loadings to be exactly zero
  and treat the identification of these elements as a variable selection problem.
This allows to identify \lq\lq simple structures\rq\rq\ where in each row only a few nonzero loadings are present \citep{and-rub:sta}, %
a long standing issue in factor analysis %
  addressed in \citet{con-etal:bay_exp}.
 It also allows to identify   irrelevant variables $y_{it}$  which are uncorrelated
with the remaining variables in $\ym_t$, since the entire  row of
the factor loading matrix  is zero  for these variables.
Among other fields, this is of relevance in economics \citep{kau-sch:ide}, where it is common practice to include as many variables as possible in factor analysis \citep{sto-wat:mac,boi-ng:are},
and in bioinformatics \citep{luc-etal:spa}, where typically only a few out of potentially ten thousands of genes may be related to a certain physiological outcome.

A challenging problem in factor analysis  is choosing the  factor dimension $\nfactrue$
which is usually unknown, see  \citet{owe-wan:bic} for an excellent review. Often information criteria
   \citep{bai-ng:det2002}  are used  also in a Bayesian context, see e.g.\ \citet{ass-etal:bay} and \citet{cha-etal:inv}, other authors
     employ marginal likelihoods \citep{lee-son:bay,lop-wes:bay}. Learning about the  factor dimension $\nfactrue$  is intrinsic in Bayesian approaches that allow
  infinitely many columns in the loading matrix \citep{gri-gha:inf,bha-dun:spa,roc-geo:fas,leg-etal:bay}, even if this approach requires careful tuning of hyperparameters   \citep{dur:not}.
  A number of authors  exploit variable selection in a finite-dimensional overfitting factor model \citep{fru-lop:par,con-etal:bay_exp,kau-sch:ide} and we will follow their lead in the present paper.
Our approach relies on
the following sparse Bayesian
exploratory factor analysis (EFA)  model:
\begin{eqnarray}  \label{fac1reg}   %
 \ym_t =  \facload_\nfac  \facmk{\nfac}{t} + \errorm_t,  \qquad  \errorm_t \sim \Normult{\dimy}{\bfz,\Vare_\nfac} ,  \quad  \facmk{\nfac}{t}   \sim  \Normult{\nfac}{\bfz,\identy{\nfac}},
\end{eqnarray}
where   $\facload_\nfac$ is an  $\dimmat{\dimy}{\nfac}$   loading matrix
 with elements $\load_{ij}$ and $\Vare_\nfac$ is a diagonal matrix with strictly positive diagonal elements. Model (\ref{fac1reg}) is potentially overfitting, since 
  we assume a finite factor dimension $\nfac$
   which is larger than true number of factors $\nfactrue$. 
   We employ  spike-and-slab priors, where  the elements  $\load_{ij}$ of $\facload_\nfac$  are allowed  to be exactly zero, with the corresponding $\dimmat{\dimy}{\nfac}$ sparsity matrix with elements $\delta_{ij}$ being denoted by  $\deltav_\nfac$.
 To learn the factor dimension $\nfactrue$, we 
   exploit a finite version  of the
 two-parameter Beta prior \citep{gha-etal:bay,fru:gen}
     to define a  shrinkage process prior on $\deltav_\nfac$ that induces increasing shrinkage 
     of the factor loadings toward zero as the column index increases, extending
        recent work by   \citet{leg-etal:bay}. 

    To achieve identification, we
   impose a UGLT structure \citep{fru-etal:whe}
 on the  loading matrix  $\facload_\nfac$ and the corresponding sparsity matrix $\deltav_\nfac$ in the EFA model (\ref{fac1reg}).
  Commonly, in  machine learning and statistics,  no constraints are imposed on  $\deltav_\nfac$; however, leaving
 the
  sparsity pattern  unconstrained makes it
 more  difficult to recover the true number of factors and reconstruct $\facloadtrue $ from $\facload_\nfac$. In econometrics as well as in statistics, $\deltav_\nfac$ is often constrained to a lower triangular matrix, however such a constraint is too restrictive \citep{joe:gen,car-etal:hig}.
 The  UGLT structure that we impose is a much weaker
 constraint which only requires the top non-zero elements in each column of the loading matrix to lie in different rows  and still leaves the model  unidentified.
 As shown in \citet{fru-etal:whe}, on the one hand it is weak enough to ensure that
 any loading matrix can be rotated into a GLT representation, on the other hand it strong enough   to ensure \lq\lq controlled unidentifiability\rq\rq\ which can be easily resolved.

  For practical Bayesian inference,   we develop a new and efficient Markov chain Monte Carlo (MCMC)  procedure that delivers posterior draws from  the EFA model (\ref{fac1reg}) under point mass mixture  priors, which is known to be particularly challenging \citep{pat-etal:pos}. As part of our algorithm, we design a (simple)
reversible jump MCMC sampler to navigate through the space of UGLT loading matrices of varying factor dimension. We achieve mathematically rigorous identification in the spirit of \citet{and-rub:sta} through post-processing the posterior draws $(\facload_\nfac , \Vare_\nfac)$ from this unidentified EFA
 model. By  ensuring variance identification through the algorithm of 
 \citet{hos-fru:cov} 
 and resolving rotational invariance
    during postprocessing, we are able recover the factor dimension
 $\nfactrue$, the idiosyncratic variances $\Varetrue$  and an ordered GLT representation $\facloadtrue$ of the loading matrix from the posterior draws. In this way, we  add 
 a mathematically rigorous approach   to a growing literature   where commonly more heuristic post-processing procedures are applied for this purpose; see  \citep{ass-etal:bay,kau-sch:bay,pow-etal:eff,pap-ntz:ide}.

Our sampling as well as our identification strategy work under arbitrary choices regarding the slab distribution, including fractional priors \citep{fru-lop:par}, the horseshoe prior \citep{zha-etal:bay_gro} and the Lasso prior \citep{roc-geo:fas} . In high-dimensional models, we work with structured priors
with column-specific shrinkage \citep{leg-etal:bay} 
and employ the triple gamma prior \citep{cad-etal:tri} to achieve good separation of signal and noise.

 The rest of the paper  is organized as follows. Section~\ref{section:basic} introduces sparse Bayesian EFA models with UGLT structures, while prior choices are discussed in Section~\ref{priorel}. Section~\ref{mcmc} introduces our MCMC sampler for this model class and Section~\ref{secGLT} discusses post-processing MCMC draws to achieve identification.
    Section~\ref{secalpp} considers applications to exchange rate data and NYSE100 returns, while   Section~\ref{secconcluse} concludes.

\section{Sparse Bayesian EFA models with UGLT structures}\label{section:basic}

\subsection{Model definition}  \label{secmodel}

Throughout the paper, we work with the exploratory factor analysis (EFA)  model  (\ref{fac1reg}),
with $\nfac$  potential common factors.
 Factor analysis based on this EFA model  yields the extended  variance decomposition
 \begin{eqnarray}
\Vary=     \facload _\nfac  \trans{\facload _\nfac } + \Vare _\nfac,   \label{fac4beta}
\end{eqnarray}
instead of the true variance decomposition (\ref{fac4}) and the
question arises how to
recover $(\nfactrue, \facloadtrue , \Varetrue)$ from $(\facload _\nfac, \Vare _\nfac)$.
Without imposing constraints on $\facload _\nfac$, this question is not easily answered,
due to the many identifiability issues in factor analysis.
In recent work, \citet{fru-etal:whe} discuss econometric identification of $(\nfactrue, \facloadtrue , \Varetrue)$
from the sparse overfitting Bayesian factor model (\ref{fac1reg}), if  the loading matrices are restricted to unordered generalized lower triangular (UGLT) structures.

 \begin{defn}{{\bf UGLT structures.}} \label{UGLT}
  Let $\deltav_{\nfactrue}$ be  an binary matrix of $\nfactrue$ non-zero columns.  Let  $l_j$  denote
          the row index (also called pivot) of the top non-zero entry in the $j$th column of $\deltav_{\nfactrue}$ (i.e.\ ${\delta}_{ij}=0, \forall \, i<l_j$).
          $\deltav_{\nfactrue}$ is said to be a UGLT structure,
         if the pivot elements $\lm_{\nfactrue}=(l_{1}, \ldots ,  l_{\nfactrue})$  lie in different rows. More generally, a  binary matrix $\deltav_\nfac$ with zero columns has an unordered GLT structure,
         if the submatrix  $\deltav_{\nfactrue}$ containing the $\nfactrue$ non-zero column of $\deltav_\nfac$ is a UGLT structure.
    \end{defn}

 \noindent   To facilitate identification in sparse Bayesian factor analysis, we assume in the present paper that the factor loading matrix  $\facload_\nfac$ and the corresponding sparsity matrix
 $\deltav_\nfac$ in  the EFA model (\ref{fac1reg}) exhibit a UGLT  structure.
  Compared to the common literature, where all elements of $\deltav_\nfac$ are left unspecified,
  this imposes the  constraint on $\deltav_\nfac$ that the top non-zero element in the various (non-zero) columns lie in different rows.
   As discussed in  \citet{fru-etal:whe}, the (weak) UGLT constraint on $\deltav_\nfac$
   is sufficient for  a  mathematically rigourous
   identification even in an overfitting factor model, see Section~\ref{secide} for more details.
    These insights will be exploited in Section~\ref{secGLT},
    where the posterior draws from a sparse EFA model with UGLT structure
    are screened in a post-processing manner to ensure identification
    of the posterior draws and  to learn
    about the unknown factor dimension $\nfactrue$, the loading matrix $\facloadtrue$ as well as  $\Varetrue$ from the data.

\subsection{Econometric identification in factor models}  \label{secide}

Consider first an EFA model (\ref{fac1reg}) that is not overfitting, i.e.~$\nfac = \nfactrue$.
A rigorous approach toward identification of factor models was first offered by \citet{and-rub:sta}
who consider  identification as a two-step procedure. The first step is
  variance identification, i.e.~identification of $\Varetrue$  from the variance
 decomposition  (\ref{fac4}).  Variance identification is easily violated for sparse Bayesian factor models, regardless whether $\deltav_\nfac$ is unconstrained or exhibits a UGLT structure.
 \citet{fru-etal:whe} prove that for UGLT structures  the so-called \CountAR\ counting rule %
 for the elements
in the $\nfactrue$ non-zero columns $\deltav_\nfactrue$ of  $\deltav_\nfac$ is sufficient for variance identification. \citet{hos-fru:cov} provide an efficient algorithm to verify this rule.

\begin{defn}{{\bf \CountAR\ counting rule}.} \label{countAR}
A binary matrix  $\deltav_\nfactrue$ satisfies the \CountAR\ counting rule, if the following condition is satisfied: for  each $q =1,\ldots,\nfactrue$ and for each submatrix consisting of $q$ column of $\deltav_\nfactrue$, the  number  of nonzero rows in this sub-matrix is at least equal to $2q+1$.
   \end{defn}

\noindent  The \CountAR\ counting rule states that every  column of  $\deltav_\nfactrue$  should have at
least 3, every pair of non-zero columns at least 5, every subset of 3 columns at least 7  elements and so forth. If an indicator matrix $\deltav_\nfactrue$ obeys the \CountAR\ counting rule, this implies that  $\Varetrue =\Vare_\nfactrue$ and hence, $\facloadtrue  \trans{\facloadtrue }=\facload _\nfactrue \trans{\facload_\nfactrue}$ is identified. A necessary condition for
 $\deltav_{\nfactrue}$ to satisfy the \CountAR\  counting rule is the following upper bound for $\nfactrue$:
 \begin{eqnarray}  \label{kboundr}
   \nfactrue \leq \frac{\dimy -1}{2}.
\end{eqnarray}

\noindent The second step of identification is  solving the rotational invariance problem provided that variance identification
 holds for $(\facload_\nfactrue, \Vare_\nfactrue)$.
Since variance identification implies that $\facloadtrue  \trans{\facloadtrue }=
\facload _\nfactrue \trans{\facload_\nfactrue}$, it follows that
$\facload _\nfactrue =\facloadtrue \Pm$ for some orthogonal  matrix  $\Pm$ \citep[Lemma~5.1]{and-rub:sta}.
 The rotational identification problem is usually solved  by imposing a structure on the loading matrices $\facload _\nfactrue$ %
that ensures identification of $\facloadtrue$ from
 $\facload _\nfactrue \trans{\facload _\nfactrue}$ in the sense that $\facload _\nfactrue =\facloadtrue \Pm$ iff $\Pm$ is equal to
 the identity.
 A weaker condition  is rotational identification up to signed permutations
 $ \facload _\nfactrue   =      \facloadtrue  \bP _{\pm} \bP _{\rho}$,
where the permutation matrix $\bP_{\rho}$ corresponds to one of   the $\nfactrue$! permutations %
 and the reflection matrix
 $\bP_{\pm}=\Diag{\pm 1, \ldots, \pm 1}$  corresponds
 to one of  the $2^\nfactrue$ ways to  switch the signs of the  $\nfactrue$  columns of $\facloadtrue$.
\citet{fru-etal:whe} show that imposing a UGLT structure on $\facload_\nfactrue$ and $\facloadtrue$ leads  to
rotational identification up to signed permutations. Provided that
 $\facload_\nfactrue$  is variance identified,
$\facloadtrue$ is recovered from $\facload_\nfactrue$ by reordering the columns of $\facload _\nfactrue$
such that the pivots  $l_1< \ldots < l_{\nfactrue}$ are increasing.

In applied factor analysis, the EFA  model  (\ref{fac1reg}) is typically  overfitting with $\nfac  > \nfactrue$ and additional identifiability issues arise. In an overfitting EFA model,
  identifiability of  $\Vare _\nfac$ and $\facload _\nfac  \trans{\facload_\nfac  } $   from (\ref{fac4beta}) is lost and infinitely many representations $(\facload _\nfac,\Vare _\nfac)$
with $\Vare _\nfac \neq \Varetrue$ exist that imply the same covariance matrix $\Vary$ as $(\facloadtrue,\Varetrue)$
\citep{gew-sin:int,tum-sat:ide}.
\citet{fru-etal:whe} show that imposing a UGLT structure on $\facload_\nfac$ %
 allows to reveal $\facloadtrue$ from $\facload_\nfac$ even in this case.

Consider, for illustration, an overfitting EFA model  with $\nfac=\nfactrue +1 $.
Factor analysis  typically yields  loading matrices $\facload_{\nfactrue +1}$  with  $\nfactrue +1$ non-zero columns rather than matrices with $\nfactrue$ non-zero and a single zero column.
 \citet[Theorem~7]{fru-etal:whe} prove that by imposing a UGLT structure on $\facload _{\nfactrue +1}$, the additional column in the overfitting model
is equal to a so-called \emph{spurious factor} $\facloadsp$, e.g.:
\begin{eqnarray} \label{adsp}
{\small \facload _{\nfactrue +1} =  \left(\begin{array}{cc}
                \facloadtrue   &\facloadsp
                  \end{array}
               \right) , \quad
                \facloadsp=  \left(\begin{array}{c}
                                           \bfz \\
                                           \loadsp_{\lsp} \\
                                           \bfz
                                         \end{array} \right), \quad
                \Vare _{\nfactrue +1} =  %
                \Diag{\idiov_1,\ldots, \idiov_{\lsp} - \loadsp_{\lsp}^2, \ldots, \idiov_\dimy} ,}
\end{eqnarray}
with  a single non-zero factor loading  $\loadsp_{\lsp}$ satisfying  $0 <  \loadsp_{\lsp}^2< \idiov_{\lsp}$
which lies in a pivot row   $\lsp$  different from the pivot rows  $l_1, \ldots, l_\nfactrue $  in $ \facloadtrue$.

This result allows to identify spurious columns $\facloadsp$ in the loading matrix $\facload_{\nfactrue+1}$ of the overfitting  EFA model (\ref{fac1reg})
and to recover   $\facloadtrue$ from the remaining (active) columns  up to a signed permutation, $\facload_{\nfactrue}   = \facloadtrue  \bP _{\pm} \bP _{\rho} $. The covariance matrix
$\Varetrue= \Vare _{\nfactrue +1} +  \facloadsp \trans{\facloadsp}$ is identified by moving the spurious column $\facloadsp$ to the idiosyncratic errors.
This results also holds for a higher degree of overfitting and is the cornerstone of our MCMC sampler which relates exploratory and confirmatory Bayesian factor analysis.

\subsection{Relating exploratory to confirmatory Bayesian factor analysis} \label{EFACFA}

 The sparsity matrix  $\deltav_{\nfac}$ in the EFA model (\ref{fac1reg}) allows to identify which factors %
 are active (the corresponding column of $\deltav_{\nfac}$ has at least two non-zero loading), which factors are spurious (the corresponding column of $\deltav_{\nfac}$ has a single non-zero loading), and which ones are inactive (the corresponding column of $\deltav_{\nfac}$ is zero).
  This allows to
split the sparsity matrix $\deltav_{\nfac}$
 into an
$\dimmat{\dimy}{\nfactrue}$  submatrix $\deltav _\nfactrue$ with $\nfactrue$ active columns,
 an
$\dimmat{\dimy}{\nsp}$  submatrix $\deltavsp$ with  $\nsp$ spurious columns, and a submatrix with $\jzero= \nfac - \nfactrue - \nsp$ zero columns. The loading matrix $\facload _\nfac$ is split accordingly into the $\dimmat{\dimy}{\nfactrue}$  submatrix $\facload _\nfactrue$, the
$\dimmat{\dimy}{\nsp}$  submatrix $\facloadsp$, and
$\jzero$ zero columns, while  the factors $\facm_t^k$ are split 
  into $\facm_t^r$, $\facm_t^{\loadsp}$ and $\facm^0_t$.

Exploiting representation (\ref{adsp}), we
 extract the following model of factor dimension $\nfactrue$
which is embedded  in any EFA model  with UGLT structure,
\begin{eqnarray}  \label{fac1CFA}
  \facmk{\nfactrue}{t}   \sim  \Normult{\nfactrue}{\bfz,\identy{\nfactrue}}, \quad \ym_t =  \facload_\nfactrue  \facmk{\nfactrue}{t} + \errorm_t,  \quad  \errorm_t \sim \Normult{\dimy}{\bfz,\Vare_\nfactrue} ,  \quad
  \Vare _\nfactrue= \Vare _\nfac + \facloadsp \trans{\facloadsp},
\end{eqnarray}
by moving  the $\nsp$ spurious columns  $\facloadsp$ to the idiosyncratic variances $\Vare _\nfactrue$. We call (\ref{fac1CFA}) the confirmatory factor analysis (CFA) model  induced
by the active columns  of the sparsity matrix $\deltav_{\nfac}$  in the EFA model.  The likelihood function is invariant
to moving from
the EFA model  to the CFA model (\ref{fac1CFA}),
since the implied covariance matrix 
$\Vary=     \facload _\nfac  \trans{\facload _\nfac } + \Vare _\nfac =
\facload _\nfactrue  \trans{\facload _\nfactrue } + \Vare _\nfactrue$ remains the same.
On the other hand, taking the CFA model (\ref{fac1CFA}) as a starting point,
 we can move to the EFA model (\ref{fac1reg}) without changing the likelihood function
 by adding $\nsp$ spurious columns $\deltavsp$ to $\deltav_{\nfactrue}$.
 The only relevant information needed for expanding the CFA model in this way
  is that $\nsp$  columns outside of
    $\deltav_\nfactrue$ have column-size one, whereas all remaining columns are zero.
  
      For $\nsp=1$, for instance,
  a single
  spurious column $\deltavsp$
  is added to $\deltav_{\nfactrue}$ to define an EFA model with $\nfac - \nfactrue - 1$ zero columns.
    The position of the only non-zero indicator  in   column $\deltavsp$, denoted by $\deltacsp_{\lsp}$,
   is not identified and can lie in any row $\lsp$
   that is  different from the pivots $\lm_{\nfactrue}$ in $\deltav_\nfactrue$.
  A spurious  column  $\facloadsp$
  is added to $\facload_{\nfactrue}$   to define $\facload_{\nfac}$,
  while the covariance matrix of the idiosyncratic errors in the EFA model is  defined as $\Vare_{\nfac} =
  \Vare_{\nfactrue} -
  \facloadsp \trans{\facloadsp}$.
  The only non-zero  loading
  $\loadsp_{\lsp}$ in $\facloadsp$ can take  any value such that 
   the $\lsp$-th diagonal element of $\Vare_{\nfac}$ remains positive, 
   i.e.~$ \Varek{\nfac}{\lsp,\lsp}=\idiov_{\lsp} -  (\loadsp_{\lsp})^2>0 $. This entire move only affects
  the $\lsp$-th row $\facloadrowr{\nfactrue}{\lsp,\cdot}$ of  $\facload_\nfactrue$. More specifically, for $t=1, \ldots,T$:
  \begin{eqnarray}
&&  y_{\lsp,t}=  \facloadrowr{\nfactrue}{\lsp,\cdot}  \facmk{\nfactrue}{t} + \error_{\lsp,t}  ,
 \, \error_{\lsp,t}  \sim \Normal{0,\idiov_{\lsp}},   %
 \label{spurious2} \\
&& y_{\lsp,t}=  \facloadrowr{\nfactrue}{\lsp,\cdot}  \facmk{\nfactrue}{t} + \loadsp_{\lsp} \fac_{t}^\loadsp    +  \tilde{\error}_{\lsp,t},
 \, \tilde{\error}_{\lsp,t}  \sim \Normal{0,\idiovsp}, \, \idiovsp =\idiov_{\lsp}- (\loadsp_{\lsp})^2 .  \nonumber
\end{eqnarray}
By integrating model (\ref{spurious2})  with respect to  the spurious factor  $\fac_{t}^\loadsp$,
it can once more be  verified that both models imply the same distribution   $p(y_{\lsp,t}| \facloadrowr{\nfactrue}{\lsp,\cdot}, \facmk{\nfactrue}{t},\idiov_{\lsp})$,
independently of the specific value of $\loadsp_{\lsp}$.
  Also for arbitrary $\nsp \in \{1, \ldots, \nfac - \nfactrue\}$,
 neither the  position of the pivots $ \lmsp$
 of $\deltavsp$ (which are the only non-zero indicator in each column) nor  the
 non-zero elements in the corresponding spurious loading matrix $\facloadsp$ are
 identified. However, the pivots in $ \lmsp$ must lie in different rows.

  Moving forth and back between  the EFA model (\ref{fac1reg}) and the  CFA model (\ref{fac1CFA})
  as described above is the cornerstone of an efficient  MCMC algorithm developed in Section~\ref{mcmc} that operates in the space of UGLT matrices with varying dimension without imposing further constraints.
   In Section~\ref{priorel},
  suitable priors are defined  for a sparse EFA model with UGLT structure that are (largely) invariant to these moves.

\section{Prior specifications}  \label{priorel}

\subsection{Column sparsity through exchangeable shrinkage process priors} \label{priordelta}

 Bayesian inference is performed in the  EFA model   (\ref{fac1reg})
with a finite number $\nfac$ of potential factors. %
Dirac-spike-and-slab prior for the factor loadings are assumed, 
 \begin{eqnarray} \label{PriorLL}
 & \load_{i j} | \tau_{j}
\sim   (1- \tau_{j})  \Delta_{0} + \tau_{j}
P_{\tiny \rm slab} (\load_{i j}),
  \end{eqnarray}
 where the columns of the loading matrix are increasingly pulled toward 0 as the column index increases. This cumulative shrinkage is  achieved indirectly  by placing an exchangeable shrinkage process  (ESP) prior on the slab probabilities $\tau_1, \ldots, \tau_k$:
   \begin{eqnarray} \label{prialt}
 \tau_j| k \sim  \Betadis{\aIBP,\bIBP}, \quad j=1, \ldots,k,
 \end{eqnarray}
 see \citep{fru:gen}.  The ESP prior turns  model (\ref{fac1reg}) into 
 a {\em sparse}  EFA model, where  the number  $\nfactrue$ of active columns in $\deltav_\nfac$ with at least two non-zero elements  is  a random variable which takes values smaller than $\nfac$ with positive probability.
 As recently shown by \citet{fru:gen}, prior (\ref{prialt})
 has a representation as a finite cumulative shrinkage process
 (CUSP) prior \citep{leg-etal:bay}.
 Prominent examples of such ESP priors are the finite 
 one-parameter-beta (1PB) prior:
    \begin{eqnarray} \label{pri1PB}
 \tau_j| k \sim  \Betadis{\frac{\alphaIBP}{\nfac},1}, \quad j=1, \ldots,k,
 \end{eqnarray}
  and the finite  two-parameter-beta (2PB) prior,  
 \begin{eqnarray} \label{prialtTeh}
 \tau_j| k \sim  \Betadis{\betaIBP \frac{\alphaIBP}{\nfac},\betaIBP}, \quad j=1, \ldots,k.
 \end{eqnarray}
 Alternative choices are $\aIBP=\alphaIBP/\nfac $ and $\bIBP= \betaIBP (\nfac-1)/\nfac$ \citep{pai-car:non}.
 For $\nfac \rightarrow \infty$,
 prior (\ref{pri1PB}) converges   to the Indian buffet process  prior \citep{teh-etal:sti}
   and has been 
 employed by \citet{roc-geo:fas} in sparse Bayesian factor analysis.  For $\nfac \rightarrow \infty$, prior (\ref{prialtTeh}) converges to the two-parameter-beta prior  introduced by \citet{gha-etal:bay}
 in Bayesian nonparametric latent feature models which can be regarded as a  factor model with infinitely many columns of which only a finite number is non-zero.

 As opposed to this literature, we stay within the framework of
 factor models with finitely many columns.
While obeying the upper bound (\ref{kboundr}),  $\nfac$ is selected large enough to encourage spurious columns and zero columns in $\deltav_\nfac $.
Spurious columns are essential for our strategy of recovering the factor dimension $\nfactrue$ through adding and deleting spurious columns
in  the overfitting  EFA model (\ref{fac1reg}). %
The hyperparameters  $\alphaIBP$ and $\betaIBP$  are instrumental in controlling prior column sparsity.   A prior with  $\alphaIBP<k$ and $\betaIBP=1$, e.g., induces sparsity, since the {\em largest} slab probability  $\tau_{(k)}  \sim \Betadis{\alphaIBP,1} $, while the smallest slab probability
 $\tau_{(1)} \sim \Betadis{\alphaIBP/k,1}$ \citep{fru:gen}. To adapt the hyperparameters to the data at hand, $\alphaIBP$ is assumed to be a random  with prior $\alpha \sim \Gammad{\aalpha, \balpha}$,while $\betaIBP$ is random with prior
$\betaIBP \sim \Gammad{\abeta, \bbeta}$  for the 2PB-prior (\ref{prialtTeh}).

\paragraph*{Imposing a UGLT structure} \label{priorUGLT}

For given numbers $\nfactrue$ and  $\nsp$ of, respectively, active and spurious columns, 
we define  a prior on the pivots $\lm_\nfactrue =(l_1, \ldots, l_\nfactrue)$ and 
$\lmsp =(l_{{\loadsp},1}, \ldots, l_{{\loadsp},\nsp}) $ 
 such that the non-zero columns of the sparsity matrix
$\deltav_\nfac$  exhibit a UGLT structure.  
 The prior $p(\lm_\nfactrue)$ in the CFA model
 is defined as follows. Let $\leadset{\lm}= \{i \in\{  1, 2, \ldots, \dimy\}:  i \notin \lm \}$.
 Given the pivots $ \lmr{\nfactrue}{-j}$ outside of any column $j$, 
   condition UGLT implies that $l_j$ has to be different from  $ \lmr{\nfactrue}{-j}$,
   and  we assume a uniform prior distribution over all admissible pivots  $l_j \in \leadset{\lmr{\nfactrue}{-j}}$: %
   \begin{eqnarray}  \label{priorLj}
 p( l_j| \lmr{\nfactrue}{-j}) = \frac{1}{|\leadset{\lmr{\nfactrue}{-j}}|}=
  \frac{1}{ \dimy - \nfactrue + 1}.
  \end{eqnarray}
The pivots
$\lmsp$ of the spurious columns $\deltavsp$  are assigned 
a uniform prior over all admissible values, given the pivots $\lm_\nfactrue $, %
using the following order-independent construction. Given $\lm_\nfactrue$,
 $l_{{\loadsp},1} $ is uniform over $\leadset{\lm_\nfactrue}$; given $l_{{\loadsp},1} $, $l_{{\loadsp},2}$  is uniform  over $\leadset{\lm_\nfactrue \cup \{l_{{\loadsp},1}\}}$, and so forth.

Given the pivots $l_j$ in all non-zero columns of $\deltav_\nfac$, %
by definition  $\delta_{l_j,j}=1$ and $\delta_{ij}=0$ for $i< l_j$,
while  the $\dimy-l_j$ indicators $\delta_{ij}$ below $l_j$ are subject to variable selection
with column-specific probabilities $\tau_{j}$ following the ESP prior (\ref{prialt}): 
  \begin{eqnarray} \label{prigen1}
\Prob{\delta_{ij}=1|l_j, \tau_{j}}=
\left\{ \begin{array}{ll}
0, & i< l_j, \\
1, & i = l_j, \\
\tau_{j}, & i=l_j+1,\ldots, \dimy. \\
\end{array}    \right.
 \end{eqnarray}
 Let $d_j=\sum_{i=1}^\dimy {\delta_{ij}}$ be the number of non-zero indicators in columns $j$.
 With  $d_j-1$ successes and  $\dimy-l_j- d_j+1$ failures in the experiment defined in (\ref{prigen1}),  we obtain the following prior for  column $\deltacolr{\nfactrue}{j}$:
   \begin{eqnarray} \label{prigen3A}
 \Prob{\deltacolr{\nfactrue}{j}|l_j, \tau_j}= \tau_j ^{d_j-1} (1-\tau_j) ^{\dimy-l_j- d_j+1}.
 \end{eqnarray}
If we integrate over $\tau_j$, then we obtain:
  \begin{eqnarray} \label{prigen3}
 \Prob{\deltacolr{\nfactrue}{j}| l_j}= \frac{\Betafun{\aIBP  + d_j-1,\bIBP + \dimy-l_j- d_j+1}}{\Betafun{\aIBP,\bIBP}}.
 \end{eqnarray}

\subsection{Choosing the slab distribution} \label{priorfl}

To define a  prior on the loading matrix $\facload_\nfac $,
 we split %
  $\facload_\nfac $
as discussed in Section~\ref{EFACFA}
and define  a prior  $p(\facload _\nfactrue|\Vare _\nfactrue, \deltav _\nfactrue )$ on the factor loading matrix $\facload _\nfactrue$ in the CFA model (\ref{fac1CFA}), conditional on
$\Vare _\nfactrue = \Diag{\idiov_1, \ldots, \idiov_\dimy}$ and $\deltav _\nfactrue$.
 When expanding the CFA model  to an EFA model with $\nsp$ columns,
 we define a prior $p(\facloadsp| \facload_\nfactrue, \Vare_{\nfactrue}, \lmsp)$ 
 on the spurious loadings   conditional on $\facload_\nfactrue$, $\Vare_{\nfactrue}$, and  $\lmsp$.
The spurious  factor loadings 
are  assigned a uniform prior
over all  values that lead to a positive definite matrix $\Vare _\nfac=
\Vare _\nfactrue - \facloadsp \trans{\facloadsp}$ in the EFA model:
 \begin{eqnarray} \label{prioradd}
\loadsp_{\lsp}^2 |\idiov_{\lsp} \sim \Uniform{0, \idiov_{\lsp}}.
\end{eqnarray}
This ensures for all $\lsp \in \lmsp$ that
$ \Varek{\nfac}{\lsp,\lsp} = \idiov_{\lsp} -  \loadsp_{\lsp}^2 >0$.
 By this definition, 
 both the likelihood and the prior are invariant to moving between the EFA and the CFA model for a given number of spurious columns $\nsp$, regardless of the chosen slab distribution.\footnote{Note that this is a major improvement compared to \citet{fru-lop:spa}.}

 The spike-and-slab prior  (\ref{PriorLL}) %
 is formulated for the factor loading matrix $\facload_\nfactrue$ in the CFA model. 
 A broad range of slab distributions has been considered for sparse Bayesian factor analysis and can be combined 
 with the RJMCMC sampler we introduce in Section~\ref{mcmc}. 
 Since the conditional likelihood function factors into a product over the rows of the loading matrix,  prior independence across the rows  is assumed. For each  row $i$ of $\deltav_\nfactrue$ with  $q_i = \sum_j %
 \delta_{ij} >0 $
  nonzero elements, we consider two different prior families %
for  the vector $\facload_{i\cdot}^{\deltav}$ 
of unconstrained elements in this row. %
First, hierarchical Gaussian priors are considered, taking  the form
  \begin{eqnarray}
\facload_{i\cdot}^{\deltav}|\idiov_i  \sim \Normult{q_i}{\bfz, \bV_{i0}^{\deltav}\idiov_i},  \label{prior1}
\end{eqnarray}
where $\bV_{i0}^{\deltav}$ is a diagonal matrix.
   The variance of prior  (\ref{prior1}) 
     is assumed to depend on   $\idiov_i$, because  this allows
joint drawing of   $\facload_\nfactrue$ and  $\idiov_1, \ldots, \idiov_\dimy$  and, even more importantly, sampling the sparsity matrix $\deltav_\nfactrue$ without conditioning on the model parameters  during MCMC estimation, see   Algorithm~\ref{Algo3}.
For models with a moderate number of features $\dimy$, the choice $\bV_{i0}^{\deltav}=A_0 \identy{q_i}$   with a fixed
hyperparameter  $A_0$ is common \citep{lop-wes:bay,gho-dun:def,con-etal:bay_exp}.
For high-dimensional models with larger $\dimy$, often a structured hierarchical prior is assumed to achieve better shrinkage of the factor loadings. One example are spike-and-slab priors with a column specific shrinkage parameter  $\taucol_j$  in the slab distribution of $\load_{ij}$,
see e.g.\ \citet{kau-sch:bay}, \citet{leg-etal:bay} and \citet{fru:gen}:
\begin{eqnarray}
\load_{ij}| \delta_{ij}=1, \taucol_j, \idiov_i, \tauglob  \sim \Normal{0, \tauglob
 \taucol_j \idiov_i }.  \label{priorEXP2}
\end{eqnarray}
The column specific shrinkage parameters 
follow an exchangeable prior, either 
an inverse gamma prior, $\taucol_j  \sim \Gammainv{\ccol,\bcol}$ \citep{leg-etal:bay},
or a triple gamma prior, 
$\taucol_j \sim \Fd{2\acol,2\ccol}$  \citep{cad-etal:tri}. The triple gamma prior
 acts as a variance selection prior and  puts considerable mass on small values of 
 $\taucol_j$ which simultaneously pulls the factors $\fac_{jt}$ toward 0 for all time points $t$.
For $\acol=\ccol=0.5$, this yields a grouped version of the horseshoe prior \citep{zha-etal:bay_gro}.
The global shrinkage parameter follows  $\tauglob \sim \Gammainv{\cglob,\bglob} $ or
$\tauglob \sim \Fd{2\aglob,2\cglob} $.
Prior (\ref{priorEXP2}) can be extended by assuming
 local shrinkage parameters $\tauloc_{ij}$ arising from an  F-distribution
for each non-zero factor loading:
 \begin{eqnarray} \label{prior0ext}
\load_{ij}| \delta_{ij}=1, \tauloc_{ij}, \taucol_j, \idiov_i, \tauglob  \sim \Normal{0,  \tauglob \taucol_j \idiov_i \tauloc_{ij}},  \quad \tauloc_{ij} \sim \Fd{2\aloc,2\cloc}. \label{priorEXP4}
\end{eqnarray}
 For  $\aloc=\cloc=0.5$, %
 reduces to the  horseshoe prior employed by \citet{zha-etal:bay_gro}.

As an alternative shrinkage prior, \cite{fru-lop:par} introduced a conditionally conjugate   fractional prior 
$p(\facload_{i\cdot}^{\deltav}|\idiov_i \addb ,\facm_\nfactrue)
\propto \displaystyle p(\tilde{\ym}_i| \facm_\nfactrue, \facload_{i\cdot}^{\deltav} ,\idiov_i)^b$
in the spirit of \citet{oha:fra}. %
It can be interpreted as the posterior of a  non-informative prior and a small fraction $b>0$  the conditional likelihood is derived from regression model %
\begin{eqnarray}
\tilde{\ym}_i= \Xb_i ^{\deltav} \facload_{i\cdot}^{\deltav} + \tilde{\errorm}_i,
 \label{regnonp}
\end{eqnarray}
where  $\tilde{\ym}_i=\trans{(y_{i1} \cdots y_{iT})}$, 
$\tilde{\errorm}_i=\trans{(\error_{i1} \cdots \error_{iT})}$ and  $\Xb_i ^{\deltav}$ is a regressor matrix constructed from the latent factors  $\facm_\nfactrue=\{\facmk{\nfactrue}{1},  \ldots, \facmk{\nfactrue}{T} \}$ (see Appendix~\ref{postdisfac} for details).
 This yields:
 \begin{eqnarray}   \label{priorfrac}
& \facload_{i\cdot}^{\deltav} | \idiov_i \addb ,\facm_\nfactrue  \sim
\Normult{q_i}{\bm_{iT}  ^{\deltav}  , \bV_{iT}^{\deltav}  \idiov_i /b_N}, & 
\end{eqnarray}
where $\bm_{iT}^{\deltav} =\bV_{iT} ^{\deltav}  \trans{(\Xb_i ^{\deltav})}\tilde{\ym}_i$ and $\bV_{iT}^{\deltav} = (\trans{(\Xb_i ^{\deltav})} \Xb_i ^{\deltav} ) ^{-1}$. 
Similar fractional priors have been applied  for variable selection in other latent variable models  \citep{fru-tue:bay,fru-wag:sto}.
In the spirit of \citet{fos-geo:ris}, $b_N=1/N$, since $N=\dimy T$  observations  are available to estimate $\facload_\nfactrue$.
 The fractional  prior often works better than hierarchical Gaussian priors for moderate $\dimy$, however is not flexible enough to handle larger dimensions  $\dimy$.

\subsection{The prior on the idiosyncratic variances} \label{priorsi}

Finally, we define  a  prior
on the idiosyncratic  variances $\idiov_1, \ldots, \idiov_\dimy$,
independently of all other parameters.
 When estimating factor models using classical statistical methods  such as %
 ML estimation,
  it frequently happens that  the optimal solution lies outside the admissible parameter space with one
   or more of the idiosyncratic variances $\idiov_i$s being negative, see e.g.
\citep{joe:som,bar:lat_fac}. 
   This difficulty became
 known as the Heywood problem. %
Using a prior on the  idiosyncratic variances $\idiov_1, \ldots, \idiov_\dimy$
 within a Bayesian framework, typically chosen from the  inverted Gamma family, %
 \begin{eqnarray}
\idiov_i \sim \Gammainv{c_0,C_{i0}}, \label{priorsiidg}
\end{eqnarray}
 naturally avoids negative values for  $\idiov_i$.
  Nevertheless, there exists a Bayesian analogue of the Heywood problem which takes the form of
  multi-modality of the posterior of  $\idiov_i$ with one mode lying at 0. 
   Heywood problems typically occur, if the  constraint
      \begin{eqnarray}
 \frac{1}{\idiov_i} \geq (\Vary^{-1})_{ii}   \quad \Leftrightarrow \quad   \idiov_i  \leq \frac{1}{(\Vary^{-1})_{ii}} \label{const1}
\end{eqnarray}
is violated, where $\Vary$ is the covariance matrix of $\ym_t$ defined in (\ref{fac4}),
see e.g.\ \citet[p.~54]{bar:lat_fac}. It is clear from  inequality (\ref{const1}) that $1/\idiov_i$  has to be bounded away
from 0. For this reason, improper priors on the idiosyncratic variances such as $p(\idiov_i)\propto 1/\idiov_i$
 \citep{mar-mcd:bay,aka:fac} are not able to prevent Heywood problems.
Similarly, proper inverted Gamma priors with small degrees of freedom such as $c_0=1.1$
 \citep{lop-wes:bay} allow values too close to 0.
 Hence,  the hyperparameters $c_0$ and  $C_{i0}$ in (\ref{priorsiidg}) have to be selected in such a way that Heywood problems are avoided.
 In particular,  $c_0$   should be chosen large enough to
 bound the prior away from 0, typically $c_0=2.5$.
 Regarding $C_{i0}$, the most common choice is using a fixed value $C_{i0}=C_0$.

 An alternative choice has been suggested in  \cite{fru-lop:par}.
 They reduce the occurrence probability of a Heywood problem  which is equal to
  $\Prob{X\leq C_{i0}(\Vary^{-1})_{ii}}$ where $X \sim \Gammad{c_0,1}$ through
  the choice of individual scalings $C_{i0}$.  The smaller $C_{i0}$, the  smaller is this probability.  On the other hand,  a downward bias is introduced, if $C_{i0}$ is too small, since
  $\Ew{\idiov_i}=C_{i0}/(c_0-1)$.
  Choosing $C_{i0}=(c_0-1)/(\widehat{\Vary^{-1}})_{ii}$
   as the largest value for which inequality
   (\ref{const1}) is fulfilled by the prior expectation $\Ew{\idiov_i}$   yields the following prior:
\begin{eqnarray}
\idiov_i \sim \Gammainv{c_0,(c_0-1)/(\widehat{\Vary^{-1}})_{ii}}, \label{priorsiid}
\end{eqnarray}
 based on an estimator  $\widehat{\Vary^{-1}}$ of $\Vary^{-1}$.
 In \citet{fru-lop:spa},
a Bayesian estimator %
is proposed  which combines the
 sample information with an inverted Wishart prior $\Vary^{-1} \sim \Wishart{\dimy}{\nu_o, \nu_o \Sm_o}$:
 \begin{eqnarray}
\widehat{\Vary^{-1}}= (\nu_o + T/2)(\nu_o \Sm_o + 0.5 \sum_{t=1}^T \ym_t \trans{\ym_t})^{-1}. \label{Varyhat}
\end{eqnarray}

\section{MCMC estimation} \label{mcmc}

We use MCMC techniques to sample from the posterior
distribution
of the EFA model (\ref{fac1reg}),
given  the priors  introduced in Section~\ref{priorel}.
 As noted by many authors,   e.g.\ \citet{pat-etal:pos}, MCMC sampling for sparse Bayesian factor models is notoriously difficult,
  since sampling the indicator matrix $\deltav_\nfac$ corresponds to navigating through an extremely high dimensional model space.
 This is even more challenging, if the sparse factor model is overfitting. %
In the present paper, an MCMC scheme summarized in Algorithm~\ref{Algo3} is developed where several steps are  designed  specifically  for sparse  Bayesian factor models with UGLT structure where  the factor dimension is unknown.
  Updating  $\deltav_\nfac$   for sparse exploratory Bayesian factor analysis  without
 imposing identification constraints on $\deltav_\nfac$
is fairly  straightforward, see e.g.\  \citet{car-etal:hig} and \citet{kau-sch:bay}, among many others.
A more refined approach is implemented in the present  paper under sparse UGLT structures
which will allow us to address  econometric  identification in a post-processing manner in Section~\ref{secGLT}.

 \begin{alg}[\textbf{MCMC estimation for sparse Bayesian factor models with unordered  GLT structures}] \label{Algo3}
  Choose initial values for
   $\deltav_\nfactrue$, $\facm_\nfactrue$ and $\nfacsp$.\footnote{See Appendix~\ref{init} for details.} Iterate $M$ times through the following steps and  discard the first $M_0$ draws as burn-in:
  \begin{itemize} \itemsep 0mm
 \item[(CFA)] %
      Update all unknowns in the CFA model (\ref{fac1CFA}) corresponding to $\deltav_\nfactrue$:
       \end{itemize} 
       \vspace*{-5mm}
 \begin{itemize}  \itemsep 0mm
\item[(H)] Update any unknown  %
hyperparameters %
in the ESP prior (\ref{prialt}) without conditioning on the slab probabilities
$\tau_1, \ldots, \tau_r$.
For %
$j=1, \ldots, \nfactrue $, %
sample %
$\tau_j | l_j, d_j \sim  \Betadis{\aIBP  + d_j-1,\bIBP + \dimy-l_j- d_j+1} $,
  where $d_j=\sum_{i=1}^\dimy {\delta_{ij}}$.

\item[(D)] Loop over all  columns  of the sparsity matrix $\deltav_\nfactrue$ in a
      random order and perform variable selection  by
      updating the $j$th column $\deltacolr{\nfactrue}{j}$  below the pivot, conditional on  the remaining columns $\deltacolr{\nfactrue}{-j}$, the factors $\facm_\nfactrue=(\facmk{\nfactrue}{1},\ldots,\facmk{\nfactrue}{T})$  and the hyperparameter $\tau_j$
       without conditioning on the model parameters   $  \facload_\nfactrue$ and $\idiov_1,\ldots,\idiov_{\dimy}$:
      \begin{itemize}
         \item[(a)] Sample all indicators $\delta_{ij}$ below the pivot  $l_j$ from
            $p(\delta_{ij}| l_j, \deltacolr{\nfactrue}{-j}, \facm_\nfactrue , \tau_j, \ym)$.
       \item[(b)] If column $\deltacolr{\nfactrue}{j}$ is spurious after this update,
       increase $\nfacsp$ by one. Remove the $j$th column  from $\deltav_\nfactrue$, the factors $f_{jt}, t=1, \ldots, T$ from $\facm_\nfactrue$ and $\tau_j$ from $\hypv_\nfactrue$
        to define, respectively,  $\deltav_{\nfactrue-1}$,  $\facm_{\nfactrue-1}$ and $\hypv_{\nfactrue-1}$      and  decrease $\nfactrue$ by one.

  \end{itemize}
       \item[(L)] Loop over all  columns  of $\deltav_\nfactrue$ in a
      random order and update the pivots without conditioning on $  \facload_\nfactrue$, $\idiov_1,\ldots,\idiov_{\dimy}$ and the slab probabilities  $\hypv_\nfactrue$:
      \begin{itemize}
         \item[(a)]
      Sample a new pivot row $l_j$ in each column $j$
        from $p(l_j| \deltacolr{\nfactrue}{-j},
        \facm_\nfactrue , \ym)$. %

        \item[(b)] If column $\deltacolr{\nfactrue}{j}$ is spurious after this update,
       increase $\nfacsp$ by one. Remove the $j$th column  from $\deltav_\nfactrue$, the factors $f_{jt}, t=1, \ldots, T$ from $\facm_\nfactrue$ and $\tau_j$ from $\hypv_\nfactrue$
        to define, respectively,  $\deltav_{\nfactrue-1}$,  $\facm_{\nfactrue-1}$ and $\hypv_{\nfactrue-1}$      and  decrease $\nfactrue$ by one.
\end{itemize}
 \item [(P)]   Sample   the model parameters  $\facload_\nfactrue$ and $\idiov_1,\ldots,\idiov_{\dimy}$   jointly conditional on the sparsity matrix   $\deltav_\nfactrue$ and the factors $\facm_\nfactrue=(\facmk{\nfactrue}{1},\ldots,\facmk{\nfactrue}{T})$ from $p(\facload _\nfactrue, \idiov_1,\ldots,\idiov_{\dimy}| \deltav_\nfactrue,\facm_\nfactrue,\ym)$.

  \item[(F)]  Sample the latent factors $\facm_\nfactrue=(\facmk{\nfactrue}{1},\ldots,\facmk{\nfactrue}{T})$ conditional on
the model parameters $\facload_\nfactrue$ and $\idiov_1,\ldots,\idiov_{\dimy}$ from $ p(\facmk{\nfactrue}{1},\ldots,\facmk{\nfactrue}{T}|\facload_\nfactrue,\idiov_1,\ldots,\idiov_{\dimy},\ym) $.

\item[(S)] For hierarchical Gaussian priors, update all unknown scaling factors,
i.e. the global shrinkage parameter $\tauglob$, the  column-specific
 shrinkage parameters $\taucol_j$, and the local shrinkage parameters $
\tauloc_{ij}$.

\item[(A)] Perform a boosting step to enhance mixing.

\end{itemize}

 \begin{itemize}
 \item[(EFA)] %
 Move from the current CFA model %
   to an EFA model (\ref{fac1reg}) with $\nfacsp$ spurious columns:
 \begin{itemize} %
 \item[(R)] Use Algorithm~\ref{AlgoRJMCMC} to change
   $\nfacsp$, while holding the number of active factors $\nfactrue$ fixed.
   Loop over all  columns  of the  spurious factors $\deltavsp$ and try to turn them into  active factors.
   \end{itemize}
Move  from the current EFA model %
   back to the CFA model (\ref{fac1CFA}) and preserve the updated number $\nfacsp$ of spurious columns.
 \end{itemize}
\end{alg}

\noindent Algorithm~\ref{Algo3}  consists of two main blocks.
 Block~(CFA) operates
 in the confirmatory factor analysis  model  (\ref{fac1CFA}) corresponding to $\deltav_\nfactrue$.  Due to the prior specification
  in Section~\ref{priorel},
  the number $\nfacsp$ of spurious columns is a  sufficient
  statistic for the remaining columns in  $\deltav_\nfac$ %
   and  no further information, such as the position of the pivots
    $\lmsp$ in $\deltavsp$ or the spurious loading matrix $\facloadsp$,  is needed
   to  update  the parameters in the CFA model, %
   namely the inclusion probabilities $\hypv_\nfactrue$, the sparsity matrix  $\deltav_\nfactrue$,
the loading matrix $\facload_\nfactrue$, the idiosyncratic variances $\idiov_1,\ldots,\idiov_{\dimy}$ in  $\Vare_\nfactrue$,   the latent factors $\facm_\nfactrue=(\facmk{\nfactrue}{1},\ldots,\facmk{\nfactrue}{T})$, and
 all unknown shrinkage factors.

In Block~(EFA), the sampler  moves from the current
 CFA model %
 to an EFA model %
 with $\nfacsp$ spurious columns.
In Step~(R),  dimension changing moves are performed
 in  the much larger space underlying the EFA model. %
      Exploiting  the results of Section~\ref{secide}, spurious factors
    are added and deleted with the help of  a  reversible jump MCMC (RJMCMC) step  described  in  more detail in Section~\ref{RJMCMC}.
     The sampler finally returns to a CFA model %
 with a potentially larger number of active factors $\nfactrue$.

To  ensure that the loading matrix in the CFA model exhibits a UGLT structure,   Step~(L) performs  MH steps  that %
 navigate  through the space of all admissible $\deltav_\nfactrue$ where the pivots $\lm_{\nfactrue}=(l_1, \ldots, l_{\nfactrue})$ lie in different rows, see Section~\ref{movelead}. These steps are  performed marginalized w.r.t. $\hypv_\nfactrue=(\tau_1, \ldots, \tau_\nfactrue)$.
 Given the pivots $\lm_{\nfactrue}$, the    hyperparameters $\aIBP$ and $\bIBP$ in the ESP prior (\ref{prialt}) are updated in Step~(H) using an MH step without conditioning on the
slab probabilities $\tau_1, \ldots, \tau_k$.
 Details are provided %
 in Appendix~\ref{detailsAppH}.
 The posterior $\tau_j | l_j, d_j $ is updated 
  by combining the  likelihood (\ref{prigen3A}) %
with the prior $\tau_j \sim \Betadis{\aIBP,\bIBP}$ for all columns  $j$.
In Step~(D), variable selection is performed in each column $j$ for all indicators $\delta_{ij}$
   below the pivot element $l_j$. This step  potentially turns an active factor into
    a spurious one and in this way decreases the number of active
  factors $\nfactrue$, while increasing  $\nfacsp$.
    All moves in Step~(D) are implemented  conditionally on $\tau_j$ (and all shrinkage parameters for hierarchical Gaussian priors), as this allows efficient multimove sampling  of all indicators $\{ \delta_{ij}, i \in \{l_j +1, \ldots, \dimy\} \}$, using Algorithm~\ref{AlgoInd} in Appendix~\ref{mcmcsmodi}.

The remaining steps are quite standard in Bayesian factor analysis, see \citet{gew-sin:int} and \citet{lop-wes:bay}.
  In Step~(F),  the conditional joint posterior $p(\facmk{\nfactrue}{1},\ldots,\facmk{\nfactrue}{T}| \facload_{\nfactrue},\idiov_1,\ldots,\idiov_{\dimy},\ym) $ factors   into $T$ independent normal distributions  given by:
 \begin{eqnarray}  \label{filtPXsim}
\facmk{\nfactrue}{t}
|\ym_t,  \facload_{\nfactrue}, \Vare_{\nfactrue} \sim \Normult{\nfactrue}{(\identy{\nfactrue} + \trans{\facload_{\nfactrue}} \Vare_{\nfactrue} ^{-1} \facload_{\nfactrue}) ^{-1} \trans{\facload_{\nfactrue}} \Vare _{\nfactrue}^{-1} \ym_t , (\identy{\nfactrue} + \trans{\facload _{\nfactrue}}  \Vare_{\nfactrue} ^{-1}  \facload_{\nfactrue}) ^{-1} }.
\end{eqnarray}
In Step~(P), we use an efficient algorithm for
 multi-move sampling of all unknown model parameters $\facload _{\nfactrue}$, and $\idiov_1,\ldots,\idiov_{\dimy}$,
see  Appendix~\ref{jointfac}. %
For the hierarchical Gaussian priors (\ref{priorEXP2}) and (\ref{prior0ext}), all unknown
shrinkage parameters are updated in Step~(S), see Appendix~\ref{sampleshr}. %
Finally,  the boosting Step~(A)  is added to improve the mixing of the MCMC scheme,  see Section~\ref{accelerate} and Appendix~\ref{boost_frac}.

\subsection{Split and merge moves for overfitting models}  \label{RJMCMC}

   Step~(R)  in Algorithm~\ref{Algo3} is based on  moving from the CFA model (\ref{fac1CFA}) to an EFA model with $\nsp$ spurious factors  in $\deltav_\nfac$.
As discussed in Section~\ref{secide},
  spurious columns $\facloadsp$ in an EFA model can be substituted by zero columns  without changing the likelihood function,
  by adding $\facloadsp$ to the covariance matrix  $\Vare_{\nfactrue}$
  of the idiosyncratic errors in the CFA model.
   On the other hand, for  any row $\lsp$ that is not a pivot row in the sparsity matrix $\deltav_{\nfactrue}$ of the CFA model, a fraction
  of the idiosyncratic variance $\idiov_{\lsp}$ can be used to turn
  a zero column in $\facload_\nfac$  into  a spurious column
  $\facloadsp $ with a non-zero loading $\loadsp_{\lsp}$ without changing the likelihood function either.
  This is the cornerstone of the dimension changing procedure in Step~(R), outlined in detail in
  Algorithm~\ref{AlgoRJMCMC}.

\begin{alg}[\textbf{Dimension changing move in an EFA model}] \label{AlgoRJMCMC}
Step~(R) in Algorithm~\ref{Algo3} is implemented in the following way:
  \begin{itemize}
      \item[(R-S)] Perform an RJMCMC step  %
  to change the number $\nfacsp$ of spurious columns %
      through a split move on a zero column or a merge move on a spurious column  in $\deltav_\nfac$.

   \item[(R-L)] Given $\nsp$,
       sample the pivot rows $\lmsp| \lm_\nfactrue$
       of all  $\nsp$ spurious columns       sequentially  from the set
$\leadset{\lm_\nfactrue}$, where $\lm_\nfactrue$  are the  pivot rows  of the active factors $\deltav_\nfactrue$. Order the spurious columns such that $l_{{\loadsp},1} <\ldots < l_{{\loadsp},\nsp} $.

     \item[(R-F)] Loop over all  spurious columns $\jsp $  and
     sample the factors $\facm_{\jsp} = (\fac_{\jsp, 1}, \ldots, \fac_{\jsp, T})$ independently for all $t=1, \ldots, T$  from 
      $\fac_{\jsp, t} | \facmk{\nfactrue}{t},\facload_\nfactrue, \idiov_{\lsp}, y_{\lsp,t} \sim \Normal{E_{\jsp, t} ,V_{\jsp}}$, where $U_{\jsp}$ is a draw from a uniform distribution on [-1,1] and 
       \begin{eqnarray} \label{mainonfsp}
 V_{\jsp} =  1- U_{\jsp}^2, \quad
\displaystyle  E_{\jsp,  t} =  U_{\jsp} \frac{y_{\lsp,t}- \facloadrowr{\nfactrue}{\lsp,\cdot} \facmk{\nfactrue}{t}}{\sqrt \idiov_{\lsp}}.
\end{eqnarray}

      \item[(R-H)]
       Sample $\tau_{\jsp} | \lsp \sim \Betadis{\aIBP ,\bIBP + \dimy-\lsp}$  for all  spurious columns $\jsp $.
 \item[(R-D)]  Loop over all spurious columns  from the last (with the largest pivot) to the first (with the smallest pivot) and try to turn  spurious columns into  active ones:
 \begin{itemize}
  \item[(R-Da)]    %
     sample  all indicators $\delta_{i,\jsp}$ with $i \in I_{\jsp}=\{\lsp +1, \ldots, \dimy \}$
       below the pivot %
       $\lsp$  %
  conditional on $\tau_{\jsp}$,
  $\deltav_\nfactrue$, $\facm_\nfactrue$ and  the spurious factors $\facm_{\jsp}$  without conditioning on %
   $  \facload_{\nfactrue}$, $\facloadsp$  and $\idiov_1,\ldots,\idiov_{\dimy}$.
   \item[(R-Db)] 
   If column $\jsp$ remains spurious,
   $\nfactrue$ and $\nsp$ are unchanged and  the EFA model is integrated
  over factor $\jsp$ by removing $\deltacol{\jsp}$ from $\deltavsp$ and  $\facm_{\jsp}$ from   $\facm_\loadsp$.
 Otherwise,  decrease  $\nsp$ by 1, increase $\nfactrue$ by 1,  add $\deltacol{\jsp}$ to  $\deltav_\nfactrue$ and  $\facm_{\jsp}$ to
   $\facm_\nfactrue$.
\end{itemize}
\end{itemize}
\end{alg}

 \noindent  Step~(R-S) in Algorithm~\ref{AlgoRJMCMC} changes $\nsp$  by adding  and deleting spurious columns in $\deltav_\nfac$.
For a given $\nsp$, the conditional prior $p(\deltav_\nfac,\facload_\nfac, \Vare_{\nfac}|\nsp)$ is  invariant to the specific choice of $\lsp$ and $\facloadsp$.
However, the prior odds that %
a zero column in
$\deltav_{\nfac}$ can be turned into an additional spurious column depends both on  $\nfactrue$ and $\nsp$ (see Appendix~\ref{RJdetails} for a proof):
\begin{eqnarray} \label{prispnew}
 \oddrat (\nfactrue, \nsp) =
  \frac{\aIBP(\dimy-\nfactrue- \nsp) }{\bIBP + m - \nfactrue - \nsp -1} .
 \end{eqnarray}
  Hence, simply adding or deleting spurious columns  would lead to an invalid  MCMC procedure and an RJMCMC step that incorporates  $\oddrat (\nfactrue, \nsp)$ is performed in Step~(R-S). %
  As opposed to other applications of RJMCMC, the acceptance rate is extremely easy to compute, see (\ref{AccRJSp}) and (\ref{AccRJMe}).

At each sweep of the sampler, a split or a merge move is performed %
with,
respectively, probability $\psplit(\nfactrue,\nfacsp)$  or  $\pmerge (\nfactrue,\nfacsp)$.
 In  a split move (which requires $\nfactrue< \nfac$)  a randomly chosen  zero column in $\deltav_\nfac$ is  turned into a spurious column. The  corresponding proposal  density reads $q_{\mbox{\rm \footnotesize split}}(\deltav_\nfac \new|\deltav_\nfac)=
     \psplit (\nfactrue,\nfacsp)/(\nfac - \nfactrue  - \nsp )$.
The merge move (which requires $\nfacsp >0 $)  is  obtained by  reversing the split move and
 turns  one of the  spurious columns 
 into a zero column. The  corresponding proposal  density reads $q_{\mbox{\rm \footnotesize merge}}(\deltav_\nfac \new|\deltav_\nfac)=
  \pmerge (\nfactrue,\nfacsp)/\nfacsp$.
  A symmetric proposal is selected  for all $0\leq \nfacsp <  \nfac -\nfactrue $ such that
$\psplit (\nfactrue,\nfacsp)= \pmerge (\nfactrue,\nfacsp+1) = p_{\nfacsp}$, where $p_{\nfacsp} \leq 0.5$ is a tuning parameter, while  $\pmerge (\nfactrue,\nfacsp)=0$ for $ \nfacsp=0$ and  $\psplit (\nfactrue,\nfacsp)=0$ for $\nfacsp = k-\nfactrue$. We found it useful to choose
a fixed probability $p_{\nfacsp}=p_s$, although other choices are possible.
 A split move is accepted  with probability
   $\min(1,A_{\mbox{\rm \footnotesize split}}(\nfactrue,\nfacsp))$, where:
 \begin{eqnarray}  \label{AccRJSp}
A_{\mbox{\rm \footnotesize split}} (\nfactrue,\nfacsp)=
\frac{q_{\mbox{\rm \footnotesize merge}}(\deltav_\nfac |\deltav_\nfac \new)}{q_{\mbox{\rm \footnotesize split}}(\deltav_\nfac \new|\deltav_\nfac)}
  \oddrat
   (\nfactrue, \nsp) =
     \frac{\aIBP(\dimy-\nfactrue- \nsp) (\nfac - \nfactrue  - \nsp )}{(\nfacsp+1)(\bIBP + m - \nfactrue - \nsp -1)} ,
 \end{eqnarray}
whereas %
a merge move is accepted  with probability
   $\min(1,A_{\mbox{\rm \footnotesize merge}}(\nfactrue, \nsp))$,  where
    \begin{eqnarray} \label{AccRJMe}
A_{\mbox{\rm \footnotesize merge}}(\nfactrue, \nsp) =
 \frac{1}{A_{\mbox{\rm \footnotesize split}}(\nfactrue, \nsp-1)} =
  \frac{  \nfacsp (\bIBP+  \dimy - \nfactrue  - \nsp )}
  {\aIBP  (\dimy-\nfactrue- \nsp + 1) (\nfac - \nfactrue -\nsp + 1)} .
\end{eqnarray}
Very conveniently,
 $A_{\mbox{\rm \footnotesize split}} (\nfactrue, \nsp)$ and $A_{\mbox{\rm \footnotesize merge}}(\nfactrue, \nsp)$ are independent of the pivots $\lmsp$  in the spurious columns.
 Note that there is a dynamic feature underlying this RJMCMC algorithm, with acceptance
 depending on the number of spurious columns $\nsp$. For $\bIBP=1$, for instance, $A_{\mbox{\rm \footnotesize split}} (\nfactrue, \nsp)$ is monotonically decreasing
and $A_{\mbox{\rm \footnotesize merge}} (\nfactrue, \nsp)$ is monotonically increasing in $\nsp$.

Once $\nfacsp$ has been updated,  Step~(R-L) is trying to turn each spurious column into an 
active one.
Since the likelihood is non-informative about spurious columns,
pivots $\lsp$  are sampled from the prior $\lmsp|\lm_\nfactrue$, %
while the spurious factor loadings   $\loadsp_{\lsp}$
 are sampled from the prior (\ref{prioradd}).
Given $\lsp$, the  idiosyncratic variance $\idiov_{\lsp}$  in the CFA model is split,
with the help of a random variable $U_{\jsp} \sim \Uniform{-1,1}$,
 between $\loadsp_{\lsp}$ and  an updated idiosyncratic
variance  $\idiovsp$. More specifically: %
\begin{eqnarray} \label{prorjitA}
\loadsp_{\lsp} = U_{\jsp} \sqrt{\idiov_{\lsp}}   , \qquad  %
\idiovsp = (1-U_{\jsp}^2) \idiov_{\lsp} .
\end{eqnarray}
Given $\loadsp_{\lsp}$ and  $\idiovsp$,
 factors $\fac_{\jsp,t} $ are
proposed in Step~(R-F)
independently for  $t=1, \ldots,T$,  from the conditional density $ p(\fac_{\jsp, t}| \facmk{\nfactrue}{t},\facload_\nfactrue,  \idiov_{\lsp}, y_{\lsp,t})$
given in (\ref{mainonfsp}), see Appendix~\ref{RJdetails} for a proof.
The slab probabilities $\tau_{\jsp}$ are sampled
in Step~(R-H) as in Algorithm~\ref{Algo3},~Step~(H) using that $d_{\jsp}=1$.
Finally, in Step~(R-Da), variable selection is  performed  in each spurious column
on all indicators below $\lsp$ as in Step~(D) of Algorithm~\ref{Algo3}, conditional on $\fac_{\jsp,t} $, but
marginalizing w.r.t.\ the idiosyncratic variances and the factor loading matrices $\facload_\nfactrue$ and $\facloadsp$. More details about this step are provided in Appendix~\ref{RJdetails}. Any spurious column that is active after this step is integrated
 into the CFA model in Step~(R-Db), increasing in this way the number of active columns $\nfactrue$.

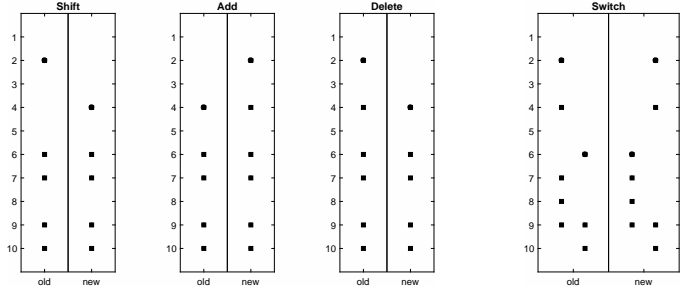
 \begin{Figure2}{MCMC moves to change the leading indices of an unordered GLT structure;
 from left to right: shifting the leading index,   adding a new leading index, deleting a leading
 index and  switching  the leading elements}{figStepL}{lead_change}{switch_lead}{0.3}
 \end{Figure2}

\subsection{Special MCMC moves for  unordered GLT structures} \label{movelead}

Step~(L) in Algorithm~\ref{Algo3} implements moves that explicitly change the position of the pivots
$\lm_{\nfactrue}=(l_1, \ldots, l_{\nfactrue})$ in the $\nfactrue$  columns  of  the UGLT indicator matrix $\deltav_\nfactrue$.
We scan all columns of $\deltav_\nfactrue$ in a random order and  propose  to change $l_j|\lmr{\nfactrue}{-j}$
given the pivots $\lmr{\nfactrue}{-j}$ in the other columns.
To this aim, we
use  one of four MH moves, namely
 shifting the pivot,   adding a new pivot, deleting a pivot and  switching  the pivots (and additional indicators) between column $j$ and a randomly selected column $j'$; see Figure~\ref{figStepL} for illustration.
 All moves are  performed  marginalized w.r.t.\ $\hypv_\nfactrue$. %
 Changing the pivot from $l_j$ to $l_j \new$ changes the number of unconstrained indicators, whereas  the prior ratio
$$\frac{p(l \new _j| \lmr{\nfactrue}{-j})}{p(l _j| \lmr{\nfactrue}{-j})}=1,$$
since $l _j$ has a uniform prior over $\leadset{\lmr{\nfactrue}{-j}}$, the set  of admissible pivot rows in column $j$.
With $d_j\new$ being the new number of non-zero elements in column $j$, the prior ratio $\oddratl_{\mbox{\rm \footnotesize move}}$ can be derived from (\ref{prigen3}):
 \begin{eqnarray} \label{prirat3A}
   \oddratl_{\mbox{\rm \footnotesize move}}= \frac{\Prob{\deltacol{j} \new | l_j \new}}{\Prob{\deltacol{j}| l_j}}=
   \frac{\Betafun{\aIBP  + d_j \new -1,\bIBP + \dimy-l_j \new - d_j \new +1}}{\Betafun{\aIBP + d_j-1,\bIBP + \dimy-l_j- d_j+1}}.
 \end{eqnarray}
Further details are provided in Appendix~\ref{updatelead}.

\subsection{Boosting MCMC}  \label{accelerate}

  Step~(F) and  Step~(P)  in Algorithm~\ref{Algo3} perform full conditional Gibbs sampling for a confirmatory factor model with sparsity matrix $\deltav_\nfactrue$, by sampling the factors $\facmk{\nfactrue}{t}$ conditionally on the loading matrix $\facload_{\nfactrue}$
      and the idiosyncratic covariance matrix $\Vare_\nfactrue$ and sampling  $\facload_{\nfactrue}$ and $\Vare_\nfactrue$
       conditionally on $\facm_\nfactrue = (\facmk{\nfactrue}{1},\ldots,\facmk{\nfactrue}{T})$.  Depending on the signal-to-noise ratio of the latent variable representation,   such full conditional Gibbs sampling   tends to be poorly mixing.
In a CFA model, where  $\facmk{\nfactrue}{t}  \sim  \Normult{\nfactrue}{\bfz,\identy{\nfactrue}}$,
the information in the data (the \lq\lq signal\rq\rq ) can be
 quantified by  the matrix $\trans{\facload_\nfactrue}  \Vare_\nfactrue ^{-1}  \facload_\nfactrue  $ in comparison to the identity matrix  $\identy{\nfactrue}$ (the \lq\lq noise\rq\rq ) in the filter for $\facmk{\nfactrue}{t}|\ym_t,\facload_\nfactrue, \Vare_\nfactrue$,
see (\ref{filtPXsim}).
In particular for large
factor models with many measurements, one would expect that the data contain ample information to estimate the factors $\facmk{\nfactrue}{t}$. However, this is true only,  if the information matrix $\trans{\facload_\nfactrue}  \Vare_\nfactrue ^{-1}  \facload _\nfactrue$  increases with $\dimy$, hence if most of the factor loadings are nonzero. For sparse factor models  many columns with quite a few  zero loadings  are present, leading to a low signal-to-noise ratio and ,consequently, to a poorly mixing Gibbs sampler, as illustrated in the left-hand panel in Figure~\ref{Boostadd} showing posterior draws of
$\trace{\trans{\facload_\nfactrue}  \Vare _\nfactrue^{-1}  \facload _\nfactrue}$  without boosting Step~(A)  for the exchange data to be discussed in Section~\ref{applicEx22}.

\begin{Figure3} %
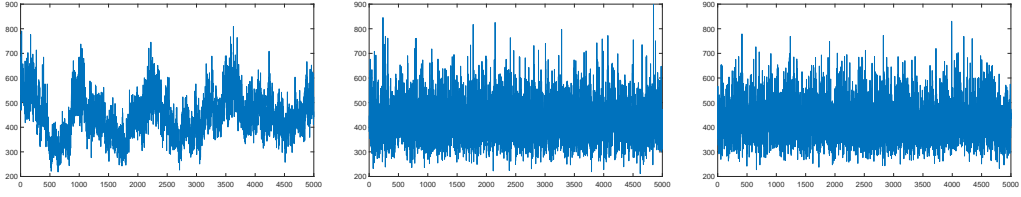
{Exchange rate data;  fractional prior. Posterior draws of  $\trace{\trans{\facload_\nfactrue}  \Vare _\nfactrue^{-1}  \facload _\nfactrue}$ without boosting (left-hand side), boosting through ASIS with the largest loading (in absolute values) in each nonzero column
serving as $\sqrt{\Psi_j}$ (middle) and boosting through MDA based on the inverted Gamma working prior $\Psi_j  \sim \Gammainv{1.5,1.5}$ (right-hand side).}{Boostadd}{ex22_noboost}{ex22_boost_asis}{ex22_boost_mda}{0.2}
\end{Figure3}  %

Hence,  boosting steps are essential  to obtain  efficient MCMC schemes for sparse factor models.
Several papers \citep{gho-dun:def,fru-lop:par,con-etal:bay_exp,pia-pap:bay} apply
marginal data augmentation (MDA) in the spirit of \citet{van-men:art}; others \citep{kas-etal:eff,fru-lop:spa} exploit
the ancillarity-suffiency interweaving strategy (ASIS) introduced by \citet{yu-men:cen}.
Some boosting strategies enhances mixing at the cost of changing the prior of the factor loading matrix $\facload_\nfactrue$ \citep{gho-dun:def}, however, this appears undesirable in a   variable selection context and is avoided by the boosting strategies applied in the present paper.

Boosting is based on moving from the CFA model  (\ref{fac1CFA}) where  $\facmk{\nfactrue}{t}  \sim  \Normult{\nfactrue}{\bfz,\identy{\nfactrue}}$ to an  expanded model
 with a more general prior:
\begin{eqnarray} \label{fac1px}
 \ym_t =  \tilde{\facload}_\nfactrue   \facmktilde{\nfactrue}{t} + \errorm_t, \quad \errorm_t \sim \Normult{\dimy}{\bfz,\Vare_\nfactrue},  \qquad
  \facmktilde{\nfactrue}{t} \sim  \Normult{\nfactrue}{\bfz,\Psiv},
\end{eqnarray}
where $\Psiv=\Diag{\Psi_1,\ldots,\Psi_{\nfactrue}}$ is a diagonal matrix. The relation between the two systems is given by the transformations
$ \facmktilde{\nfactrue}{t} = (\Psiv)^{1/2} \facmk{\nfactrue}{t} $ and $ \tilde{\facload}_\nfactrue = \facload_\nfactrue  (\Psiv)^{-1/2}$, where the nonzero elements in  $\tilde{\facload}_\nfactrue  $  have the same position as the nonzero elements in $\facload_\nfactrue$ and the sparsity matrix $\deltav_\nfactrue$ is not affected by the transformation.
 The main difference between MDA  and ASIS  lies in the choice of $\Psiv$.
 While  $\Psi_j $ is  sampled from a working prior for MDA, $\Psi_j $  is chosen in a deterministic fashion for ASIS, 
 see  Appendix~\ref{boost_frac} for details. 
For  illustration, Figure~\ref{Boostadd} shows considerable efficiency gain in  the posterior draws of  $\trace{\trans{\facload_\nfactrue}  \Vare _\nfactrue^{-1}  \facload _\nfactrue}$, when a boosting strategy such as ASIS (middle panel) or  MDA (right-hand panel) is applied. 

For hierarchical Gaussian priors with column specific shrinkage parameters,
such as (\ref{priorEXP2})  and (\ref{prior0ext}), we found it more useful to apply
  {\em column boosting}  and interweave $\taucol_j$
  into the state equation by choosing $\Psi_j= \taucol_j$. Column boosting and an additional boosting step %
which interweaves the global
shrinkage parameter $\tauglob$ into the prior of the shrinkage parameter
$\taucol_j$ is discussed in detail in Appendix~\ref{boost_frac}.

\section{Post-processing posterior draws} \label{secGLT}

Algorithm~\ref{Algo3} delivers posterior draws $(\deltav_\nfactrue, \facload_\nfactrue ,  \Vare_\nfactrue)$ in a CFA model with a  varying number $\nfactrue$  of active columns.
Instead of sampling these parameters without any constraints, our sampler imposes the (mild) constraint that the pivots (the first non-zero loading in each column) lie in different rows, ensuring that posterior draws of the loading matrix exhibit a UGLT structure.
While these draws are not
identified in the rigorous sense discussed in Section~\ref{secide},
the UGLT structure allows identification during post-processing.

We use
the \CountAR\ counting rule and  the algorithm of \citet{hos-fru:cov} to check for each draw $\deltav_\nfactrue$, if  the variance decomposition is unique, and remove all posterior draws that are not variance identified.  
In our experience, the fraction $M_V$ of variance identified draws is relatively high for reasonably chosen priors. A low fraction of  variance identified draws can be the result of poorly chosen shrinkage priors and
is a hint to consider another prior family. The fractional prior on the loadings $\beta_{ij}$, for instance, tends to deliver a low fraction of variance identified draws in high-dimensional factor models, 
see Section~\ref{secalpp}.

Many quantities can be inferred from the variance identified posteriors draws with varying factor dimension $\nfactrue $, such as  the marginal covariance matrix $\Vary=\facload_\nfactrue \facload_\nfactrue ^T +  \Vare_\nfactrue$ and  the idiosyncratic variances  $\idiov_1, \ldots, \idiov_\dimy$ appearing in the diagonal of  $\Vare_\nfactrue$. 
In addition,  overall sparsity  in terms of the number $d=\sum_{j=1}^\nfactrue  \sum_{i=1}^\dimy \delta_{ij}$ of nonzero elements in $\deltav_\nfactrue$,
 can be evaluated.
 Posterior draws  of $d$ are particularly  useful to check convergence and assessing efficiency of the MCMC sampler, as $d$ captures the ability of the sampler to move across (variance identified) UGLT factor models of different dimensions.
 Functionals of $\Vary$ and $\Vare_\nfactrue$,  such as the trace,  the  log determinant,
   and  $ \trans{\unit{\dimy}} \Vary^{-1}  \unit{\dimy} $ are further useful means to assess MCMC convergence.
Furthermore, for each variable  $y_{it}$, inference with respect to  the proportion of the variance
 explained by the common factors (also known as communalities $R^2_i$) is possible:
  \begin{eqnarray}
R^2_i =   \sum_{j=1}^{\nfactrue} R_{ij}^2,  %
\qquad  R_{ij}^2 = \frac{\load_{ij}^2 }{\sum_{l=1}^{\nfactrue} \load_{il}^2+\idiov_i} .
 \label{faccum}
\end{eqnarray}
Most importantly, variance identified draws
  are instrumental for  estimating  the number of factors from an overfitting factor model with
  cumulative shrinkage process priors.
As shown by \citet[Theorem~7]{fru-etal:whe}, the number $\nfactrue$ of nonzero columns in $\deltav_\nfactrue$ is equal to the factor dimension,  if  the variance decomposition is unique for $\nfactrue$. In this way, posterior draws of
 $\nfactrue$  can be used 
 for variance identified factor loading matrices
 to draw posterior
 inference on the factor dimension $\nfactrue$.
 The mode $\tilde{r}$ of the posterior distribution $p(\nfactrue|\ym)$, e.g., %
 can be used to estimate the factor dimension.

Due to the UGLT structure imposed on $\facload_\nfactrue $, %
 rotational invariance  reduces for variance identified draws to sign and column switching \citep[Theorem~1]{fru-etal:whe} and $\facloadtrue$ and $\deltavlam$ are easily recovered.
First,   the columns of $\deltav_\nfactrue$ and $\facload_\nfactrue $ are ordered such that the pivots $\lm_\nfactrue=(l_1, \ldots , l_\nfactrue)$
  obey $l_1 < \ldots < l_\nfactrue$; i.e.\ $\deltavlam =\deltav_\nfactrue \bP _{\rho}$.
   Then,   the sign of all columns in $\facload_\nfactrue \bP _{\rho}$ is switched if the leading element
   is negative; i.e.\ $\facloadtrue= \facload_\nfactrue   \bP _{\rho} \bP _{\pm}  $.
  In addition,  the  factors $\facmk{\nfactrue}{t}$   are reordered
  through $ \trans{\bP _{\pm}} \trans{\bP _{\rho}} \facmk{\nfactrue}{t}$ for  $t=1,\ldots,T$.
  Finally, $\bP _{\rho}$ is used to reorder the draws of  $\hypv_\nfactrue=(\tau_1, \ldots,
 \tau_\nfactrue) $ and any local and column-specific shrinkage parameters in a hierarchical Gaussian shrinkage prior.

  The posterior draws of  $\deltavlam$  are exploited in various ways.
    The
    highest probability model (HPM),~i.e.\ the indicator matrix  $\deltavlam_H$ visited most often,
    its frequency $p_{H}$  (an estimator of the posterior probability of the HPM), 
    its factor dimension $r_H$,
    its model size $d_H$,  and its sequence of pivots $\lm  ^H$  are  of interest.
For each draw of $\deltavlam$,  the ordered
  pivots $l_1 < \ldots < l_\nfactrue$  are draws  from  the marginal
  posterior distribution $p(l_1, \ldots , l_\nfactrue|\ym)$,  allowing
 additional posterior inference w.r.t.\  $\lm_\nfactrue $.
   First of all, the posterior probability that a specific feature (row)  serves as a
pivot, i.e.\ $\Prob{i \in \lm_r|\ym}$ for $i=1, \ldots, m$  is of interest.
  Second, the sequence of pivots
    $\lm  ^\star=(l_1^\star, \ldots, l_{\rstar}^\star)$ visited most often is determined
     together with its frequency $p_{L}$ which reflects posterior uncertainty with respect to
     choosing the pivots.  The number  $\rstar$ of elements in $ \lm  ^\star$ provides yet another estimator of the number of factors.
If the frequencies $p_{H}$ and $p_{L}$ are small,
  the estimator $\rstar$ and $r_H$ might not coincide with the posterior mode $\tilde{r}$ and
  $\lm  ^H$  will different from $\lm  ^\star$. %

  To estimate the factor loading matrix %
  for a chosen number of factors $r$, 
  Bayesian model averaging is performed conditionally on an estimator  $\hat{\lm}_r $ of the pivots.
    Averaging over  all
  posterior draws 
  with these specific pivots  avoids column switching and provides an estimate  of  the factor loading matrix $\facloadtrue$
and the marginal inclusion probabilities
$\Prob{\delta^\Lambda_{ij}=1|\ym, \hat{\lm}_r}$ for all elements of the corresponding sparsity matrix. %
     The median probability model (MPM) $\deltavlam_M$
   is  obtained by setting each indicator to one
 whenever $\Prob{\delta_{ij}^\Lambda=1|\ym, \hat{\lm}_r}\geq 0.5$. %
As for  any Bayesian approach, the reliability of any of these estimators  might depend on how informative the data are. In settings where the data are not very informative, the chosen degree of sparsity in the prior  might exhibit considerable influence on the result.

  \section{Applications}  \label{secalpp}

  \subsection{Sparse factor analysis for exchange rate data}  \label{applicEx22}

As a first application, we analyze 
log returns from $m = 22$  exchange rates with respect to the Euro, observed for $T=96$ months.\footnote{The data was obtained from
the European Central Bank’s Statistical Data Warehouse and ranges from January 3, 2000
to December 3, 2007. It contains the 22  exchange rates listed in Table~\ref{abbrev} in Appendix~\ref{Dataex}  from which we derived monthly returns based on the first trading day in a month. The data are demeaned and standardized.}%
 A sparse EFA with UGLT structure is fitted  with
$\nfac=10$ equals the upper bound (\ref{kboundr}).
Regarding  the %
prior  on the sparsity matrix $\deltav_\nfac$, we consider the  1PB prior (\ref{pri1PB})   and the 2PB prior (\ref{prialtTeh}), where $\alphaIBP \sim \Gammad{6,3}$ and $\betaIBP\sim \Gammad{6,6}$ 
for the 2PB prior. 
We combine  the   fractional prior  (\ref{priorfrac}) as slab distribution for the non-zero loadings in $\facload_\nfactrue$  with the  prior   (\ref{priorsiid})  on the idiosyncratic variances $\idiov_1, \ldots, \idiov_\dimy$,  where  $c_0=2.5$ and  $\widehat{\Vary^{-1}}$ is estimated  from (\ref{Varyhat}) with $\nu_o =3 $ and   $\Sm_o=\identy{\dimy}$.

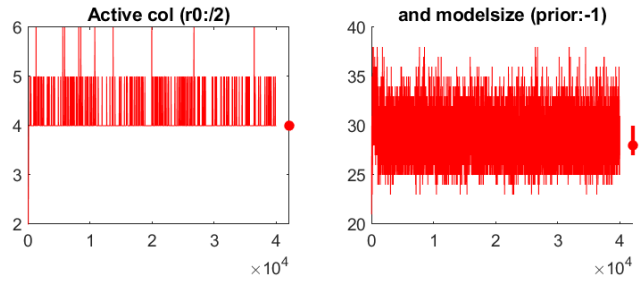
\begin{Figurepng}{Exchange rate data;  posterior draws  of the factor dimension $\nfactrue$  (left-hand side) and  model size $d$ (right-hand side) including burn-in, starting from  $\nfactrue=2$ and $\nsp=3$.} 
{fig_ken_0}{euro_ex_GLTRJ_bN_ALL}{0.7} \end{Figurepng}

\begin{Tabelle}{Exchange rate data; %
posterior distribution $p(\nfactrue|\ym)$ of the number of factors %
 and posterior mean $\Ew{d | \ym}$ of model size (based on $100p_V$ percent variance identified draws).}{ken_tab1}
{ \small \begin{tabular}{lcccccccccc} \hline
 &  & \multicolumn{6}{c}{$p(\nfactrue|\ym)$} &   \\ \cline{4-9}
  &  $\Ew{\alphaIBP|\ym}$ &$\Ew{\betaIBP|\ym}$  &  0-2  & 3 & 4 & 5 & 6 & 7-10&  $100 \cdot p_V$ & $\Ew{d | \ym}$ \\   \hline
1PB  & 2.0 & - &       0  &  0.128 &   \textbf{0.867}  &  0.005 &   $\approx 0$ & 0&   93.3 & 28 \\
2PB  &  2.0 & 1.1 &    0  &  0  &  
\textbf{0.988} &   0.012 &  0  & 0&    95.2  & 28 \\
 \hline
\end{tabular}

\footnotesize{Note: non-zero probabilities  smaller
than $10^{-3}$ are indicated by $\approx 0$.}}
\end{Tabelle}

   \begin{Tabelle}{Exchange rate data;  posterior probability of the event $\Prob{q_i=0|\ym}$, where  $q_i$ is the row sum of  $\deltav_\nfactrue$,  for various currencies.}{uncorex}
{\small \begin{tabular}{lccccccc} \hline
& \multicolumn{7}{c}{$\Prob{q_i=0|\ym}$}\\ \cline{2-8}
    Currency & CHF  & CZK & MXN& NZD & RON& RUB & remaining  \\
  \hline
1PB prior & 0.87 &  0.73 &  0.81 & 0.48 & 0.62 & 0.62   &  0   \\
2PB prior  &      0.88 &  0.76 & 0.82 & 0.49 & 0.62 & 0.62  &  0   \\
\hline
\end{tabular}}
\end{Tabelle}

Algorithm~\ref{Algo3} is run for $M=40{,}000$ %
after a burn-in  of %
$20{,}000$ draws.   
As discussed in Section~\ref{mcmc}, this sampler navigates in
the space of all unordered GLT structures with an unknown number of nonzero columns and unknown pivots without forcing any further constraint.
To verify convergence, independent MCMC chains are started  with, respectively, $\nfactrue=2$
and $\nfactrue=7$, and $\nsp=3$ spurious columns.
The  sampler shows good mixing across  models of different dimension, in particular for the 2PB prior, 
  with an  inefficiency factor of  roughly  5. For illustration, Figure~\ref{fig_ken_0} shows  all posterior
 draws of $\nfactrue$  and the model size $d$, including burn-in, for the first run  under the 2PB prior.

   \begin{Tabelle}{Exchange rate data; %
   total number of visited models $N_v$;
 frequency  $p_H$ (in percent),  pivots $\lm _\nfactrue ^H$ and model size $d_H$ of the HPM; %
 pivots $\lm  ^\star$ visited most often,  corresponding frequency  $p_L$ (in percent) and corresponding number of factors $\rstar$;   pivots $\lm _H$, model size $d_H$ 
 and frequency  $p_H$ (in percent) of the HPM;
 model size $d_M$ of the MPM.}{ken_tab3}
{\small \begin{tabular}{ccccccccc}
 \hline Prior               & $N_v$   &   $100p_H$  & $\lm_H$ & $d_H$  & $\lm^{\star}$ &  $100p_L$  & $\rstar$ & $d_M$  \\
 \hline
1PB prior  & 16508 & 4.5 & (1,2,5,7) &  26& (1,2,5,7) & 84.8 &4 & 26 \\
2PB prior  & 11933  &  5.0 & (1,2,5,7) & 26 & (1,2,5,7) & 91.0 & 4 & 26\\
\hline
\end{tabular}}
\end{Tabelle}

As outlined in Section~\ref{secGLT}, we resolve identification during post-processing.
First, we screen for variance identified draws. The  fraction $p_V$ %
 of  variance identified  draws, reported in
Table~\ref{ken_tab1}, is very high for both EPS priors.
For the variance identified draws,
 the number $\nfactrue$ of columns of the sparsity matrix $\deltav_\nfactrue$ in the CFA model is regarded as a draw of the number $\nfactrue$ of factors under the specific prior choice.
 Table~\ref{ken_tab1} reports the posterior distribution $p(\nfactrue|\ym)$  for both   ESP priors. 
 For the 2PB prior,
 this posterior  is highly concentrated at a four factors. Also under the 1PB prior, 
 the posterior mode equals 4, but also 
 three factors receive some posterior probability.
The indicator matrix $\deltav_r$ is pretty sparse, with an average posterior  model size 
of 28.

Furthermore, the variance identified draws are used to explore if some  measurements are uncorrelated with the remaining measurements. This is investigated in Table~\ref{uncorex}
 through the posterior probability  $\Prob{q_i=0|\ym}$, where  $q_i$ is the row sum of  $\deltav_\nfactrue$. Regardless of the chosen prior, the  Swiss franc (CHF),  the Mexican peso (MXN) and the Czech koruna (CZK) have considerable probability  to be uncorrelated with the rest, while the situation is less clear for  the New Zealand dollar (NZD), the Romania fourth leu (RON), and the  Russian ruble (RUB). The remaining currencies are clearly correlated.

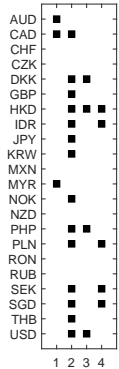
\begin{Figure}{Exchange rate data;   sparsity matrix  $\deltav_4$ corresponding both to the HPM and the MPM which are identical for the 1PB and the 2PB prior.}{fig_deltaplot}{fig_delta}{0.4} \end{Figure}

To proceed with identification for all variance identified draws,  rotation indeterminacy is resolved by ordering the pivots (and all column-specific variables) such that $l_1 < \ldots < l_\nfactrue$ and an ordered GLT structure is imposed. Several posterior summaries for the ordered GLT draws are reported in Table~\ref{ken_tab3}.
  For both ESP priors, $\lm^\star=\lm ^H=(1,2,5,7)$ is the
most likely sequence of pivots.
As a  final step, all variance identified, ordered GLT draws where the pivots coincide with   $\lm ^\star=\lm ^H=(1, 2,5,7)$ are used  for both ESP priors to identify  
the marginal inclusion probabilities
$\Prob{\delta_{ij}=1|\ym, \lm ^\star}$, the corresponding MPM,
its  model size $d_M$ %
and the factor loading matrix $\facloadtrue$ for a 4-factor model.  Sign switching in the posterior draws is resolved  by imposing
 the constraint $\loadtrue_{11} >0$, $\loadtrue_{22} >0$, $\loadtrue_{53} >0$,   and  $\loadtrue_{74} >0$ on $\facloadtrue$. 
Both ESP priors yield the same  
 MPM which coincides with the sparsity matrix of both HPMs, see  Figure~\ref{fig_deltaplot} for illustration. Further results are reported in  Appendix~\ref{Dataex}. 
The resulting model indicates considerable sparsity, with  many factor loadings being shrunk toward zero.  Factor~2 is a  common factor  among  the correlated currencies, while the remaining factors are three group specific, for the most part dedicated  factors.

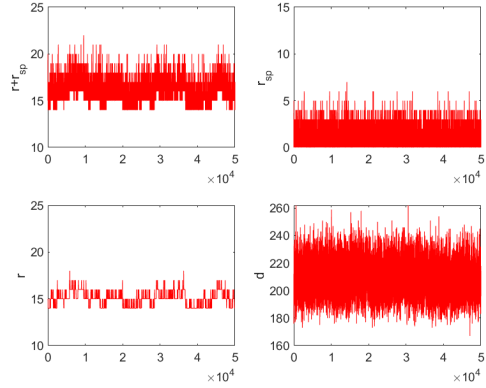
\begin{Figurepng}{NYSE data;  posterior draws  of 
 the total number of non-zero columns $\nfactrue+\nsp$ (top left),
 the number of spurious columns $\nsp$ (top right), the extracted number of factors $\nfactrue$ (bottom left), and the model dimension $d$ (bottom right).} 
{fig_nyse_0}{BA_NYSE_r}{0.5} \end{Figurepng}

 \subsection{Sparse factor analysis for NYSE stock returns}

As a second application, we consider  monthly log returns from $m=63$
firms from the NYSE
observed for $T=247$ months from February 1999 till August 2019.\footnote{The top 150 companies (as of September 13, 2019) listed on the NYSE were downloaded from Bloomberg on September 13, 2019.
Since many of the companies entered the NYSE after 1990, we picked
February 1999  as a starting date and, after removing all companies that were founded later, we use monthly return (determined on the last trading day in each month) for the 103 remaining companies listed on the NYSE for all $T = 247$ observation periods till August 2019.
For our case study, we consider the 63 firms belonging to the following  five sectors: basic industries (1-9),
non-durable consumer goods (8-17), 
energy (18-27), finance (28-45) and 
health care (46-63). The data are ordered according to industry and are The data are demeaned and standardized.}
An EFA model with UGLT structure is fitted with
$\nfac=31$ being equal to the upper bound  (\ref{kboundr}).
Regarding  the ESP prior  on the sparsity matrix $\deltav_\nfac$, we consider the 2PB prior (\ref{prialtTeh}) with $\alphaIBP \sim \Gammad{6,12}$ and $\betaIBP\sim \Gammad{6,6}$. 
As in Section~\ref{applicEx22}, we tried to apply a fractional prior as slab distribution,
however the fraction of variance identified posterior draws was extremely low (less than 1\%). Hence, the hierarchically structured Gaussian shrinkage prior
(\ref{prior0ext}) is chosen in the slab.  
To introduce 
 aggressive shrinkage for the factor loadings,   the local scaling parameters are assumed to follow 
 a triple gamma prior with $\aloc=\cloc=0.2$, i.e.
 $\tauloc_{ij} \sim \Fd{0.4,0.4}$.
 Also  the column specific  shrinkage parameters follow a triple gamma prior, $\taucol_j \sim \Fd{5,5}$, whereas the global shrinkage parameter follows an inverse gamma distribution, $\tauglob \sim \Gammainv{10,50}$. 
   Regarding the idiosyncratic variances, we choose the prior $\idiov_i \sim \Gammainv{2.5,1.5}$.
 The  fraction  of  variance identified MCMC draws under this prior is roughly 20\%.

\begin{Tabelle}{NYSE data; posterior distribution $p(\nfactrue|\ym)$ of the number of factors %
 and the  mean %
 and the quartiles (in parenthesis) of the posterior  distribution $ p(d | \ym)$  of model size ($100p_V$ percent variance identified draws).}{tabNYSEDH2}
{ \small \begin{tabular}{cccccccccc} \hline
  & \multicolumn{7}{c}{$p(\nfactrue|\ym)$} &   \\ %
 \cline{3-9}
   $\Ew{\alphaIBP|\ym}$ &$\Ew{\betaIBP|\ym}$  &  0-12  & 13 & 14 & 15 & 16 & 17& 18-31
  & $ p(d | \ym)$ \\   \hline
2.5 & 0.8 &    0  &  $\approx 0$  &  
 0.413  &  \textbf{0.488} &  0.098  & 0.001&
0
& 203 (210,218) \\
 \hline
\end{tabular}

\footnotesize{Note: non-zero probabilities  smaller
than $10^{-3}$ are indicated by $\approx 0$.}}
\end{Tabelle}

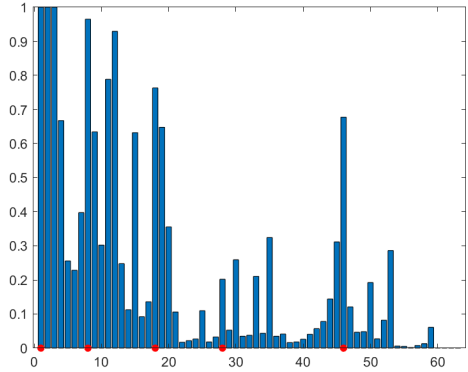
  \begin{Figurepng}{NYSE data;  posterior probability $\Prob{i \in \lm_r|\ym}$ that a specific firm serves as pivot; the red dots indicate the first firm in each sector.} 
{posli}{BA_NYSE_li}{0.5} \end{Figurepng}

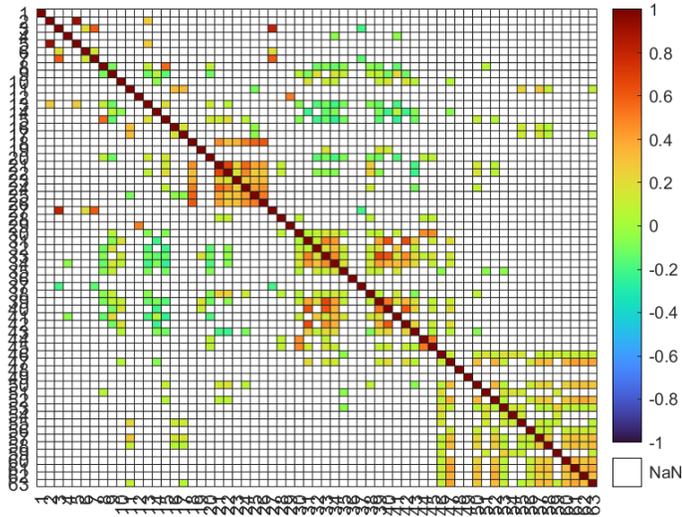
\begin{Figurepng}{NYSE data;  estimated marginal correlation matrix $\Ew{\Vary^\star|\ym}$, where $\Omega^\star_{i \ell}=\Corr{(y_{it}- \Lambda_{i1} f_{1t}) (y_ {\ell t} - \Lambda_{\ell 1} f_{1t} )}$.} {NYSE:corr2}{BA_NYSE_Corr2}{0.7} \end{Figurepng}

Algorithm~\ref{Algo3} was applied  to obtain  $M=50{,}000$ posterior draws  after a burn-in of $40{,}000$ draws, starting with  $\nfactrue=10$ factors 
and $\nsp=3$ spurious columns.
  The  MCMC scheme shows relatively good mixing, despite the high dimensionality, as illustrated by Figure~\ref{fig_nyse_0} showing    posterior draws of the total number of non-zero columns, $\nfactrue + \nsp$, the number of spurious columns $\nsp$, the extracted number of factors $\nfactrue$, and the model dimension $d$.
As shown in Table~\ref{tabNYSEDH2}, the posterior distribution $p(\nfactrue|\ym)$ 
derived from the variance identified draws yields 
a posterior mode of $\tilde{\nfactrue}=15$, but also 14 factors receive considerable posterior evidence.

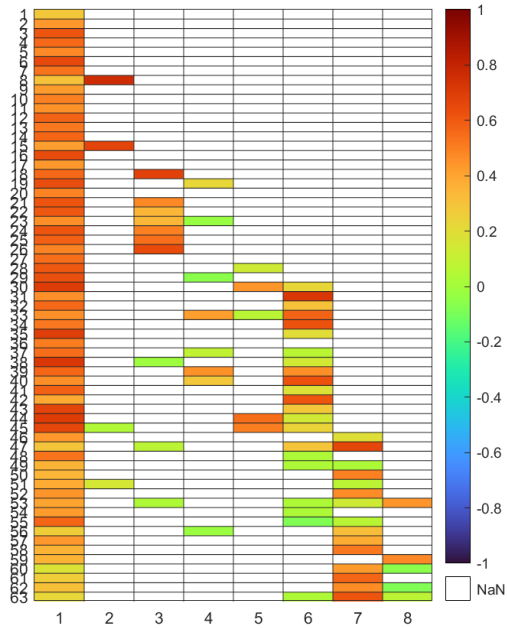
 \begin{Figurepng}{NYSE data; 
factors with pivots equal to
(1, 8, 18, 19, 28, 30, 46, 53)}{NYSE:cfac}{BA_NYSE_Load}{0.6} \end{Figurepng}

For further inference, an ordered GLT structure is imposed on all variance identified draws with $\nfactrue=15$. %
Since each draw of $\facloadtrue$ turned out to have a different shrinkage matrix $\deltav^\Lambda$,  choosing a unique  set of pivots is challenging.
In Figure~\ref{posli}, the posterior probability $\Prob{i \in \lm_r|\ym}$ that a specific firm serves as a pivot is displayed. While the first three factors use the pivots 1, 2, and 3 (as a PLT structure would do),  we find that the remaining factors clearly exhibit a GLT structure and  the first firm listed in a specific sectors (indicated by a red dot) typically serves as a  pivot for new factor.

Since the pivot of the first factor is equal to 1 for all posterior draws, we can estimate the first column of the indicator matrix and the corresponding factor loadings as the average of all posterior draws. This factor is a market factor that loads on all 63 firms,  see also Figure~\ref{NYSE:cfac}.
The remaining 14 factors mainly capture industry specific correlations as well as cross-sectional correlations between specific firms, see the estimated marginal correlation matrix $\Vary^\star$
that remains after extracting the first factor  in Figure~\ref{NYSE:corr2}.  
For further illustration, factors with pivots equal to
(1, 8, 18, 19, 28, 30, 46, 53) are extracted from all
GLT draws with  $\nfactrue=15$, where the sequence of
pivots $\lm_\nfactrue$ contains these pivots. Beyond the market factor 1, the sector specific  factors 3, 6 and 7 mainly load on firms in, respectively, the energy, the finance and the health care sector. 
The remaining factors are weak factors with very sparse loadings.

\section{Concluding remarks} \label{secconcluse}

We have estimated (from a Bayesian viewpoint) a fairly important and highly implemented class of sparse factor models when the number of common factors is unknown.  Our framework leads to a natural, efficient and simultaneous coupling of model estimation and selection on one hand and model identification and  rank estimation (number of factors) on the other hand.  More precisely, by combining point-mass mixture priors with overfitting sparse factor modelling, in an unordered generalised lower triangular loadings representation \citep{fru-etal:whe}, we obtain posterior summaries regarding factor loadings, common factors as well as the factor dimension
 via postprocessing draws from our highly efficient and customised MCMC scheme.  %

The new framework is readily available for some straightforward extensions.  Relatively immediate extensions are
           (i) idiosyncratic errors following Student's $t$-distributions or more general Gaussian mixtures
            and (ii) dynamic sparse factor models with stationary common factors; both extensions commonly found in econometrics applications, see e.g.\ the recent papers    by \citet{pia-pap:bay} and     \citet{kau-sch:bay}.
 A further interesting extension would be to design a prior on the sparsity matrix that a priori distinguishes between pervasive factors that loads on (nearly) all measurements,
 group specific factors that load on selected measurements
 and factors that mainly capture (weak) cross-sectional heterogeneity which is not built into the basic factor model. Such approximate factor models are very popular in non-Bayesian factor analysis, see e.g.~\citet{cha-rot:arb} and \citet{bai-ng:det2002} and would deserve more attention from the Bayesian community. However, we leave this interesting idea for future research.

\bibliographystyle{chicago}
\bibliography{references}

\begin{thebibliography}{}

\bibitem[\protect\citeauthoryear{Akaike}{Akaike}{1987}]{aka:fac}
Akaike, H. (1987).
\newblock Factor analysis and {AIC}.
\newblock {\em Psychometrika\/}~{\em 52}, 317--332.

\bibitem[\protect\citeauthoryear{Anderson}{Anderson}{2003}]{and:int}
Anderson, T.~W. (2003).
\newblock {\em An Introduction to Multivariate Statistical Analysis\/} (3 ed.).
\newblock Chichester: Wiley.

\bibitem[\protect\citeauthoryear{Anderson and Rubin}{Anderson and
  Rubin}{1956}]{and-rub:sta}
Anderson, T.~W. and H.~Rubin (1956).
\newblock Statistical inference in factor analysis.
\newblock In {\em Proceedings of the Third Berkeley Symposium on Mathematical
  Statistics and Probability}, Volume~V, pp.\  111--150.

\bibitem[\protect\citeauthoryear{A{\ss}mann, Boysen-Hogrefe, and
  Pape}{A{\ss}mann et~al.}{2016}]{ass-etal:bay}
A{\ss}mann, C., J.~Boysen-Hogrefe, and M.~Pape (2016).
\newblock {B}ayesian analysis of static and dynamic factor models: {An} ex-post
  approach toward the rotation problem.
\newblock {\em Journal of Econometrics\/}~{\em 192}, 190--206.

\bibitem[\protect\citeauthoryear{Bai and Ng}{Bai and Ng}{2002}]{bai-ng:det2002}
Bai, J. and S.~Ng (2002).
\newblock Determining the number of factors in approximate factor models.
\newblock {\em Econometrica\/}~{\em 70}, 191--221.

\bibitem[\protect\citeauthoryear{Bai and Ng}{Bai and Ng}{2013}]{bai-ng:pri}
Bai, J. and S.~Ng (2013).
\newblock Principal components estimation and identification of static factors.
\newblock {\em Journal of Econometrics\/}~{\em 176}, 18--29.

\bibitem[\protect\citeauthoryear{Bartholomew}{Bartholomew}{1987}]{bar:lat_fac}
Bartholomew, D.~J. (1987).
\newblock {\em Latent Variable Models and Factor Analysis}.
\newblock London: Charles Griffin.

\bibitem[\protect\citeauthoryear{Bhattacharya and Dunson}{Bhattacharya and
  Dunson}{2011}]{bha-dun:spa}
Bhattacharya, A. and D.~Dunson (2011).
\newblock Sparse {B}ayesian infinite factor models.
\newblock {\em Biometrika\/}~{\em 98}, 291--306.

\bibitem[\protect\citeauthoryear{Boivin and Ng}{Boivin and
  Ng}{2006}]{boi-ng:are}
Boivin, J. and S.~Ng (2006).
\newblock Are more data always better for factor analysis?
\newblock {\em Journal of Econometrics\/}~{\em 132}, 169--194.

\bibitem[\protect\citeauthoryear{Cadonna, Fr{\"u}hwirth-Schnatter, and
  Knaus}{Cadonna et~al.}{2020}]{cad-etal:tri}
Cadonna, A., S.~Fr{\"u}hwirth-Schnatter, and P.~Knaus (2020).
\newblock Triple the gamma -- {A} unifying shrinkage prior for variance and
  variable selection in sparse state space and {TVP} models.
\newblock {\em Econometrics\/}~{\em 8}, 20.

\bibitem[\protect\citeauthoryear{Carvalho, Chang, Lucas, Nevins, Wang, and
  West}{Carvalho et~al.}{2008}]{car-etal:hig}
Carvalho, C.~M., J.~Chang, J.~E. Lucas, J.~Nevins, Q.~Wang, and M.~West (2008).
\newblock High-dimensional sparse factor modeling: Applications in gene
  expression genomics.
\newblock {\em Journal of the American Statistical Association\/}~{\em 103},
  1438--1456.

\bibitem[\protect\citeauthoryear{Chamberlain and Rothschild}{Chamberlain and
  Rothschild}{1983}]{cha-rot:arb}
Chamberlain, G. and M.~Rothschild (1983).
\newblock Arbitrage, factor structure, and mean-variance analysis on large
  asset markets.
\newblock {\em Econometrica\/}~{\em 51}, 1281--1304.

\bibitem[\protect\citeauthoryear{Chan, Leon-Gonzalez, and Strachan}{Chan
  et~al.}{2018}]{cha-etal:inv}
Chan, J., R.~Leon-Gonzalez, and R.~W. Strachan (2018).
\newblock Invariant inference and efficient computation in the static factor
  model.
\newblock {\em Journal of the American Statistical Association\/}~{\em 113},
  819--828.

\bibitem[\protect\citeauthoryear{Conti, Fr\"uhwirth-Schnatter, Heckman, and
  Piatek}{Conti et~al.}{2014}]{con-etal:bay_exp}
Conti, G., S.~Fr\"uhwirth-Schnatter, J.~J. Heckman, and R.~Piatek (2014).
\newblock Bayesian exploratory factor analysis.
\newblock {\em Journal of Econometrics\/}~{\em 183}, 31--57.

\bibitem[\protect\citeauthoryear{Durante}{Durante}{2017}]{dur:not}
Durante, D. (2017).
\newblock A note on the multiplicative gamma process.
\newblock {\em Statistics and Probability Letters\/}~{\em 122}, 198--204.

\bibitem[\protect\citeauthoryear{Fan, Fan, and Lv}{Fan
  et~al.}{2008}]{fan-etal:hig_je}
Fan, J., Y.~Fan, and J.~Lv (2008).
\newblock High dimensional covariance matrix estimation using a factor model.
\newblock {\em Journal of Econometrics\/}~{\em 147}, 186--197.

\bibitem[\protect\citeauthoryear{Forni, Giannone, Lippi, and Reichlin}{Forni
  et~al.}{2009}]{for-etal:ope}
Forni, M., D.~Giannone, M.~Lippi, and L.~Reichlin (2009).
\newblock Opening the black box: {S}tructural factor models with large cross
  sections.
\newblock {\em Econometric Theory\/}~{\em 25}, 1319--1347.

\bibitem[\protect\citeauthoryear{Foster and George}{Foster and
  George}{1994}]{fos-geo:ris}
Foster, D.~P. and E.~I. George (1994).
\newblock The risk inflation criterion for multiple regression.
\newblock {\em The Annals of Statistics\/}~{\em 22}, 1947--1975.

\bibitem[\protect\citeauthoryear{Fr{\"u}hwirth-Schnatter}{Fr{\"u}hwirth-Schnatter}{2022}]{fru:gen}
Fr{\"u}hwirth-Schnatter, S. (2022).
\newblock Generalized cumulative shrinkage process priors with applications to
  sparse bayesian factor analysis.

\bibitem[\protect\citeauthoryear{Fr{\"u}hwirth-Schnatter, Hosszejni, and
  Lopes}{Fr{\"u}hwirth-Schnatter et~al.}{2022}]{fru-etal:whe}
Fr{\"u}hwirth-Schnatter, S., D.~Hosszejni, and H.~Lopes (2022).
\newblock When it counts - {E}conometric identification of factor models based
  on {GLT} structures.
\newblock {\em ArXiv\/}~{\em forthcoming soon}.

\bibitem[\protect\citeauthoryear{Fr{\"u}hwirth-Schnatter and
  Lopes}{Fr{\"u}hwirth-Schnatter and Lopes}{2010}]{fru-lop:par}
Fr{\"u}hwirth-Schnatter, S. and H.~Lopes (2010).
\newblock {Parsimonious {B}ayesian Factor Analysis when the Number of Factors
  is Unknown}.
\newblock Research report, Booth School of Business, University of Chicago.

\bibitem[\protect\citeauthoryear{Fr{\"u}hwirth-Schnatter and
  Lopes}{Fr{\"u}hwirth-Schnatter and Lopes}{2018}]{fru-lop:spa}
Fr{\"u}hwirth-Schnatter, S. and H.~Lopes (2018).
\newblock {Sparse {B}ayesian Factor Analysis when the Number of Factors is
  Unknown}.
\newblock {\em arXiv\/}~{\em 1804.04231}.

\bibitem[\protect\citeauthoryear{Fr{\"u}hwirth-Schnatter and
  T{\"u}chler}{Fr{\"u}hwirth-Schnatter and T{\"u}chler}{2008}]{fru-tue:bay}
Fr{\"u}hwirth-Schnatter, S. and R.~T{\"u}chler (2008).
\newblock Bayesian parsimonious covariance estimation for hierarchical linear
  mixed models.
\newblock {\em Statistics and Computing\/}~{\em 18}, 1--13.

\bibitem[\protect\citeauthoryear{Fr{\"u}hwirth-Schnatter and
  Wagner}{Fr{\"u}hwirth-Schnatter and Wagner}{2010}]{fru-wag:sto}
Fr{\"u}hwirth-Schnatter, S. and H.~Wagner (2010).
\newblock Stochastic model specification search for {G}aussian and partially
  non-{G}aussian state space models.
\newblock {\em Journal of Econometrics\/}~{\em 154}, 85--100.

\bibitem[\protect\citeauthoryear{Geweke and Singleton}{Geweke and
  Singleton}{1980}]{gew-sin:int}
Geweke, J.~F. and K.~J. Singleton (1980).
\newblock Interpreting the likelihood ratio statistic in factor models when
  sample size is small.
\newblock {\em Journal of the American Statistical Association\/}~{\em 75},
  133--137.

\bibitem[\protect\citeauthoryear{Geweke and Zhou}{Geweke and
  Zhou}{1996}]{gew-zho:mea}
Geweke, J.~F. and G.~Zhou (1996).
\newblock Measuring the pricing error of the arbitrage pricing theory.
\newblock {\em Review of Financial Studies\/}~{\em 9}, 557--587.

\bibitem[\protect\citeauthoryear{Ghahramani, Griffiths, and Sollich}{Ghahramani
  et~al.}{2007}]{gha-etal:bay}
Ghahramani, Z., T.~L. Griffiths, and P.~Sollich (2007).
\newblock Bayesian nonparametric latent feature models (with discussion and
  rejoinder).
\newblock In J.~M. Bernardo, M.~J. Bayarri, J.~O. Berger, A.~P. Dawid,
  D.~Heckerman, A.~F.~M. Smith, and M.~West (Eds.), {\em Bayesian Statistics
  8}, pp.\  ADD--ADD. Oxford: Oxford University Press.

\bibitem[\protect\citeauthoryear{Ghosh and Dunson}{Ghosh and
  Dunson}{2009}]{gho-dun:def}
Ghosh, J. and D.~B. Dunson (2009).
\newblock Default prior distributions and efficient posterior computation in
  {B}ayesian factor analysis.
\newblock {\em Journal of Computational and Graphical Statistics\/}~{\em 18},
  306--320.

\bibitem[\protect\citeauthoryear{Griffiths and Ghahramani}{Griffiths and
  Ghahramani}{2006}]{gri-gha:inf}
Griffiths, T.~L. and Z.~Ghahramani (2006).
\newblock Infinite latent feature models and the {I}ndian buffet process.
\newblock In Y.~Weiss, B.~Sch\"olkopf, and J.~Platt (Eds.), {\em Advances in
  Neural Information Processing Systems}, Volume~18, pp.\  475--482. Cambridge,
  MA: MIT Press.

\bibitem[\protect\citeauthoryear{Hosszejni and
  Fr{\"u}hwirth-Schnatter}{Hosszejni and
  Fr{\"u}hwirth-Schnatter}{2022}]{hos-fru:cov}
Hosszejni, D. and S.~Fr{\"u}hwirth-Schnatter (2022).
\newblock Cover it up! {B}ipartite graphs uncover identifiability in sparse
  factor analysis.
\newblock {\em arXiv\/}~{\em 2211.00671}.

\bibitem[\protect\citeauthoryear{J\"{o}reskog}{J\"{o}reskog}{1967}]{joe:som}
J\"{o}reskog, K.~G. (1967).
\newblock Some contributions to maximum likelihood factor analysis.
\newblock {\em Psychometrika\/}~{\em 32}, 443--482.

\bibitem[\protect\citeauthoryear{J\"{o}reskog}{J\"{o}reskog}{1969}]{joe:gen}
J\"{o}reskog, K.~G. (1969).
\newblock A general approach to confirmatory maximum likelihood factor
  analysis.
\newblock {\em Psychometrika\/}~{\em 34}, 183--202.

\bibitem[\protect\citeauthoryear{Kastner}{Kastner}{2019}]{kas:spa}
Kastner, G. (2019).
\newblock Sparse {B}ayesian time-varying covariance estimation in many
  dimensions.
\newblock {\em Journal of Econometrics\/}~{\em 210}, 98--115.

\bibitem[\protect\citeauthoryear{Kastner, Fr\"{u}hwirth-Schnatter, and
  Lopes}{Kastner et~al.}{2017}]{kas-etal:eff}
Kastner, G., S.~Fr\"{u}hwirth-Schnatter, and H.~F. Lopes (2017).
\newblock Efficient {B}ayesian inference for multivariate factor stochastic
  volatility models.
\newblock {\em Journal of Computational and Graphical Statistics\/}~{\em 26},
  905--917.

\bibitem[\protect\citeauthoryear{Kaufmann and Schuhmacher}{Kaufmann and
  Schuhmacher}{2017}]{kau-sch:ide}
Kaufmann, S. and C.~Schuhmacher (2017).
\newblock Identifying relevant and irrelevant variables in sparse factor
  models.
\newblock {\em Journal of Applied Econometrics\/}~{\em 32}, 1123--1144.

\bibitem[\protect\citeauthoryear{Kaufmann and Schuhmacher}{Kaufmann and
  Schuhmacher}{2019}]{kau-sch:bay}
Kaufmann, S. and C.~Schuhmacher (2019).
\newblock Bayesian estimation of sparse dynamic factor models with
  order-independent and ex-post identification.
\newblock {\em Journal of Econometrics\/}~{\em 210}, 116--134.

\bibitem[\protect\citeauthoryear{Koopmans and Reiers{\o}l}{Koopmans and
  Reiers{\o}l}{1950}]{koo-rei:ide}
Koopmans, T.~C. and O.~Reiers{\o}l (1950).
\newblock The identification of structural characteristics.
\newblock {\em The Annals of Mathematical Statistics\/}~{\em 21}, 165--181.

\bibitem[\protect\citeauthoryear{Lee and Song}{Lee and
  Song}{2002}]{lee-son:bay}
Lee, S.-Y. and X.-Y. Song (2002).
\newblock {B}ayesian selection on the number of factors in a factor analysis
  model.
\newblock {\em Behaviormetrika\/}~{\em 29}, 23--39.

\bibitem[\protect\citeauthoryear{Legramanti, Durante, and Dunson}{Legramanti
  et~al.}{2020}]{leg-etal:bay}
Legramanti, S., D.~Durante, and D.~B. Dunson (2020).
\newblock {Bayesian cumulative shrinkage for infinite factorizations}.
\newblock {\em Biometrika\/}~{\em 107}, 745--752.

\bibitem[\protect\citeauthoryear{Lopes and West}{Lopes and
  West}{2004}]{lop-wes:bay}
Lopes, H.~F. and M.~West (2004).
\newblock Bayesian model assessment in factor analysis.
\newblock {\em Statistica Sinica\/}~{\em 14}, 41--67.

\bibitem[\protect\citeauthoryear{Lucas, Carvalho, Wang, Bild, Nevins, and
  West}{Lucas et~al.}{2006}]{luc-etal:spa}
Lucas, J., C.~Carvalho, Q.~Wang, A.~Bild, J.~R. Nevins, and M.~West (2006).
\newblock Sparse statistical modelling in gene expression genomics.
\newblock In K.~Do, P.~M\"uller, and M.~Vannucci (Eds.), {\em Bayesian
  Inference for Gene Expression and Proteomics}, pp.\  155--176. Cambridge, UK:
  Cambridge University Press.

\bibitem[\protect\citeauthoryear{Martin and {McD}onald}{Martin and
  {McD}onald}{1975}]{mar-mcd:bay}
Martin, J.~K. and R.~P. {McD}onald (1975).
\newblock Bayesian estimation in unrestricted factor analysis: a treatment for
  {H}eywood cases.
\newblock {\em Psychometrika\/}~{\em 40}, 505--517.

\bibitem[\protect\citeauthoryear{Neudecker}{Neudecker}{1990}]{neu:ide}
Neudecker, H. (1990).
\newblock On the identification of restricted factor loading matrices: {A}n
  alternative condition.
\newblock {\em Journal of Mathematical Psychology\/}~{\em 34}, 237--241.

\bibitem[\protect\citeauthoryear{{O'H}agan}{{O'H}agan}{1995}]{oha:fra}
{O'H}agan, A. (1995).
\newblock Fractional {B}ayes factors for model comparison.
\newblock {\em Journal of the Royal Statistical Society, Ser. B\/}~{\em 57},
  99--138.

\bibitem[\protect\citeauthoryear{Owen and Wang}{Owen and
  Wang}{2016}]{owe-wan:bic}
Owen, A.~B. and J.~Wang (2016).
\newblock {Bi-Cross-Validation for Factor Analysis}.
\newblock {\em Statistical Science\/}~{\em 31}, 119--139.

\bibitem[\protect\citeauthoryear{Paisley and Carin}{Paisley and
  Carin}{2009}]{pai-car:non}
Paisley, J.~W. and L.~Carin (2009).
\newblock Nonparametric factor analysis with beta process priors.
\newblock In {\em Proceedings of the 26th International Conference on Machine
  Learning {(ICML '09)}}, pp.\  777--784.

\bibitem[\protect\citeauthoryear{Papastamoulis and Ntzoufras}{Papastamoulis and
  Ntzoufras}{2022}]{pap-ntz:ide}
Papastamoulis, P. and I.~Ntzoufras (2022).
\newblock On the identifiability of {B}ayesian factor analytic models.
\newblock {\em Statistics and Computing\/}~{\em 32}.

\bibitem[\protect\citeauthoryear{{Pati}, {Bhattacharya}, {Pillai}, and
  {Dunson}}{{Pati} et~al.}{2014}]{pat-etal:pos}
{Pati}, D., A.~{Bhattacharya}, N.~S. {Pillai}, and D.~B. {Dunson} (2014).
\newblock {Posterior contraction in sparse Bayesian factor models for massive
  covariance matrices}.
\newblock {\em Annals of Statistics\/}~{\em 42}, 1102--1130.

\bibitem[\protect\citeauthoryear{Piatek and Papaspiliopoulos}{Piatek and
  Papaspiliopoulos}{2018}]{pia-pap:bay}
Piatek, R. and O.~Papaspiliopoulos (2018).
\newblock A bayesian nonparametric approach to factor analysis.
\newblock {\em Submitted\/}.

\bibitem[\protect\citeauthoryear{Poworoznek, Ferrari, and Dunson}{Poworoznek
  et~al.}{2021}]{pow-etal:eff}
Poworoznek, E., F.~Ferrari, and D.~Dunson (2021).
\newblock Efficiently resolving rotational ambiguity in {B}ayesian matrix
  sampling with matching.
\newblock {\em ArXiv\/}~{\em 2107.13783}.

\bibitem[\protect\citeauthoryear{Reiers{\o}l}{Reiers{\o}l}{1950}]{rei:ide}
Reiers{\o}l, O. (1950).
\newblock On the identifiability of parameters in {T}hurstone's multiple factor
  analysis.
\newblock {\em Psychometrika\/}~{\em 15}, 121--149.

\bibitem[\protect\citeauthoryear{Ro\v{c}kov\'{a} and George}{Ro\v{c}kov\'{a}
  and George}{2017}]{roc-geo:fas}
Ro\v{c}kov\'{a}, V. and E.~I. George (2017).
\newblock Fast {B}ayesian factor analysis via automatic rotation to sparsity.
\newblock {\em Journal of the American Statistical Association\/}~{\em 111},
  1608--1622.

\bibitem[\protect\citeauthoryear{Rue and Held}{Rue and
  Held}{2005}]{rue-hel:gau}
Rue, H. and L.~Held (2005).
\newblock {\em {G}aussian {M}arkov Random Fields: {T}heory and Applications},
  Volume 104 of {\em Monographs on Statistics and Applied Probability}.
\newblock London: Chapman {\rm \&} Hall/CRC.

\bibitem[\protect\citeauthoryear{Stock and Watson}{Stock and
  Watson}{2002}]{sto-wat:mac}
Stock, J.~H. and M.~W. Watson (2002).
\newblock Macroeconomic forecasting using diffusion indexes.
\newblock {\em Journal of Business {\rm \&} Economic Statistics\/}~{\em 20},
  147--162.

\bibitem[\protect\citeauthoryear{Teh, G\"{o}r\"{u}r, and Ghahramani}{Teh
  et~al.}{2007}]{teh-etal:sti}
Teh, Y.~W., D.~G\"{o}r\"{u}r, and Z.~Ghahramani (2007).
\newblock Stick-breaking construction for the {I}ndian buffet process.
\newblock In M.~Meila and X.~Shen (Eds.), {\em Proceedings of the Eleventh
  International Conference on Artificial Intelligence and Statistics}, Volume~2
  of {\em Proceedings of Machine Learning Research}, San Juan, Puerto Rico,
  pp.\  556--563. PMLR.

\bibitem[\protect\citeauthoryear{Thurstone}{Thurstone}{1947}]{thu:mul}
Thurstone, L.~L. (1947).
\newblock {\em Multiple factor analysis}.
\newblock Chicago: University of Chicago.

\bibitem[\protect\citeauthoryear{Tumura and Sato}{Tumura and
  Sato}{1980}]{tum-sat:ide}
Tumura, Y. and M.~Sato (1980).
\newblock On the identification in factor analysis.
\newblock {\em TRU Mathematics\/}~{\em 16}, 121--131.

\bibitem[\protect\citeauthoryear{van {D}yk and Meng}{van {D}yk and
  Meng}{2001}]{van-men:art}
van {D}yk, D. and X.-L. Meng (2001).
\newblock The art of data augmentation.
\newblock {\em Journal of Computational and Graphical Statistics\/}~{\em 10},
  1--50.

\bibitem[\protect\citeauthoryear{West}{West}{2003}]{wes:bay_fac}
West, M. (2003).
\newblock Bayesian factor regression models in the \lq\lq large p, small
  n\rq\rq\ paradigm.
\newblock In J.~M. Bernardo, M.~J. Bayarri, J.~O. Berger, A.~P. Dawid,
  D.~Heckerman, A.~F.~M. Smith, and M.~West (Eds.), {\em Bayesian Statistics
  7}, pp.\  733--742. Oxford: Oxford University Press.

\bibitem[\protect\citeauthoryear{Williams}{Williams}{2020}]{wil:ide}
Williams, B. (2020).
\newblock Identification of the linear factor model.
\newblock {\em Econometric Reviews\/}~{\em 39}, 92--109.

\bibitem[\protect\citeauthoryear{Yu and Meng}{Yu and Meng}{2011}]{yu-men:cen}
Yu, Y. and X.-L. Meng (2011).
\newblock To center or not to center: {T}hat is not the question - {A}n
  ancillarity-suffiency interweaving strategy {(ASIS)} for boosting {MCMC}
  efficiency.
\newblock {\em Journal of Computational and Graphical Statistics\/}~{\em 20},
  531--615.

\bibitem[\protect\citeauthoryear{Zhao, Gao, Mukherjee, and Engelhardt}{Zhao
  et~al.}{2016}]{zha-etal:bay_gro}
Zhao, S., C.~Gao, S.~Mukherjee, and B.~E. Engelhardt (2016).
\newblock Bayesian group factor analysis with structured sparsity.
\newblock {\em Journal of Machine Learning Research\/}~{\em 17}, 1--47.

\end{thebibliography}

\newpage
\appendix

\setcounter{equation}{0}
\setcounter{figure}{0}
\setcounter{table}{0}
\setcounter{page}{1}

\renewcommand{\thesection}{\Alph{section}}
\renewcommand{\thetable}{\Alph{section}.\arabic{table}}
\renewcommand{\thefigure}{\Alph{section}.\arabic{figure}}
\renewcommand{\theequation}{\Alph{section}.\arabic{equation}}

\begin{center}
\sffamily\LARGE{\bfseries Supplementary material for:\\
  \lq\lq Sparse Bayesian Factor Analysis when the Number of Factors is Unknown\rq\rq\ }
\end{center}

\section{Details on Step (P)} %

\subsection{Posterior distributions in a confirmatory factor model} \label{postdisfac}

Step~(P) of Algorithm~\ref{Algo3}  updates the parameters in the  confirmatory sparse factor model
factor model
\begin{eqnarray}  \label{fac1CFAapp}   %
 \facmk{\nfactrue}{t}   \sim  \Normult{\nfactrue}{\bfz,\identy{\nfactrue}}, \quad \ym_t =  \facload_\nfactrue  \facmk{\nfactrue}{t} + \errorm_t,  \quad  \errorm_t \sim \Normult{\dimy}{\bfz,\Vare_\nfactrue} ,  \quad  \Vare_\nfactrue=\Diag{\idiov_1,\ldots,\idiov_{\dimy}},
\end{eqnarray}

where  the  indicator matrix $\deltav_\nfactrue$ imposes
a certain zero structure on the loading matrix $\facload_\nfactrue$.
The joint posterior  distribution $p(\facload_{i\cdot}^{\deltav}, \idiov_i| \ym,\facm_\nfactrue ,\deltav_\nfactrue)$
of the nonzero factor loadings $\facload_{i\cdot}^{\deltav}$ and the idiosyncratic variance  $\idiov_i$ is derived for each row $i$ ($i=1, \ldots, \dimy$)
conditional on the factors  $\facm_\nfactrue = (\facmk{\nfactrue}{1},\ldots,\facmk{\nfactrue}{T})$ and the  indicator matrix $\deltav_\nfactrue$
from  the  following  regression model:
\begin{eqnarray}
\tilde{\ym}_i= \Xb_i ^{\deltav} \facload_{i\cdot}^{\deltav} + \tilde{\errorm}_i,
 \label{regnonpAPP}
\end{eqnarray}
where  $\tilde{\ym}_i=\trans{(y_{i1} \cdots y_{iT})}$ and
$\tilde{\errorm}_i=\trans{(\error_{i1} \cdots \error_{iT})} \sim \Normult{T}{0,  \idiov_i \identm}$.   $\Xb_i ^{\deltav}$ is a regressor matrix  for
 $\facload_{i\cdot}^{\deltav}$  constructed from the $\dimmat{T}{\nfactrue}$ dimensional latent factor   matrix $\Fm=\trans{(\facmk{\nfactrue}{1} \cdots \facmk{\nfactrue}{T})}$ in the following way.
 If no element  in row $i$ of $\facload_\nfactrue$ is restricted to 0, then  $\Xb_i ^{\deltav}=\Fm$.
 If some elements are restricted to 0, then $ \Xb_i ^{\deltav}$ is obtained from  $\Fm$  by deleting all columns $j$ where $\delta_{ij}=0$, i.e. $ \Xb_i ^{\deltav}=   \Fm \Pim_i ^{\deltav} $, where $\Pim_i ^{\deltav}$ is a $\dimmat{\nfactrue}{\sum_{j=1}^{\nfactrue} \delta_{ij} }$ selection matrix, selecting those columns $j$ of $\Fm$ where $\delta_{ij}\neq 0$.
The likelihood derived from (\ref{regnonpAPP}) is combined with
 the inverted Gamma prior (\ref{priorsiidg}) on $\idiov_i$ and, respectively,
 the hierarchical Gaussian prior (\ref{prior1}) or
  the  fractional prior (\ref{priorfrac}) for  $\facload_{i\cdot}^{\deltav}|  \idiov_i$.
 In a sparse factor model,  the dimension of this posterior depends
on the number of nonzero elements in the $i$th row of  $\facload_\nfactrue$,  i.e.  $q_i=\sum_{j=1}^{\nfactrue} \delta_{ij}$.
There are basically three types of rows, when  it comes to updating the parameters: zero rows, dedicated rows and
rows with multiple loadings.

 For zero rows (i.e.~$q_i=0$),  (\ref{regnonpAPP}) reduces to a \lq\lq null\rq\rq\ model without regressors $\Xb_i ^{\deltav}$,
 that is  $\tilde{\ym}_i=  \tilde{\errorm}_i$. Hence,   the posterior of $ \idiov_i$
is  simply given by
 \begin{eqnarray} \label{idiozero} %
\idiov_i|\tilde{\ym}_i,\facm_\nfactrue, \deltav_\nfactrue  \sim \Gammainv{c_{T} ^{\nullmod},C_{iT}^{\nullmod}}, %
\qquad c_{T}^{\nullmod}= c_0 +\frac{T}{2}, \qquad  C_{iT}^{\nullmod}=C_{i0} +\frac{1}{2}\sum_{t=1}^T  y_{it}^2 .
\end{eqnarray}
For all nonzero rows (i.e.~$q_i>0)$, the posterior    $(\facload_{i\cdot}^{\deltav}, \idiov_i)$ for  a specific row $i$ is given by:
 \begin{eqnarray}  \label{idiopos}   \label{rowpos}
 \idiov_i|  \tilde{\ym}_i,\facm_\nfactrue, \deltav_\nfactrue  \sim \Gammainv{c_{T},C_{iT}^{\deltav}},  \qquad %
\facload_{i\cdot}^{\deltav} | \idiov_i, \tilde{\ym}_i, \facm _\nfactrue, \deltav _\nfactrue
 \sim \Normult{q_i}{\bV_{iT} ^{\deltav} \cv_{iT} ^{\deltav}, \bV_{iT} ^{\deltav}\idiov_i}.
\end{eqnarray}
For the hierarchical Gaussian prior (\ref{prior1}),  the moments are given by:
\begin{eqnarray}  \label{postmomunit}
& ( \bV_{iT} ^{\deltav} ) ^{-1}  =  (\bV_{i0} ^{\deltav} ) ^{-1} + \trans{(\Xb_i^{\deltav})} \Xb_i^{\deltav} , \qquad
 \cv_{iT} ^{\deltav} =   \trans{(\Xb_i ^{\deltav})}\tilde{\ym}_i, &\\
 & c_T = c_0 +\frac{T}{2}, \qquad  %
  C_{iT}^{\deltav} =C_{i0} +  \frac {1}{2} \SSR_i , \quad
   \SSR_i= \ \trans{\tilde{\ym}_i} \tilde{\ym}_i - \trans{(\cv_{iT} ^{\deltav})}
  \bV_{iT}^{\deltav}  \cv_{iT} ^{\deltav}. & \nonumber %
\end{eqnarray}
 For the fractional prior (\ref{priorfrac}),  the moments are given by:
 \begin{eqnarray}%
&  (\bV_{iT} ^{\deltav}) ^{-1}   = \label{postmomfrac}
 \trans{(\Xb_i ^{\deltav})} \Xb_i ^{\deltav}  , \quad  \cv_{iT} ^{\deltav} =   \trans{(\Xb_i ^{\deltav})}\tilde{\ym}_i, &\\[1mm]
 &  c_T  = c_0 +\frac{(1-b)T}{2},  \quad%
C_{iT}^{\deltav} = C_{i0}+ \frac {(1-b)}{2}  \SSR_i ,  \nonumber %
\end{eqnarray}
where $\SSR_i$ is   the same as in  (\ref{postmomunit}) and,
 for the fractional prior, identical to  the residual sum of squares errors.\footnote{If  the residual $\errorm_i = \tilde{\ym}_i -   \Xb_i^{\deltav}   \bV_{iT} ^{\deltav} \cv_{iT} ^{\deltav}$ is defined in the usual way, then:
\begin{eqnarray*} %
 \trans{\errorm}_i\errorm _i &=&  \trans{\tilde{\ym}_i} \tilde{\ym}_i -    \trans{(\cv_{iT} ^{\deltav})}   \bV_{iT} ^{\deltav} \trans{(\Xb_i^{\deltav} )}  \tilde{\ym}_i-
\trans{ \tilde{\ym}_i }  \Xb_i^{\deltav}  \bV_{iT} ^{\deltav}   \cv_{iT} ^{\deltav}
+   \trans{(\cv_{iT} ^{\deltav})}    \bV_{iT} ^{\deltav}    \trans{(\Xb_i^{\deltav})}  \Xb_i^{\deltav}   \bV_{iT} ^{\deltav} \cv_{iT} ^{\deltav}\\
&=&   \trans{\tilde{\ym}_i} \tilde{\ym}_i -
 \trans{(\cv_{iT} ^{\deltav})}   \bV_{iT} ^{\deltav}  \cv_{iT} ^{\deltav}
-   \trans{(\cv_{iT} ^{\deltav})}  \bV_{iT} ^{\deltav}   \cv_{iT} ^{\deltav}
+   \trans{(\cv_{iT} ^{\deltav})}    \bV_{iT} ^{\deltav}    \cv_{iT} ^{\deltav} = \SSR_i.
\end{eqnarray*}}
For dedicated rows (i.e.~$q_i=1$)  only a single nonzero factor loading $\load_{i,\ji}$ is present in a particular column $\ji$ and the posterior
given in (\ref{idiopos}) simplifies considerably:
  \begin{eqnarray}  \label{rowdedi}
 \idiov_i|  \tilde{\ym}_i,\facm_\nfactrue,\deltav_\nfactrue  \sim \Gammainv{c_{T},C_{iT}}, \qquad %
\load_{i,\ji} | \idiov_i, \tilde{\ym}_i, \facm _\nfactrue,\deltav_\nfactrue
 \sim \Normal{B_{iT} m_{iT} ,  B_{iT}  \idiov_i}.
\end{eqnarray}
For a hierarchical Gaussian prior,
 the posterior moments   are given  by:
 \begin{eqnarray}
&   B_{iT}  = 1  /(B_{i0,\ji \ji}^{-1} +  \sum_{t=1}^T \fac_{\ji,t}^2),
 \quad  m_{iT} = \sum_{t=1}^T \fac_{\ji, t} y_{it} ,  & \label{singloadfi}\\
& c_T = c_0 +\frac{T}{2}, \qquad  C_{iT}=C_{i0} + \frac{1}{2}\left(\sum_{t=1}^T  y_{it}^2
-m_{iT}^2 B_{iT} \right) , & \nonumber
\end{eqnarray}
where
  $B_{i0,\ji \ji}$ is the prior variance of the dedicated factor loading in row $i$,
   whereas  for  a fractional prior:
 \begin{eqnarray}  \label{singloadfr}
&  B_{iT} = 1/ \left(\sum_{t=1}^T \fac_{\ji,t}^2\right),  \quad  m_{iT} = \sum_{t=1}^T \fac_{\ji,t} y_{it}, &\\
&   c_T  = c_0 +\frac{(1-b)T}{2},  \quad C_{iT}= C_{i0}+ \frac {(1-b)}{2}\left( \sum_{t=1}^T  y_{it}^2 - m_{iT}^2 B_{iT}\right). & \nonumber
\end{eqnarray}

\subsection{Block sampling of idiosyncratic variances and factor loadings} \label{jointfac}

Step~(P)  in Algorithm~\ref{Algo3} could be implemented as in \citet{lop-wes:bay},  by sampling $\facload_{i\cdot}^{\deltav}$ and $ \idiov_i$ from  the posterior  distribution $p(\facload_{i\cdot}^{\deltav}, \idiov_i| \ym,\facm_\nfactrue ,\deltav_\nfactrue )$ derived in Section~\ref{postdisfac} row by row.
However,   an important improvement  is  feasible
through block sampling of all idiosyncratic variances and all nonzero factor loadings, summarized in Algorithm~\ref{algoP}.\footnote{This algorithm has been implemented for the first time in the unpublished research report by \citet{fru-lop:par}.}
The use of  the Cholesky decomposition of the information matrix (instead of the covariance matrix) to sample  from a high-dimensional
density  is  fashioned  after  \citet[Theorem~2.5 and Algorithm~2.5]{rue-hel:gau} who consider  Gaussian random fields.

\begin{alg}{\textbf{Sampling parameters for a sparse Bayesian factor model }} \label{algoP}
\begin{itemize}
  \item[(P-a)] For all zero rows, sample  $ \idiov_i$ from (\ref{idiozero}), which can  be trivially vectorized.

    \item[(P-b)] If the remaining rows are all dedicated  with  a single nonzero loading in column $j_i$ (which can be different  for different rows), then sampling from  (\ref{rowdedi}) is easily vectorized, since all posterior  moments
      are  univariate.%

   \item[(P-c)] Even if some of the nonzero  rows are not dedicated,  joint sampling of all idiosyncratic variances and all factor loadings  is  feasible for all nonzero rows. Let $i_1, \ldots,i_n$ be the indices of  all $n=\dimy - m_0$ nonzero rows of $\facload_\nfactrue $, i.e.  $q_{i_l}>0$ for  $l=1,\ldots,n $.
       Let $d=\sum_i q_i$ be the total number of nonzero elements in $\facload_\nfactrue $
       and let $\facload^{\deltav}_\nfactrue =(\facload_{i_1\cdot}^{\deltav} \cdots  \facload_{i_n\cdot}^{\deltav})'$ be the $d$-dimensional vector obtained by stacking row by row all nonzero elements in each  row of
       $\facload_\nfactrue $.
 To sample the idiosyncratic variances $\idiov_{i_1}, \ldots,\idiov_{i_n}$ and the nonzero factor loadings $\facload^{\deltav}_\nfactrue$ in $\facload_\nfactrue $ jointly, proceed in the following way:
\begin{enumerate}
  \item[(P-c1)] Construct  the information matrix $\Omegav$ and the covector $\cv$ of the joint posterior $$\facload^{\deltav}_\nfactrue | \idiov_{i_1}, \ldots,\idiov_{i_n} ,\facm_\nfactrue,\ym  \sim \Normult{d}{ \Omegav^{-1} \cv,   \Omegav^{-1} \mathbf{D} }.$$
  The matrix   $\mathbf{D} = \Diag{ \idiov_{i_1} \unit{1 \times q_{i_1}} \cdots \idiov_{i_n} \unit{1 \times q_{i_n}}}$,
with $\unit{1 \times l}$  being a $\dimmat{1}{l}$ row vector of ones, is a $\dimmat{d}{d}$ diagonal matrix containing the idiosyncratic variances, while  the  $\dimmat{d}{d}$  matrix  $\Omegav$ and  the $\dimmat{d}{1}$  vector  $\cv$ are given by:
\begin{eqnarray*}
&&\Omegav = \left(
              \begin{array}{cccc}
                (\bV_{i_1,T} ^{\deltav}) ^{-1}  & \bfzmat & \cdots & \bfzmat \\
                \bfzmat &   (\bV_{i_2,T}^{\deltav}) ^{-1} & \ddots & \vdots \\
                   \vdots     &     \ddots               & \ddots & \bfzmat\\
                \bfzmat &  \cdots & \bfzmat &  (\bV_{i_n,T} ^{\deltav}) ^{-1} \\
              \end{array}
            \right),
 \qquad \cv= \left(
              \begin{array}{c}
                \cv_{i_1,T} ^{\deltav} \\
                \vdots  \\
                \cv_{i_n,T} ^{\deltav} \\
              \end{array}
            \right) ,
\end{eqnarray*}
where $(\bV_{i_l,T} ^{\deltav}) ^{-1}$ and $\cv_{i_l,T} ^{\deltav}$ are
 the information matrix and the covector  appearing in
the    posterior (\ref{rowpos}) of the nonzero elements in row $i_l$. $\Omegav$ is a sparse  band matrix with maximal band width equal to $\max q_{i_l}$.

  \item[(P-c2)]     Compute the Cholesky decomposition
  $\Omegav=\Lv \trans{\Lv}$, where $\Lv$ is lower triangular, using a special algorithm developed for
   band matrices.   Next, solve $\Lv \xm = \cv$ for $\xm$ using  an algorithm  specially designed for triangular matrices.
  Evidently,  $\xm$ is a $\dimmat{d}{1}$  vector.

   \item[(P-c3)] Sample  $\idiov_{i_1},\ldots,  \idiov_{i_n}$  jointly from  (\ref{idiopos}).
  The squared sum $\trans{\xm} \xm$  can be used to vectorize the computation of  $C_{i_l,T}^{\deltav}$  for  each $l=1, \ldots, n$, since
    \begin{eqnarray} \label{citfrac}
   \trans{\xm_{i_l}} \xm_{i_l} =   \trans{(\cv_{i_l,T} ^{\deltav})} \bV_{i_l,T}^{\deltav } \cv_{i_l,T} ^{\deltav},
    \end{eqnarray}
     where  $ \xm_{i_l}$  is the $q_{i_l}$-dimensional sub vector  of $\xm$ corresponding to  $\facload_{i_l,\cdot}^{\deltav}$.

 \item[(P-c4)]   Finally, define the   diagonal matrix $\mathbf{D}$  from $ \idiov_{i_1}, \ldots,
\idiov_{i_n}$  as described above and draw $\zm \sim \Normult{d}{\bfz, \mathbf{D}}$. Solving the system
 \begin{eqnarray} \label{pposbb}
 \trans{\Lv} \facload^{\deltav}_\nfactrue= \xm +   \zm
\end{eqnarray}
for $\facload^{\deltav}_\nfactrue$ leads to a  draw  from  the joint posterior
$ \facload^{\deltav}_\nfactrue| \idiov_{i_1}, \ldots, \idiov_{i_n},\ym,\facm _\nfactrue$.
\end{enumerate}
 \end{itemize}
\end{alg}

\vspace*{1mm}  \noindent
 To derive  (\ref{citfrac}),  let $\Lv_{i_l}$ be the $\dimmat{q_{i_l}}{q_{i_l}}$
 submatrix of $\Lv$ corresponding to  $\facload_{i_l,\cdot}^{\deltav}$.  Evidently, $\Lv_{i_l}$  is
 equal to the Cholesky decomposition of the individual
 information matrix $(\bV_{i_l,T}^{\deltav}) ^{-1}$. Furthermore,
 the $q_{i_l}$-dimensional sub vector
   $\xm_{i_l}$ corresponding to  $\facload_{i_l,\cdot}^{\deltav}$ satisfies $\Lv_{i_l}\xm_{i_l}=\cv_{i_l,T} ^{\deltav}$.
   Therefore:
   \begin{eqnarray*}
&&  \trans{\xm_{i_l}} \xm_{i_l} =   \trans{(\cv_{i_l,T} ^{\deltav})}
  (\trans{\Lv_{i_l}}) ^{-1} \Lv_{i_l} ^{-1}  \cv_{i_l,T} ^{\deltav}=
 \trans{(\cv_{i_l,T} ^{\deltav})} (\Lv_{i_l} \trans{\Lv_{i_l}} ) ^{-1}  \cv_{i_l,T} ^{\deltav} =
 \trans{(\cv_{i_l,T} ^{\deltav})} \bV_{i_l,T}^{\deltav } \cv_{i_l,T} ^{\deltav}.
\end{eqnarray*}
 It is easy to  prove that the solution $\facload^{\deltav}_\nfactrue$ of (\ref{pposbb}) is a draw from the posterior
 $p(\facload^{\deltav}_\nfactrue|\idiov_{i_1}, \ldots, \idiov_{i_n},\ym,\facm)$.  Note that
$\Lv \trans{\Lv} \facload^{\deltav}_\nfactrue= \Lv \xm + \Lv  \zm  = \cv + \Lv  \zm $.
Therefore
\begin{eqnarray*}
&&   \facload^{\deltav}_\nfactrue= ( \Lv \trans{\Lv})^{-1} \cv + ( \Lv \trans{\Lv})^{-1} \Lv  \zm=  \Omegav ^{-1} \cv + (\trans{\Lv})^{-1}  \zm.
\end{eqnarray*}
Evidently,  $\Ew{\facload^{\deltav}_\nfactrue}= \Omegav ^{-1} \cv$.   Since
   for each $l=1, \ldots,n$,  $  \Lv_{i_l} \idiov_{i_l}  = \idiov_{i_l}\Lv_{i_l} $, it holds that $  \Lv \mathbf{D} =  \mathbf{D} \Lv $
   and therefore $\mathbf{D}  \Lv  ^{-1}=   \Lv ^{-1}\mathbf{D}$. Since $\V{\facload^{\deltav}_\nfactrue}=  (\trans{\Lv})^{-1}  \mathbf{D}  \Lv  ^{-1}=  (\trans{\Lv})^{-1}   \Lv  ^{-1}\mathbf{D} = \Omegav ^{-1}\mathbf{D}$,
it follows  that
   $ \facload^{\deltav}_\nfactrue \sim \Normult{d}{\Omegav ^{-1} \cv,\Omegav ^{-1}\mathbf{D}}$.

 \section{Details on Step~(D)} \label{appSecD}

\subsection{Marginal likelihoods when the factors are known} \label{marfac}

Although we work throughout  this paper with a factor model where the factors $\facmk{\nfactrue}{t}$ are latent, several steps of Algorithm~\ref{Algo3} perform model selection with respect to $\deltav_\nfactrue$  conditional
on the most recent draw of  the factors $\facm_\nfactrue=(\facmk{\nfactrue}{1}, \ldots, \facmk{\nfactrue}{T})$ in the confirmatory factor model (\ref{fac1CFAapp}).
Hence, to sample new indicators $\deltav_{i \cdot}$ in row $i$,
 the marginal likelihood $p(\tilde{\ym}_i|  \facm_\nfactrue, \deltav_\nfactrue)$
of  regression model  (\ref{regnonpAPP}) is  needed.

If   $\deltav_{i \cdot}$ is a zero row (i.e $q_i=0$), then the marginal likelihood  simplifies to
\begin{eqnarray}
 p(\tilde{\ym}_i|  \facm_\nfactrue, \deltav_\nfactrue)  =  p(\tilde{\ym}_i) =
   \frac{\Gamfun{c_{T}^{\nullmod} } (C_{i0})^{c_0} }
{(2\pi)^{T/2}\Gamfun{c_0} (C_{iT}^{\nullmod})^{c_{T}^{\nullmod}}} ,  \label{marnull}
\end{eqnarray}
where $c_{T}^{\nullmod}$ and $C_{iT}^{\nullmod}$ are the posterior moments of $\idiov_i$ under
 the \lq\lq null\rq\rq\ model given by (\ref{idiozero}).

 If at least one element of $\deltav_{i \cdot}$ is different from zero,
then the marginal likelihood  computation differs between the hierarchical Gaussian prior  (\ref{prior1})  and the fractional prior (\ref{priorfrac}).

  \paragraph{Marginal likelihoods for a hierarchical Gaussian prior.}
  For a hierarchical Gaussian prior,  a  well-known  exercise in Bayesian regression analysis yields:
\begin{eqnarray}
 p(\tilde{\ym}_i|  \deltav_\nfactrue, \facm_\nfactrue )  =
  \frac{1}{(2\pi)^{T/2} }   \frac{|\bV_{iT}^{\deltav}|^{1/2}}{|\bV_{i0}^{\deltav } |^{1/2}}
   \frac{\Gamfun{c_{T}} (C_{i0})^{c_0} }
{\Gamfun{c_0} (C_{iT}^{\deltav})^{c_{T}}} ,  \label{ADDGAD}
\end{eqnarray}
where $\bV_{iT} ^{\deltav}$, $c_{T}$ and $C_{iT}^{\deltav}$ are the posterior moments of
 $p(\facload^{\deltav}_{i\cdot},\idiov_i|\deltav_\nfactrue,\tilde{\ym}_i, \facm_\nfactrue)$
 given by (\ref{postmomunit}).

  \paragraph{Marginal likelihoods for a fractional prior.}
 For  a fractional prior, the derivation of the marginal likelihood if at least one element of $\deltav_{i \cdot}$ is different from zero, is less standard  and can be
   obtained in a  similar way as in \citet{fru-wag:sto}.
   A fraction $b$ of the full conditional likelihood  of regression model  (\ref{regnonpAPP})
   is used to  define  the fractional prior $p(  \facload_{i\cdot}^{\deltav} | \idiov_i \addb ,\facm_\nfactrue )$ in (\ref{priorfrac}):
    \begin{eqnarray*}
          p(\tilde{\ym}_i|  \facm_\nfactrue, \facload_{i\cdot}^{\deltav} ,\idiov_i)=   p(\tilde{\ym}_i|  \facm_\nfactrue,\facload_{i\cdot}^{\deltav} ,\idiov_i)^{1-b}
           p(\tilde{\ym}_i|  \facm_\nfactrue, \facload_{i\cdot}^{\deltav} ,\idiov_i)^b \propto  p(\tilde{\ym}_i|  \facm_\nfactrue, \facload_{i\cdot}^{\deltav} ,\idiov_i)^{1-b}
              p(  \facload_{i\cdot}^{\deltav} | \idiov_i \addb ,\facm_\nfactrue ) .
    \end{eqnarray*}
    The remaining part of  the likelihood, that is $ p(\tilde{\ym}_i|  \facm_\nfactrue,\facload_{i\cdot}^{\deltav} ,\idiov_i)^{1-b} $, is used for model selection and is combined with the   prior  $p(\idiov_i)$  defined  in  (\ref{priorsiidg})  and the  {\em normalized} fractional prior  $ p(  \facload_{i\cdot}^{\deltav} | \idiov_i \addb ,\facm_\nfactrue )$, given by:
  \begin{eqnarray*}
  p(  \facload_{i\cdot}^{\deltav} | \idiov_i \addb ,\facm_\nfactrue )  =\frac{ p(\tilde{\ym}_i|  \facm_\nfactrue, \facload_{i\cdot}^{\deltav} ,\idiov_i)^b}{c_i (\idiov_i, \facm_\nfactrue ,b)}.
    \end{eqnarray*}
 The normalizing constant $c_i (\idiov_i, \facm_\nfactrue ,b)$ is   given by:
\begin{eqnarray} \label{normfrac}
  c_i (\idiov_i, \facm_\nfactrue ,b)=  \int p(\tilde{\ym}_i| \facm_\nfactrue, \facload_{i\cdot}^{\deltav} ,\idiov_i)^b  \, d \,  \facload_{i\cdot}^{\deltav} =
( 2\pi \idiov_i)^{\frac{ q_i-Tb}{2}}
b^{-\frac{ q_i}{2}}  |\bV_{iT}^{\deltav}|^{1/2} \exp\left(  -  \frac {b}{2  \idiov_i} \SSR_i \right),
 \end{eqnarray}
 where $\bV_{iT} ^{\deltav}$  and $\SSR_i $  are the posterior moments of
 $p(\facload^{\deltav}_{i\cdot},\idiov_i|  \facm_\nfactrue, \deltav_{i\cdot},\tilde{\ym}_i )$
 given by (\ref{postmomfrac}). %
   Integrating the  fractional posterior
  \begin{eqnarray*}
 p(\tilde{\ym}_i|  \facm_\nfactrue,\facload_{i\cdot}^{\deltav} ,\idiov_i)^{1-b}   p(  \facload_{i\cdot}^{\deltav} | \idiov_i \addb ,\facm_\nfactrue ) p(\idiov_i)
  \end{eqnarray*}
   over $\facload_{i\cdot}^{\deltav}$, yields the fractional likelihood  $p(\tilde{\ym}_i|  \facm_\nfactrue,\idiov_i,b)$:
 \begin{eqnarray*}
 \displaystyle p(\tilde{\ym}_i|  \facm_\nfactrue,\idiov_i,b) &=&  \int p(\tilde{\ym}_i|  \facm_\nfactrue,\facload_{i\cdot}^{\deltav} ,\idiov_i)^{1-b}   p(  \facload_{i\cdot}^{\deltav} |  \facm_\nfactrue, \idiov_i \addb  )  \, d \,  \facload_{i\cdot}^{\deltav} = \frac{ 1}{ c_i (\idiov_i, \facm_\nfactrue ,b) } \int p(\tilde{\ym}_i|  \facm_\nfactrue,\facload_{i\cdot}^{\deltav} ,\idiov_i)   \, d \,  \facload_{i\cdot}^{\deltav} \\
 &=& \displaystyle  \left(  \frac {1}{2\pi \idiov_i}\right) ^{\frac{(T - q_i) -(Tb - q_i)}{2} }  b^{ \frac{q_i}{2}}  \frac{ |\bV_{iT}^{\deltav}|^{1/2}} { |\bV_{iT}^{\deltav}|^{1/2}}  \exp \left(  -  \frac {(1-b)}{2  \idiov_i} \SSR_i \right)  \\
&=& \displaystyle  \left(  \frac {1}{2\pi \idiov_i}\right) ^{\frac{T(1 - b)}{2} }  b^{\frac{q_i}{2}}  \exp\left(  -  \frac{(1-b)}{2  \idiov_i} \SSR_i \right).
 \end{eqnarray*}
   When we combine  $p(\tilde{\ym}_i|  \facm_\nfactrue,\idiov_i,b)$  with  the prior $p(\idiov_i)$, then  we obtain:
\begin{eqnarray*}
p(\tilde{\ym}_i|  \facm_\nfactrue, \idiov_i,b)  p(\idiov_i) =   \frac{C_{i0}^{c_0} }{\Gamfun{c_0}}
 \left( \frac {1}{2\pi } \right) ^{\frac {T(1 - b)}{2} }  b^{\frac{q_i}{2}} \left(  \frac {1}{\idiov_i}\right) ^{ \frac{c_0 + T(1 - b)}{2} }
 \exp\left(  -  \frac{ C_{i0} +   \SSR_i (1-b)/2 }{\idiov_i} \right),
 \end{eqnarray*}
which  is the kernel of  the  inverted Gamma distribution in (\ref{idiopos}). Integrating the right hand side with respect to $\idiov_i$ yields
the marginal likelihood  under the fractional prior: %
 \begin{eqnarray}
p(\tilde{\ym}_i|\deltav_\nfactrue, \facm_\nfactrue)  =
 \frac {b^{q_i /2}\Gamma(c_T)(C_{i0})^{c_0}}{(2\pi)^{T(1-b)/2}\Gamma(c_0) (C_{iT}^{\deltav})^{c_T}}.
 \label{frac_marlik}
\end{eqnarray}

\subsection{Multimove sampling of a set of  indicators in a column}   \label{mcmcsmodi}

Another important building block of MCMC inference for sparse Bayesian factor models is sampling all
indicators $\delta_{ij}$ in column ${j}$   for a set of  rows $i \in I_j  \subseteq  \{1, \ldots, m\}$, %
conditional on  the factors $\facm_\nfactrue=(\facmk{\nfactrue}{1}, \ldots, \facmk{\nfactrue}{T})$, the remaining indicator $ \deltav_\nfactrue^\uncon$  and
the hyperparameter $\tau_j$,  without conditioning on the model parameters $  \facload_\nfactrue$ and $\idiov_1,\ldots,\idiov_{\dimy}$, see Step~(D) of Algorithm~\ref{Algo3}.

 According to the prior   (\ref{prialt}),  the  indicators $\delta_{ij}$  are independent apriori
  conditional on the hyperparameter $\tau_j$, with the log prior odds $\oddpr_{ij}$ of  $\delta_{ij}=1$ versus
$\delta_{ij}=0$ being given by:
\begin{eqnarray} \label{samdelpr}
 \oddpr_{ij}= \log  \frac{\Prob{\delta_{ij}=1| \tau_j}}{\Prob{\delta_{ij}=0| \tau_j}} =  \log \frac{\tau_j}{1- \tau_j}.
\end{eqnarray}
    To sample $\delta_{ij}$ conditional $ \deltav_\nfactrue^\uncon$ and $\facm_\nfactrue$, without conditioning on   $  \facload_{\nfactrue}$ and $(\idiov_1,\ldots,\idiov_{\dimy})$,
 the   log posterior odds $\oddpost_{ij}$, given by
    \begin{eqnarray}
\oddpost_{ij} &= &\log  \frac{\Prob{\delta_{ij}=1| \deltav_\nfactrue^\uncon, \tau_j,  \tilde{\ym}_i, \facm_\nfactrue}}
{\Prob{\delta_{ij}=0| \deltav_\nfactrue^\uncon,\tau_j, \tilde{\ym}_i, \facm_\nfactrue}}  = \log  \frac{p(\tilde{\ym}_i|\delta_{ij}=1, \deltav_\nfactrue^\uncon ,  \facm_\nfactrue)}
{p(\tilde{\ym}_i|\delta_{ij}=0, \deltav_\nfactrue^\uncon,  \facm_\nfactrue)} +
 \log  \frac{\Prob{\delta_{ij}=1| \tau_j}}{\Prob{\delta_{ij}=0| \tau_j}} \nonumber
 \\ &=&   \odd_{ij} + \oddpr_{ij},  \label{samdelmult}
\end{eqnarray}
  is required which combines  the log prior odds  $\oddpr_{ij}$ given in (\ref{samdelpr}) with the  log likelihood ratio $ \odd_{ij}$, given by:
     \begin{eqnarray}  \label{samplpost}
  \odd_{ij} = \log  \frac{p(\tilde{\ym}_i|\delta_{ij}=1, \deltav_\nfactrue^\uncon ,  \facm_\nfactrue)}
{p(\tilde{\ym}_i|\delta_{ij}=0, \deltav_\nfactrue^\uncon,  \facm_\nfactrue)}.
\end{eqnarray}
 The  likelihood ratio  $ \odd_{ij}$  is easily  computed from  the
 marginal likelihoods $p(\tilde{\ym}_i|\delta_{ij}, \deltav_\nfactrue^\uncon,  \facm_\nfactrue)$ where,  %
respectively,  $\delta_{ij}=1$  and $\delta_{ij}=0$.  As discussed in  Section~\ref{marfac}, these marginal likelihoods are available in closed form   and
marginal likelihood computation  can be done  individually for each row $i \in I_j$,  separately for  $\delta_{ij}=0$ and  $\delta_{ij}=1$.

However, this procedure is likely to be  inefficient, in particular, if  the set $ I_j$  is large.
To achieve greater efficiency,   Algorithm~\ref{AlgoInd} outlined below provides a technique to
  compute  directly the log likelihood  ratio $\odd_{ij}$ %
   (rather than the individual marginal likelihoods) {\em simultaneously}  for all rows  $i \in I_j$.
  This allows  joint sampling of all indicators $\delta_{ij}$ in column $j$ for all  rows  $i \in I_j$.

The precise form of    the log  likelihood ratio $\odd_{ij}$ of  $\delta_{ij}=1$ versus  $\delta_{ij}=0$  defined  in (\ref{samplpost})
  depends on the remaining indicators  $\deltav_{i,-j}$ in row $i$.
  The computation of
 $\odd_{ij}$   is easily vectorized for all rows $i \in I_j$   where all elements of $\deltav_{i,-j}$ are zero. In this case,   a model  where observation $y_{it}$ is
 dedicated to factor $j$ ($\delta_{ij}=1$) is compared to a  model  where  $y_{it}$ is
 uncorrelated   with all remaining observations ($\delta_{ij}=0$). In this case,   $\odd_{ij}$ is easily
obtained from the marginal likelihood  of   a  dedicated model with $\ji=j$
 and the  \lq\lq null\rq\rq\ model. %
 As shown in Algorithm~\ref{AlgoInd}, it is   possible (but less straightforward)  to vectorize   the computation of the  log  likelihood ratio $\odd_{ij} $
  also for the remaining rows $i \in I_j$
  where at least one element of $\deltav_{i,-j}$ is different from
  zero.  \\  %

\begin{alg}{\textbf{Multimove sampling of indicators in a column.}} \label{AlgoInd}
Sample all
indicators $\delta_{ij}$ in column $\deltacol{j}$  jointly for all rows $i \in I_j  \subseteq  \{1, \ldots, m\}$ %
conditional on  the factors $\facm_\nfactrue=(\facmk{\nfactrue}{1}, \ldots, \facmk{\nfactrue}{T})$, the remaining indicators $\deltav_\nfactrue^\uncon$  and the hyperparameter $\tau_j$,  without conditioning on the model parameters $  \facload_\nfactrue$ and $\idiov_1,\ldots,\idiov_{\dimy}$ using the following steps:
\begin{itemize}
  \item[(I-a)]   Compute the log likelihood  ratio $\odd_{ij} $ for  all rows $i \in I_j$  where  all elements of $\deltav_\nfactrue^\uncon$ are zero  as
      \begin{eqnarray}  \label{saliklmult}
&& \odd_{ij} = \log  \frac{p(\tilde{\ym}_i|\delta_{ij}=1, \deltav_\nfactrue^\uncon,   \facm _\nfactrue)}
{p(\tilde{\ym}_i|\delta_{ij}=0, \deltav_\nfactrue^\uncon,  \facm_\nfactrue)}  =
\log  \frac{\Gamma(c_T) ( C_{iT} ^{\nullmod} )^{c_T ^{\nullmod}}}
{\Gamma(c_T^{\nullmod}) (C_{iT})^{c_T }}
  +  \Bodd_{ij}.
\end{eqnarray}
$c_T ^{\nullmod}$ and $C_{iT} ^{\nullmod}$ are the posterior moments of the null model  given in (\ref{idiozero}).
$c_T$ and $ C_{iT}$  are the  posterior moments  of $\sigma_i^2$ for a dedicated measurement with $\ji=j$, given in (\ref{singloadfi}) and (\ref{singloadfr}),
respectively,  for a hierarchical Gaussian and a fractional prior.
 For a hierarchical Gaussian prior,  $   \Bodd_{ij}  = 0.5  \log(B_{iT}/B_{i0,jj})$, where
  $B_{i0,jj}$ is $j$th diagonal element
 of the prior variance  $\bV_{i0}$ and  $B_{iT}$  is the posterior scale factor  for a dedicated measurement with $\ji=j$, given in  (\ref{singloadfi}).
    For  a fractional prior, $  \Bodd_{ij} = 0.5 \log(b(2\pi)^{bT})$.

 This step is  trivial  to vectorize.

   \item[(I-b)] For all  rows   $i \in \{i_1,\ldots,i_n\} \subseteq I_j$  where  $\deltav_{i,-j}$ is non-zero,   compute
\begin{eqnarray}  \label{saliklmult2}
\odd_{ij}=  \log \frac{p(\tilde{\ym}_i |\delta_{ij}=1, \deltav_\nfactrue^\uncon,  \facm_\nfactrue)}
{p(\tilde{\ym}_i |\delta_{ij}=0, \deltav_\nfactrue^\uncon,  \facm _\nfactrue)}   =
c_T   \log  \frac{C_{iT} ^{0}} { C_{iT} ^1}   +  \Bodd_{ij},
\end{eqnarray}
where  $c_T$ and $C_{iT} ^{\delta_{ij}}$  are the posterior moments of $\sigma^2_i|\delta_{ij},\cdot  $
given in   (\ref{idiopos}) and %
$C_{iT}^0$ refers to  a model with $\delta_{ij}=0$,  while $C_{iT}^1$ refer to  a model
 with  $\delta_{ij}=1$.
    For a hierarchical Gaussian  prior,
  \begin{eqnarray}
   \Bodd_{ij} = 0.5  \log(|\bV_{iT}^1 |/|\bV_{iT}^0|)- 0.5\log ( |\bV_{i0} ^1|/|\bV_{i0}^0| ), \label{B38A}
   \end{eqnarray}
  where  $\bV_{i0}^{\delta_{ij}}$ and $\bV_{iT}^{\delta_{ij}}$ refer to the prior and posterior moments
  of  $ \facload_{i\cdot}^{\deltav} |  \delta_{ij}, \cdot $  given in  (\ref{postmomunit}).
  $\bV_{i0}^1$ and $\bV_{iT}^1$ refer to the prior and posterior moments for  a model
where   $\delta_{ij}=1$,
  while $\bV_{i0}^0$ and $\bV_{iT}^0$   refer to the prior and posterior moments for  a model
where $\delta_{ij}=0$.
For a fractional prior,  $ \Bodd_{ij} \equiv 0.5 \log b $.

Use Algorithm~\ref{Algoratio} to determine
 $C_{iT}^1$, $C_{iT}^0$,  as well as $\Bodd_{ij}$   for a hierarchical Gaussian prior,   simultaneously for all rows  $i \in \{i_1,\ldots,i_n\} \subseteq I_j$.

   \item[(I-c)]   Determine the vector  of the log posterior odds $\oddpost_{ij}=\odd_{ij} + \oddpr_{ij} $ for all rows $i \in I_j$.
    Joint sampling of   $\delta_{ij}| \tau_j,\cdot $  is easily vectorized:
       \begin{itemize}
         \item[(I-c1)] Propose  $\delta_{ij}\new=1-\delta_{ij}$  for   $i \in I_j$.
         \item[(I-c2)] Draw a   vector  of  $| I_j|$   random variables $U_i \sim \Uniform{0,1}$, indexed by   $i \in I_j$.
         \item[(I-c3)] For all rows $i \in I_j$, where $\delta_{ij}=0$, accept the proposal $\delta_{ij}\new=1$, iff  $\log  U_i \leq \oddpost_{ij}$;
         \item[(I-c4)] For all rows $i \in I_j$,  where $\delta_{ij}=1$, accept the proposal $\delta_{ij}\new=0$, iff  $\log  U_i  \leq - \oddpost_{ij}$.
       \end{itemize}
       \end{itemize}
\end{alg}

\vspace*{1mm}  \noindent
 Using, respectively,  (\ref{ADDGAD}) and   (\ref{frac_marlik}),  the expression for $ \odd_{ij}$ in (\ref{saliklmult2}) is easily derived.
  Since the indicators   in column $j$ are  independent  given $\tau_j$, Step~(I-c)  is based  on $| I_j|$
 independent  Metropolis-Hastings (MH) steps  each of which proposes to update $\delta_{ij}$ by flipping  the   indicator,
 i.e. $\delta_{ij}\new=1-\delta_{ij}$.\footnote{Alternatively,  a Gibbs step  may be used, i.e.
 set $\delta_{ij} \new=1$, iff $\log (U_i/(1-U_i)) \leq \oddpost_{ij}$, otherwise  $\delta_{ij} \new=0$. However,  simulation experiments  indicate that the MH step is more efficient.}  It easy to verify
 that the acceptance rules formulated in Step~(I-c3) and  (I-c4)  are equivalent to the more convential
rule  to accept  $\delta_{ij}\new$ with probability
$$
 \min\left\{1, \frac{\Prob{\delta_{ij}\new| \deltav_\nfactrue^\uncon, \tau_{j},  \tilde{\ym}_i , \facm_\nfactrue}}
{\Prob{\delta_{ij} | \deltav_\nfactrue^\uncon,\tau_{j}, \tilde{\ym}_i , \facm_\nfactrue}} \right\} = \min\left\{1,  \exp( \oddpost_{ij} )  \right\}.$$

\begin{alg} %
 \label{Algoratio}
 To compute  all relevant posterior moments in (\ref{saliklmult2}) simultaneously for all rows $\{i_1,\ldots,i_n\}$,  proceed as follows:
\begin{itemize}
     \item[(a)] Set  the indicator $\delta_{i_l,j}=1$ in each row $i_l \in \{i_1,\ldots,i_n\}$. Reorder the columns of the factor loading matrix
 in such a way, that the $j$th column appears
last.  This is simply done by permuting the column of $\Fm$ appropriately before defining
 $\Xb_{i_l} ^{\deltav}$.\footnote{While the fractional prior is not affected by this, it might be necessary to reorder
 the prior mean and the prior  covariance matrix for a hierarchical Gaussian prior.}

\item[(b)]   Set up the information matrix $\Omegav$ and the covector $\cv$ of the corresponding joint
posterior of all nonzero factor loadings in the rows  $i_1,\ldots,i_n$ as described in Algorithm~\ref{algoP}.
Compute the Cholesky decomposition $\Lv$  of $\Omegav$   and the corresponding vector $\xm$  solving
$ \Lv  \xm= \cv$.

    \item[(c)] Knowing $ \Lv$ and  $\xm$,  a   vectorized computation of  the
log  likelihood ratio   (\ref{saliklmult})   for  all rows $i_l \in \{i_1,\ldots,i_n\}$  is possible.
    The posterior moments $C_{i_l,T} ^{1}$ are  directly available from the appropriate
sub vectors $\xm_{i_l}$ of $\xm$, defined in  (\ref{citfrac}).  When we switch from  $\delta_{i_l,j}=1$   to a model where $\delta_{i_l,j}=0$, then
for a hierarchical Gaussian prior,
  \begin{eqnarray} \label{Citpostb}
  C_{i_l,T}^{0} =  C_{i_l,T}  ^{1} + \frac{1}{2 } (x^{\star}_{i_l})^2.
\end{eqnarray}
 Furthermore,
  \begin{eqnarray} \label{Citpostc}
  0.5 \log(|\bV_{i_l,T} ^{1} |/|\bV_{i_l,T}^0|)= -\log L^{\star}_{i_l},
\end{eqnarray}
 where $L^{\star}_{i_l}=(\Lv_i)_{q_{i_l},q_{i_l}}$ is the last diagonal element of the submatrix $\Lv_{i_l}$.
 Therefore,
$$D_{ij}=  -\log L^{\star}_{i_l} - 0.5 \log  B_{i0,jj}. $$
 For the fractional prior,
  \begin{eqnarray} \label{Citposta}
  C_{i_l,T}^{0} =  C_{i_l,T} ^{1}  + \frac{1-b}{2 }  (x^{\star}_{i_l})^2,
\end{eqnarray}
where   $x^{\star}_{i_l}=(\xm_{i_l})_{q_{i_l}}$ is the last element of  $\xm_{i_l}$.
\end{itemize}
\end{alg}

\noindent
{\em Derivation  of Step~(c).}  When we switch from a model  where all indicator $\delta_{i_1,j}= \ldots = \delta_{i_n,j}=1$ are equal to one
to a model where all   indicators $\delta_{i_1,j}= \ldots = \delta_{i_n,j}=0$ are zero, then the information matrix $\Omegav^0$
 and the covector $\cv^0$ of the   joint
posterior of  the remaining  nonzero factor loadings is obtained from $\Omegav$ and $\cv$ simply by deleting all
rows and columns corresponding to $\delta_{i_1,j}, \ldots, \delta_{i_n,j}$, and
the Cholesky decomposition $\Lv^0$ of  $\Omegav^0$ is  obtained from
 $\Lv$ in the same way. Also the  vector $\xm^0$
 solving
$ \Lv ^0 \xm^0 = \cv^0$   is obtained from $\xm$ simply by deleting the
rows corresponding to $\delta_{i_1,j}, \ldots, \delta_{i_n,j}$.  This last result is easily seen by considering the subsystem $\Lv_{i_l} \xm_{i_l} = \cv_{i_l,T} ^{\deltav}$ corresponding to the
$i_l$th row. Because
\begin{eqnarray} \label{deflio}
&& \Lv_{i_l}  = \left(\begin{array}{cc}
                      \Lv_{i_l} ^0 & \bfzmat  \\
                      \mathbf{l}_{i_l} & (\Lv_i)_{q_{i_l},q_{i_l}}  \\
                    \end{array} \right) =   \left(\begin{array}{cc}
                      \Lv_{i_l} ^0 & \bfzmat  \\
                      \mathbf{l}_{i_l} & L^{\star}_{i_l}  \\
                    \end{array} \right) , \qquad
\end{eqnarray}
we obtain $\Lv_{i_l} ^0 \xm_{i_l}^0= \cv^0_{i_l}$, where  $\xm_{i_l}^0$ is obtained from $\xm_{i_l}$ by deleting the  $q_{i_l}$th
element $x^{\star}_{i_l}=(\xm_{i_l})_{q_{i_l}}$.
 Hence, $\xm_{i_l}^0$ defines the desired subvector of  $\xm^0$  to compute  $C_{i_l,T}^{0}$     as in  (\ref{citfrac}).
 Since  $ \trans{(\xm_{{i_l}}^0)} \xm_{i_l}^0 = \trans{\xm_{i_l}} \xm_{i_l}- (\xm_{i_l})_{q_{i_l}}^2$  we obtain  from (\ref{idiopos}) that  (\ref{Citposta})  holds. Note, however, that this simple relationship would not hold
 without reordering the columns as described above.

Finally, to compute the log   likelihood ratio for a hierarchical Gaussian prior,   the ratio of the determinants $|\bV^{1}_{i_l,T}|/|\bV_{i_l,T}^{0}|$ is required. Since
  the lower triangular matrices $\Lv_{i_l}$ and $\Lv_{i_l} ^0$ are, respectively, the Cholesky decomposition of $(\bV_{i_l,T}^{1})^{-1}$ and $(\bV_{i_l,T}^{0})^{-1}$,   we obtain:%
 \begin{eqnarray} \label{detvchol}
&&  1/|\bV_{i_l,T} ^{1} |^{1/2} = |(\bV_{i_l,T}^{1} )^{-1} |^{1/2} = |\Lv_{i_l} |,
\end{eqnarray}
 where   $|\Lv_{i_l} |$  is   the product of the diagonal elements of  $\Lv_{i_l}$.  Computing $|\bV_{i_l,T}^{0}|$ in the same way and
    using   (\ref{deflio})  proves  (\ref{Citpostc}).

 \section{Updating shrinkage parameters in Step~(S)} \label{sampleshr}

 Under a hierarchical Gaussian prior on
 the factor loadings $\load_{ij}$,
  Steps~(P), (D) and (L) 
 are performed conditional on all hyperparameters of this prior. 
 All local, column specific and global scaling parameters in the prior of the loadings $\load_{ij}$
  are updated  in Step~(S) of Algorithm~\ref{Algo3}.
 This step relies on a representation of an F-distribution as a Gamma scale mixture of inverse gamma distributions and avoids the GIG-distribution which was used in  \citet{cad-etal:tri}.\footnote{A r.v. $X  \sim \Fd{2a,2c}$ has a representation as a Gamma scale mixture of inverse gamma distributions:
\begin{eqnarray}
 X| b   \sim \Gammainv{c,b}, \quad b \sim \Gammad{a,\frac{a}{c}}.
 \end{eqnarray}
 Hence,
 \begin{eqnarray}
 b| X   \sim \Gammad{a+c,\frac{a}{c}+ \frac{1}{X}}.
 \end{eqnarray}
 If $Y|X \sim \Normal{0,dX}$, then
 \begin{eqnarray}
 X| b,Y   \sim \Gammainv{c+\frac{1}{2},b + \frac{Y^2}{2d}}.
 \end{eqnarray}
 }

 \paragraph*{Step~(S-a).}  Under  prior (\ref{priorEXP4}), sample the local shrinkage parameters
 $\tauloc_{ij}$ using the representation of the F-distribution as a scale mixture of inverse gamma distributions:
 \begin{eqnarray*}
 \tauloc_{ij}|  \bloc_{ij}    \sim \Gammainv{\cloc , \bloc_{ij}}, \quad \bloc_{ij}  | \tauloc_{ij}  \sim \Gammad{\aloc,\frac{\aloc}{\cloc}},
 \end{eqnarray*}
This yields a two step sampler, where $\bloc_{ij} |\tauloc_{ij} $
is  imputed from
 \begin{eqnarray} \label{impibij}
 \bloc_{ij}  | \tauloc_{ij}  \sim \Gammad{\aloc+\cloc,\frac{\aloc}{\cloc}+ \frac{1}{\tauloc_{ij}}},
 \end{eqnarray}
and   $\tauloc_{ij}|  \bloc_{ij}, {\load}_{ij}, \delta_{ij}  $ given $\bloc_{ij}$ 
is sampled  from
 \begin{eqnarray}
\tauloc_{ij}|  \bloc_{ij}, {\load}_{ij} , \delta_{ij}
   \sim \Gammainv{\cloc  + \delta_{ij} \frac{1}{2}, \bloc_{ij} +\delta_{ij} \frac{{\load}_{ij}^2}{2 \tauglob \taucol_j \idiov_i}}  . \label{postlocshr}
\end{eqnarray}
This has to be done for the entire matrix $\taulocv_{\nfactrue}$, as  $\tauloc_{ij}$ is needed
to compute the odd likelihood ratio $\odd_{ij}$ of  $\delta_{ij}=1$ versus
$\delta_{ij}=0$  (see (\ref{samplpost})) also above the current
pivot.

 \paragraph*{Step~(S-b).} Sample the 
 column specific  shrinkage parameters $\taucolv_{\nfactrue}$ (if any). For $\taucol_j  \sim \Gammainv{\ccol,\bcol_j}$, this
yields for all $j=1$ to $\nfactrue$:
    \begin{eqnarray}
 \taucol_j | \tauglob  ,  \facload_{\nfactrue}, \bcol_j    \sim \Gammainv{\ccol  + \frac{d_j}{2}, \bcol _j +
\frac{1}{2 \tauglob}
\sum_{i:\delta_{ij}=1} ^\dimy \frac{{\load}_{ij}^2}{\idiov_i \tauloc_{ij}}} . \label{postExppinv}
\end{eqnarray}
For $\taucol_j  \sim \Fd{2\acol,2\ccol}$, we use again the representation of the F-distribution as a scale mixture of inverse gamma distributions:
 \begin{eqnarray*}
\taucol_j  | \bcol_j   \sim \Gammainv{\ccol,\bcol_j}, \quad \bcol_j \sim \Gammad{\acol,\frac{\acol}{\ccol}},
 \end{eqnarray*}
 and impute $\bcol_j | \taucol_j$ from
 \begin{eqnarray} \label{imputeb}
 \bcol_j | \taucol_j   \sim \Gammad{\acol+\ccol,\frac{\acol}{\ccol}+ \frac{1}{\taucol_j}},
 \end{eqnarray}
 before we update  $\taucol_j | \tauglob  ,  \facload_{\nfactrue}$ from (\ref{postExppinv}).

 \paragraph*{Step~(S-c).} If $\tauglob$ is random  with prior $\tauglob  \sim \Gammainv{\cglob,\bglob}$,
 then  $\tauglob| {\facload}_{\nfactrue}, \Vare_{\nfactrue}, \taulocv_{\nfactrue}$ is updated
 from
  \begin{eqnarray}
 \tauglob| {\facload}_{\nfactrue}, \Vare_{\nfactrue},  \taulocv_{\nfactrue}   \sim \Gammainv{\cglob  + \frac{d}{2}, \bglob +  1 /2  \sum_{i=1} ^\dimy \frac{1}{\idiov_i}   \sum_{j=1} ^\nfactrue \delta_{ij}\frac{{\load}_{ij}^2}{\taucol_j \tauloc_{ij}}}, \label{ptauppinv}
\end{eqnarray}
where $d=\sum_{i=1} ^\dimy \sum_{j=1} ^\nfactrue \delta_{ij}$ is the total number of non-zero loadings in the  CFA model.
If
$\tauglob  \sim \Fd{2\aglob,2\cglob}$, then using  the same representation of the F-distribution,
we impute $\bglob | \tauglob $ from
 \begin{eqnarray} \label{imputebkappa}
\bglob | \tauglob  \sim \Gammad{\aglob + \cglob,\frac{\acol}{\ccol}+ \frac{1}{\tauglob}},
 \end{eqnarray}
 before we update  $\tauglob | {\facload}_{\nfactrue}, \Vare_{\nfactrue},  \taulocv_{\nfactrue}  $ from (\ref{ptauppinv}).

\section{Details on split and merge in Step~(R)} %
 \label{RJdetails}

\paragraph*{Proof of (\ref{prispnew}).}
For any zero column $\jsp$ in a sparsity matrix $\deltav_{\nfac}$ with
$\nfactrue$ active and $\nsp$ spurious columns, there are
$\dimy-(\nfactrue+\nsp)$ unrestricted elements that can be subjected to variable selection
and take the value one with probability $\tau_{\jsp} | k \sim  \Betadis{\aIBP,\bIBP}$ following the prior (\ref{prialt}).
Hence,
\begin{eqnarray} \label{gggzero}
& \Prob{d_{\jsp}=0| \nfactrue,\nsp} = \int (1-\tau_{\jsp})^{\dimy-(\nfactrue+\nsp)} p(\tau_{\jsp})
d \, \tau_{\jsp} = & \\ \nonumber
& \displaystyle \frac{\Betafun{\aIBP,\bIBP+\dimy-(\nfactrue+\nsp)}}{\Betafun{\aIBP,\bIBP}} &
 \end{eqnarray}
whereas
\begin{eqnarray} \nonumber
& \Prob{d_{\jsp}=1| \nfactrue,\nsp} =
 \displaystyle  (\dimy-(\nfactrue+\nsp))
 \int \tau_{\jsp} (1-\tau_{\jsp})^{\dimy-(\nfactrue+\nsp)-1} p(\tau_{\jsp})
d \, \tau_{\jsp} = &\\   \label{gggsp}
&  \displaystyle  \frac{(\dimy-(\nfactrue+\nsp)) \Betafun{\aIBP+1,\bIBP+\dimy-(\nfactrue+\nsp)-1}}{\Betafun{\aIBP,\bIBP}} .&
 \end{eqnarray}
Therefore,
\begin{eqnarray*} %
 \frac{\Prob{d_{\jsp}=1| \nfactrue,\nsp}}{\Prob{d_{\jsp}=0|\nfactrue,\nsp}}=
 \frac{\aIBP(\dimy-\nfactrue- \nsp) }{\bIBP + m - \nfactrue - \nsp -1} .
 \end{eqnarray*}

\paragraph*{Proposing  factors in a spurious column - Proof of (\ref{mainonfsp}).}

Whenever a zero column is turned into a spurious column (with columns index $\jsp$),
   factors $\facm_{\jsp} =
(\fac_{\jsp,1}, \ldots, \fac_{\jsp, T}) $ are proposed, while holding the factors $\facmk{\nfactrue}{t}, t=1, \ldots,T,$ in all active columns  fixed.
 Draws of   $\fac _{\jsp, t}$  could be proposed from the prior $\fac_{\jsp, t}  \sim \Normal{0,1}$, since column $\jsp$ was a zero column before splitting. However, with $\lsp$ being the pivot row,   $y_{\lsp,t}$  is a measurement that contains information about $f_{\jsp, t}$ in a spurious column
 and its likelihood can be combined with the prior to define  the  conditional posterior density $ p(\fac_{\jsp, t}|  \facmk{\nfactrue}{t},\facload_\nfactrue, \idiov_{\lsp}, y_{\lsp,t})$  given $y_{\lsp,t}$ in addition to $\facmk{\nfactrue}{t},\facload_\nfactrue, \idiov_{\lsp}, y_{\lsp,t}$.

 It is easy to verify from the filter given in (\ref{filtPXsim}) that for a spurious column $\jsp$ with leading element
 $\loadsp_{\lsp}$,  the conditional density $ p(\fac_{\jsp, t}|  \facmk{\nfactrue}{t},\facload_\nfactrue, \idiov_{\lsp}, y_{\lsp,t})$ of $\fac_{\jsp, t}$  is given by:
 \begin{eqnarray*} %
&\fac_{\jsp, t}|  \facmk{\nfactrue}{t},\facload_\nfactrue, \idiov_{\lsp}, y_{\lsp,t}  \sim \Normal{E_{\jsp, t},V_{\jsp} }, &\\
&\displaystyle V_{\jsp}  = \left(1+ \frac{\loadsp_{\lsp} ^2 }{\idiovsp}\right)^{-1} =
\frac{\idiovsp}{\idiovsp + \loadsp_{\lsp}^2 }, \quad
 E_{\jsp, t} = \frac{V_{\jsp}  \loadsp_{\lsp}  }{\idiovsp} \tilde{y}_{\lsp,t} =
 \frac{\loadsp_{\lsp} }{\idiovsp + \loadsp_{\lsp} ^2 } \tilde{y}_{\lsp,t} , &\nonumber
\end{eqnarray*}
where  the pseudo outcome $\tilde{y}_{\lsp,t}$ is given by
$\tilde{y}_{\lsp,t}= y_{\lsp,t}- \facloadrowr{\nfactrue}{\lsp,\cdot} \facmk{\nfactrue}{t}$.
   Using the definition of $\loadsp_{\lsp}$ and
   $\idiovsp$  in terms of $\idiov_{\lsp}$ and a uniform random variable $U_{\jsp} \in [-1,1]$
   given in
   (\ref{prorjitA}), we obtain
 the  simple expressions for the posterior moments given in (\ref{mainonfsp}):
  \begin{eqnarray*}
 V_{\jsp}  = \frac{\idiovsp}{\idiovsp + \loadsp_{\lsp}^2} = 1- U_{\jsp} ^2, \quad
 E_{\jsp, t} = \frac{\loadsp_{\lsp}}{\idiovsp + \loadsp_{\lsp}^2} \tilde{y}_{\lsp,t} =
 \frac{U_{\jsp}}{\sqrt{\idiovsp}} \tilde{y}_{\lsp,t}  . \nonumber
\end{eqnarray*}

\paragraph*{Turning spurious into active columns.}
 In Step~(R-D), variable selection is  performed  on the spurious columns of the current EFA model  by marginalizing over the idiosyncratic variances and the factor loading matrix $(\facload_\nfactrue, \, \facloadsp)$ as in Step~(D) of Algorithm~\ref{Algo3}.
 By starting with the last
  spurious column, we ensure that the application of Step~(D)  is valid.
  Indeed, ordering the spurious pivots in Step~(L) by size guarantees that all rows of the current EFA model below $\lsp$ are equal to the corresponding rows of the current CFA model in the first
  $\nfactrue$  columns and zero in all other columns. Hence, updating
   the zero  column $\nfactrue+1$ below $\lsp$ can be regarded as an attempt to introduce an
   additional active column in the current CFA model and to increase $\nfactrue$, while $\nsp$ decreases. If a column remains spurious, then we integrate
in the current EFA model immediately over the corresponding factor,
while the current number of spurious columns $\nsp$  is preserved.

\paragraph*{Step (R-Da) for hierarchical
Gaussian priors with unknown shrinkage factors.}
  Step (R-Da) has to be adjusted slightly for hierarchical
Gaussian priors with unknown shrinkage factors.
A column specific shrinkage parameter $\taucol_{\jsp}$ is needed to compute the odds of turning zero loadings in the spurious column into non-zero ones. In addition, under  the  shrinkage prior (\ref{priorEXP4}), local shrinkage parameters $\tauloc_{i,\jsp}$ are needed for all rows $i=1, \ldots, \dimy$ in the spurious column $\jsp$ (note that $\tauloc_{i,\jsp}$ will be needed later in step~(L), even if $i<\lsp$ lies above the spurious pivot element).
 All these quantities could be simply simulated from the prior. However, we found considerably gain in exploited the information
contained in $\loadsp_{\lsp}$ and $\idiovsp$, when sampling $\taucol_{\jsp}$ and $\tauloc_{\lsp,\jsp}$, which basically results from conditioning on the r.v. $U_{\jsp}$.

For the Student-$t$ slab prior (\ref{priorEXP2}),
the information
contained in $\loadsp_{\lsp}$ and $\idiovsp$ is exploited by sampling $\taucol_{\jsp}|
\loadsp_{\lsp}, \idiovsp$ from (\ref{postExppinv}):
 \begin{eqnarray}
 \taucol_{\jsp} | \tauglob  , U_{\jsp}   \sim \Gammainv{\ccol  + \frac{1}{2}, \bcol _{\jsp}
 + \frac{U_{\jsp} ^2}{2 \tauglob (1-U_{\jsp} ^2) }},  \label{RJcol}
\end{eqnarray}
where the scale  parameter is obtained from combining (\ref{prorjitA}) and (\ref{postExppinv}):
\begin{eqnarray*}
 \frac{\load_{\lsp, \jsp}^2}{(\Vare_{\nfactrue+1})_{\lsp, \lsp} } = \frac{\loadsp_{\lsp} ^2}{\idiovsp }= \frac{U_{\jsp} ^2}{1-U_{\jsp} ^2}.
\end{eqnarray*}
Under the prior $\taucol_{\jsp} \sim \Gammainv{\ccol,\bcol}$, the scale $\bcol _{\jsp}=\bcol$,
whereas under the F-prior on $\taucol_{\jsp}$,
  $\bcol _{\jsp}$ is imputed from the prior. This is achieved by sampling
 $\taucol_{\jsp} \sim \Fd{2\acol,2\ccol} $ and $\bcol _{\jsp}| \taucol_{\jsp}$ from (\ref{imputeb}):
 \begin{eqnarray*} %
\bcol _{\jsp}| \taucol_{\jsp} \sim \Gammad{\acol+\ccol,\frac{\acol}{\ccol}+ \frac{1}{\taucol_{\jsp}}}.
 \end{eqnarray*}
For the  shrinkage prior (\ref{priorEXP4}),
first local shrinkage parameter $\tauloc_{i,\jsp}$ are sampled from the prior for all rows, i.e. $\tauloc_{i,\jsp} \sim \Fd{2\aloc,2\cloc}$, $i=1, \ldots, \dimy$.
The information 
contained in $\loadsp_{\lsp}$ and $\idiovsp$ is then exploited by sampling $\taucol_{\jsp}|
\loadsp_{\lsp}, \idiovsp, \tauloc_{\lsp,\jsp} $ from (\ref{postExppinv}):
 \begin{eqnarray*}
 \taucol_{\jsp} | \tauglob  , U_{\jsp}, \tauloc_{\lsp,\jsp}   \sim \Gammainv{\ccol  + \frac{1}{2}, \bcol_{\jsp}
 + \frac{U_{\jsp} ^2}{2  \tauglob \tauloc_{\lsp,\jsp}  (1-U_{\jsp} ^2) }}, %
\end{eqnarray*}
where $\bcol _{\jsp}$ is defined as for the  Student-$t$ slab prior (\ref{priorEXP2}).
Finally, we resample $\tauloc_{\lsp,\jsp}$  by first imputing $ \bloc_{\lsp,\jsp}$ from
the prior by sampling  $\bloc_{\lsp,\jsp}|  \tauloc_{\lsp,\jsp}$ from (\ref{impibij}):
\begin{eqnarray*}
 \bloc_{\lsp,\jsp}  | \tauloc_{\lsp,\jsp}  \sim \Gammad{\aloc+\cloc,\frac{\aloc}{\cloc}+ \frac{1}{\tauloc_{\lsp,\jsp}}},
 \end{eqnarray*}
 (where $\tauloc_{\lsp,\jsp}$ has been sampled from the prior in the first step)
 and then sampling from (\ref{postlocshr}):
\begin{eqnarray*}
\tauloc_{\lsp,\jsp}| \bloc_{\lsp,\jsp} ,  \taucol_{\jsp} , \tauglob  , U_{\jsp}   \sim
 \Gammainv{\cloc  +  \frac{1}{2}, \bloc_{\lsp,\jsp} +
 \frac{U_{\jsp} ^2}{2 \tauglob \taucol_{\jsp}   (1-U_{\jsp} ^2) }}.
 \end{eqnarray*}
In this way, we exploit the information  contained in $\loadsp_{\lsp}$ and $\idiovsp$ for both shrinkage parameters.

\section{Step~(L) - Updating the pivots}    \label{updatelead}

This subsection provides details on Step~(L) of Algorithm~\ref{Algo3}, which was briefly discussed in  Section~\ref{movelead}.  Four local moves are applied which were illustrated  in Figure~\ref{figStepL}  in  Section~\ref{movelead}.

\paragraph{Shifting the pivot.}   A  shift move  is selected with probability $\pshift$.
Let   $l_\star$ denote the index of the first nonzero row below $l_j$ (i.e. $\delta_{l_\star,j}= 1$, $\delta_{ij}= 0, l_j< i < l_\star$). If $l_\star>2$, then  a new  pivot
  $l \new _j$ is proposed
by  sampling  uniformly from
the set ${\cal M} (  l_\star , \lmr{\nfactrue}{-j} )=\{1, \ldots, l_\star-1\} \cap \leadset{\lmr{\nfactrue}{-j}}$,
 where $\leadset{\lmr{\nfactrue}{-j}}$ is the set of pivots outside of column $j$. %
If   ${\cal M} (  l_\star , \lmr{\nfactrue}{-j})$ is empty, then no shift  move is performed.  Otherwise, two indicators in column $j$ are changed, namely
 $\delta_{l_j \new,j} $ from zero to one and  $\delta \new_{l_j ,j}$ from one to zero,  while the remaining elements of $\deltav_\nfactrue$ are unchanged.
The new indicator matrix $\deltav_\nfactrue \new $ is accepted
   with probability   $\min(1,\alpha_{\mbox{\rm \footnotesize shift}})$, where
 \begin{eqnarray*}
\alpha_{\mbox{\rm \footnotesize shift}}
= \exp(\odd_{l_j \new,j} - \odd_{l_j,j})  \oddratl _{\mbox{\rm \footnotesize shift}},
\end{eqnarray*}
with $\odd_{ij}$  being the log likelihood  ratio of $\delta_{ij}=1$ versus $\delta_{ij}=0$  defined in (\ref{samplpost}).
A shift move does not change the number of nonzero elements $d_j$ in column $j$ and the prior ratio
(\ref{prirat3A}) of this move simplifies to:
 \begin{eqnarray} \label{prirat3}
\oddratl _{\mbox{\rm \footnotesize shift}} =
   \frac{\Betafun{\aIBP  + d_j  -1,\bIBP + \dimy-l_j \new - d_j+1}}{\Betafun{\aIBP  + d_j-1,\bIBP + \dimy-l_j- d_j+1}}.
   \end{eqnarray}
Since $l \new _j$ is sampled from a set ${\cal M} (  l_\star , \lmr{\nfactrue}{-j} )$
that does not depend on $l_j$,
the proposal density is symmetric   and  the proposal ratio
   $q(l \new _j|   l_\star, \lmr{\nfactrue}{-j})/ q(l  _j|    l_\star, \lmr{\nfactrue}{-j})$
   cancels from $\alpha_{\mbox{\rm \footnotesize shift}}$.

 \paragraph{Switching  pivots.}

 This move is selected with probability $p_{\mbox{\rm \footnotesize switch}}$
 and is performed only if  $\nfactrue>1$.
A nonzero column $\jtwo \neq j$ is selected randomly and  all indicators between (and including) row  $l_j$  and $l_ \jtwo$
that are different  are switched between the two columns, i.e. $\delta_{ij}\new=1-\delta_{ij}$
and $\delta_{i \jtwo}\new=1-\delta_{i \jtwo}$ for all $i \in  {\cal S}_{j,\jtwo} =\{i: \min (l _\jtwo ,  l_j) \leq i \leq  \max (l _\jtwo ,  l_j), \delta_{ij} \neq \delta_{i \jtwo}\}$.
 Evidently, this move  switches pivots between the two columns. Since the corresponding proposal density satisfies $q(\deltav _\nfactrue\new|\deltav _\nfactrue )=
   q(\deltav_\nfactrue | \deltav _\nfactrue\new)$,  $\deltav_\nfactrue \new $  is accepted
   with probability   $\min(1,\alpha_{\mbox{\rm \footnotesize switch}})$, where
\begin{eqnarray*}
\alpha_{\mbox{\rm \footnotesize  switch}} = \prod _{i \in {\cal S}_{j,\jtwo}}
\frac{p (\tilde{\ym}_i| \delta_{ij}\new,\delta_{i\jtwo}\new , \deltav_\nfactrue^\uncon, \facm_\nfactrue)}
{p (\tilde{\ym}_i| \delta_{ij} ,\delta_{i\jtwo} , \deltav_\nfactrue^\uncon,  \facm_\nfactrue)}
\oddratl _{\mbox{\rm \footnotesize  switch}},
\end{eqnarray*}
where the prior odds ratio of this move %
 is  derived from (\ref{prirat3A}):
 \begin{eqnarray} \label{prirat3AB}
  \oddratl_{\mbox{\rm \footnotesize  switch}}= \prod_{{\tilde l}=j,l}
   \frac{\Betafun{\aIBP  + d_{\tilde l} \new -1,\bIBP + \dimy-l_{\tilde l} \new - d_{\tilde l} \new +1}}{\Betafun{\aIBP  + d_{\tilde l}-1,\bIBP + \dimy-l_{\tilde l}- d_{\tilde l}+1}}.
 \end{eqnarray}
If $\delta_{ij}=0$ (and consequently $ \delta_{i \jtwo}=1$), then we obtain:
 \begin{eqnarray*}
 && \frac{p (\tilde{\ym}_i| \delta_{ij}\new =1 ,\delta_{i\jtwo} \new =0 , \deltav_\nfactrue^\uncon,  \facm_\nfactrue)}
{p (\tilde{\ym}_i| \delta_{ij}=0 ,\delta_{i\jtwo}=1 , \deltav_\nfactrue^\uncon,  \facm_\nfactrue)} =\\
&& \frac{p (\tilde{\ym}_i| \delta_{ij}\new =1 ,\delta_{i\jtwo} \new =0 , \deltav_\nfactrue^\uncon,  \facm_\nfactrue)}
{p (\tilde{\ym}_i| \delta_{ij}=0 ,\delta_{i\jtwo}\new =0 , \deltav_\nfactrue^\uncon,  \facm_\nfactrue)}
\frac{p (\tilde{\ym}_i| \delta_{ij}=0 ,\delta_{i\jtwo}\new =0 , \deltav_\nfactrue^\uncon,  \facm_\nfactrue)}
{p (\tilde{\ym}_i| \delta_{ij}=0 ,\delta_{i\jtwo}=1 , \deltav_\nfactrue^\uncon,  \facm_\nfactrue)}=
\exp( \odd_{ij|\jtwo}  - \odd_{i\jtwo|j}),
 \end{eqnarray*}
 where $\odd_{i,  j_1| j_2}$ is the log likelihood ratio  of $\delta_{i,j_1}=1$ versus $\delta_{i,j_1} =0$  provided that the indicator $ \delta_{i,j_2}=0$.
 It can be obtained  as  the likelihood ratio   $\odd_{i , j_1 }$ given in (\ref{samplpost}),  with  $ \delta_{i,j_2}=0$   for both models.
 On the other hand, if $\delta_{ij}=1$ (and consequently $\delta_{i  \jtwo}=0$), then
  \begin{eqnarray*}
 && \frac{p (\tilde{\ym}_i| \delta_{ij}\new =0 ,\delta_{i\jtwo} \new =1 , \deltav_\nfactrue^\uncon,  \facm_\nfactrue)}
{p (\tilde{\ym}_i| \delta_{ij}=1 ,\delta_{i\jtwo}=0 , \deltav_\nfactrue^\uncon,  \facm_\nfactrue)} =\\
&& \frac{p (\tilde{\ym}_i| \delta_{ij}\new =0 ,\delta_{i\jtwo} \new =1 , \deltav_\nfactrue^\uncon,  \facm_\nfactrue)}
{p (\tilde{\ym}_i| \delta_{ij} \new =0 ,\delta_{i\jtwo} =0 , \deltav_\nfactrue^\uncon,  \facm_\nfactrue)}
\frac{p (\tilde{\ym}_i| \delta_{ij} \new =0 ,\delta_{i\jtwo}  =0 , \deltav_\nfactrue^\uncon,  \facm_\nfactrue)}
{p (\tilde{\ym}_i| \delta_{ij}=1 ,\delta_{i\jtwo}=0 , \deltav_\nfactrue^\uncon,  \facm_\nfactrue)}=
\exp(  \odd_{i\jtwo|j} -  \odd_{ij| \jtwo}).
 \end{eqnarray*}
 Therefore
 \begin{eqnarray} \label{propospdd}
\alpha_{\mbox{\rm \footnotesize  switch}} = \oddratl _{\mbox{\rm \footnotesize  switch}}
\exp \left(  \sum_{i \in {\cal S}_{j,\jtwo}: \delta_{ij}=0 } ( \odd_{ij | \jtwo}  - \odd_{i\jtwo|j}) +   \sum_{i \in {\cal S}_{j,\jtwo}: \delta_{ij}=1 }   (\odd_{i\jtwo|j} - \odd_{ij| \jtwo}  ) \right).
\end{eqnarray}
 This move allows  changes in $d_j$ and $d_\jtwo$, but leaves the overall number $d$ of nonzero elements unchanged.

 \paragraph{Adding or deleting a pivot.}
 Finally, a reversible pair of moves is selected  with probability $1- \pshift - p_{\mbox{\rm \footnotesize switch}}$. The add move
introduces  a new pivot $l_j \new $ in a row
   above the current pivot $l_j$ which is not occupied by the pivots
   of the other columns.
   Hence,  $l \new _j$ is selected randomly from
the set ${\cal A} (l_j,\lmr{\nfactrue}{-j} )=\{1, \ldots, l_j-1\} \cap \leadset{\lmr{\nfactrue}{-j}}$,  i.e. $\delta_{l_j \new,j} \new= 1$,  while the remaining elements of   $\deltav_{\nfactrue}$ are unchanged (in particular    $\delta_{l_j,j} \new=\delta_{l_j,j}=1$). An add move is only possible, if   $|{\cal A} (l_j,\lmr{\nfactrue}{-j} )| >0$.\footnote{The number of  rows in ${\cal A} (l_j,\lmr{\nfactrue}{-j} )$ is equal to %
$l_j- \rankl_{j}$, where  $\rankl_{j}=\Count{l_{j'} \in \lm: l_{j'} < l_j}$  is the rank of $l_j$ among the leading indices.
Hence, an add move is possible, whenever  $l_j > \rankl_{j}$.} %

The corresponding reverse move  is deterministic and  deletes the current pivot $l_j$,  turning
$l_j \new=l_\star$  into the new pivot where $l_\star$ is the row index of the first nonzero element in $\deltav_{\nfactrue}$ below $l_j$. Hence,  $\delta_{l_j,j} \new=0$, while all other elements  of $\deltav_{\nfactrue}$ remain unchanged.
A delete move is only performed,  if  $l_\star$ is not a pivot
in any other column of $\deltav_{\nfactrue}$ (that is  $ l_\star  \in  \leadset{\lmr{\nfactrue}{-j}}$).

 If  neither an add nor a delete move is possible, then $l_j$ remains unchanged.
 Otherwise,   either an add or a delete move is selected with probability $p_{\mbox{\rm \footnotesize add}}(\deltav_\nfactrue)$ and $1-p_{\mbox{\rm \footnotesize add}}(\deltav_\nfactrue)$.
 If both add and delete moves are possible, then $p_{\mbox{\rm \footnotesize add}}(\deltav_\nfactrue)=p_a$, with $p_a$ being a tuning parameter;
 if only an add move is possible,  then   $p_{\mbox{\rm \footnotesize add}}(\deltav_\nfactrue)=1$, whereas
   $p_{\mbox{\rm \footnotesize add}}(\deltav_\nfactrue)=0$, if only a delete move is possible.
Note that whenever an add move is selected, the reverse delete move is always possible; and similarly,  the reverse add move is always possible,  whenever a delete move is selected.
    This move changes $d_j$ and increases or decreases the overall number of nonzero elements by one.

    The acceptance probability for an add  move is equal to $\min(1,\alpha_{\mbox{\rm \footnotesize add}})$, with
\begin{eqnarray*} %
\alpha_{\mbox{\rm \footnotesize add}} =
  \exp(\odd_{l_j \new,j}) \oddratl _{\mbox{\rm \footnotesize add}}
 \frac{|{\cal A} (l_j,\lmr{\nfactrue}{-j} ) | (1-  p_{\mbox{\rm \footnotesize add}}(\deltav_\nfactrue \new ) )}{p_{\mbox{\rm \footnotesize add}}(\deltav _\nfactrue)},
\end{eqnarray*}
where $\odd_{ij}$ is the log likelihood ratio of $\delta_{ij}=1$ versus $\delta_{ij}=0$ defined in (\ref{samplpost}). Since $d_j \new= d_j+1$,
the prior ratio
(\ref{prirat3A}) of this move simplifies to:
 \begin{eqnarray} \label{Radd}
 \oddratl_{\mbox{\rm \footnotesize add}} =   \frac{\Betafun{\aIBP  + d_j ,\bIBP + \dimy-l_j \new - d_j}}{\Betafun{\aIBP  + d_j-1,\bIBP + \dimy-l_j- d_j+1}}.
\end{eqnarray}
The acceptance probability for a  delete  move is equal to
$\min(1,\alpha_{\mbox{\rm \footnotesize del}})$,
  with
 \begin{eqnarray*}
\alpha_{\mbox{\rm \footnotesize del}} = \exp(-\odd_{l_j,j}) \oddratl_{\mbox{\rm \footnotesize del}} \frac{p_{\mbox{\rm \footnotesize add}}(\deltav _\nfactrue\new)}{  |{\cal A} (l_j \new ,\lmr{\nfactrue}{-j})| (1-p_{\mbox{\rm \footnotesize add}}(\deltav _\nfactrue)) }.
\end{eqnarray*}
Since $d_j \new= d_j-1$, the prior ratio
(\ref{prirat3A}) of this move simplifies to:
 \begin{eqnarray} \label{Rdel}
\oddratl _{\mbox{\rm \footnotesize del}} =   \frac{\Betafun{\aIBP  + d_j -2 ,\bIBP + \dimy-l_j \new - d_j + 2}}{\Betafun{\aIBP  + d_j-1,\bIBP + \dimy-l_j- d_j+1}}.
\end{eqnarray}

\paragraph{Tuning parameters.}
  These four moves involve three tuning probabilities, namely  $\pshift$, $p_{\mbox{\rm \footnotesize switch}}$, and $p_a$, with $1- \pshift - p_{\mbox{\rm \footnotesize switch}}>0$ and $0 < p_a <1$.

\section{Details on Step~(H)}\label{detailsAppH}

In this section, we provide details on 
 updating the  hyperparameters $\alphaIBP$ and $\betaIBP$
for the 2PB  prior (\ref{prialtTeh}) without conditioning on the slab probabilities $\tau_1, \ldots, \tau_k$. Inference is performed conditionally on 
the current columns sizes $d_1, \ldots, d_{\nfac}$  and the number of spurious columns $\nsp$ in the EFA model and the current pivots in the CFA model:
\begin{eqnarray*} \label{likabet}
 p(\alphaIBP, \betaIBP|\cdot) \propto p(\alphaIBP) p( \betaIBP) 
 \prod_{j: d_j>1} \Prob{\deltacolr{\nfactrue}{j}|\alphaIBP, \betaIBP, l_j}  \, 
\prod_{j: d_j \leq 1}  \Prob{d_j|\alphaIBP, 
 \betaIBP},
 \end{eqnarray*}
  where  $\alphaIBP \sim \Gammad{\aalpha, \balpha}$ and 
 $\betaIBP \sim \Gammad{\abeta, \bbeta}$.

The conditional posterior $p(\alphaIBP, \betaIBP|\cdot)$ is based on the likelihood (\ref{prigen3}) for active columns, the  likelihood
(\ref{gggsp}) 
for spurious columns and the likelihood
(\ref{gggzero}) for zero columns:\footnote{Note that in  (\ref{gggsp}), $\nsp$ is the number of spurious columns excluding column $\jsp$. Hence, the likelihood is computed sequentially from the first to the last spurious column.}
\begin{eqnarray*} %
& \displaystyle  p(\alphaIBP, \betaIBP|\cdot)  \propto  \frac{p(\alphaIBP) p( \betaIBP)}{\Betafun{\frac{\alphaIBP \betaIBP }{\nfac},\betaIBP}^\nfac} \cdot \Betafun{\frac{\alphaIBP \betaIBP}{\nfac},\betaIBP+\dimy-\nfactrue-\nsp)}
^{\nfac-\nfactrue-\nsp} &\\
& \displaystyle \cdot \prod_{j: d_j>1}\Betafun{ \frac{\alphaIBP \betaIBP}{\nfac}  + d_j-1,\betaIBP + \dimy-l_j- d_j+1} \cdot
 \prod_{\jsp=1}^{\nsp} \Betafun{\frac{\alphaIBP \betaIBP}{\nfac}+1,\betaIBP+\dimy-\nfactrue-\jsp} .
 &
 \end{eqnarray*}
 An MH-step  based on a random walk proposal for, respectively, $\log \alphaIBP$ and $\log \betaIBP$ 
 is implemented  
 to sample  $p(\alphaIBP |\betaIBP, \cdot)$  and $p(\betaIBP| \alphaIBP , \cdot)$,. 
\section{More about boosting in  Step~(A)} \label{boost_frac}

 Algorithm~\ref{AlgoA} described a generic
 boosting strategy in  Step~(A) in Algorithm~\ref{Algo3}.

 \begin{alg}[\textbf{Boosting MCMC}]\label{AlgoA}
  Step~(A) in Algorithm~\ref{Algo3} is implemented in three steps:
  \begin{itemize}
   \item[(A-a)]    Choose a  (current) value $\Psiv $  %
and  move from the CFA  model (\ref{fac1CFA}) to the expanded  model (\ref{fac1px})   using
the following transformation: %
\begin{eqnarray} \label{fac5px}
 \facmktilde{\nfactrue}{t} = (\Psiv)^{1/2} \facmk{\nfactrue}{t} ,  \quad  \tilde{\facload}_\nfactrue = \facload_\nfactrue  (\Psiv)^{-1/2}.
\end{eqnarray}

\item[(A-b)] Sample a new value   $\Psiv \new$ in model (\ref{fac1px}) conditional on   $\tilde{\facm}_\nfactrue=(\facmktilde{\nfactrue}{1}, \ldots, \facmktilde{\nfactrue}{T})$ and   $\tilde{\facload}_\nfactrue$  from the conditional posterior $p(\Psiv | \tilde{\facm}_\nfactrue, \tilde{\facload}_\nfactrue ) $ given by
    \begin{eqnarray}
 p(\Psiv | \tilde{\facm}_\nfactrue, \tilde{\facload}_\nfactrue )
 \propto p(\Psiv) p(\tilde{\facload} _\nfactrue | \Psiv )
  \prod_{j=1}^\nfactrue   \Psi_j ^{-T/2} \exp \left\{ - \frac{1}{2 \Psi_j} \sum_{t=1}^T \tilde{\fac}_{jt}^2 \right\} .   \label{fac6px}
\end{eqnarray}

\item[(A-c)] Move   back from   model (\ref{fac1px}) to  the CFA   model (\ref{fac1CFA}) using
  the inverse of transformation (\ref{fac5px}) for $j=1, \ldots, \nfactrue$:
\begin{eqnarray*}
 \load_{ij}\new  = \load_{ij}  \sqrt{\Psi_j \new/\Psi_j }, %
 \quad i=1,\ldots,\dimy, %
 \quad \fac_{jt}\new = \fac_{jt}  \sqrt{ \Psi_j /\Psi_j \new} %
  \quad t=1,\ldots,T.
\end{eqnarray*}
\end{itemize}
\end{alg}

 \noindent Boosting  affects the  factor loading matrix $\facload_\nfactrue$ and all factors $\facm_\nfactrue =(\facmk{\nfactrue}{1},
\ldots, \facmk{\nfactrue}{T})$, see Step~(A-c).
 The main difference between ASIS and MDA  lies in the choice of  the current value of $\Psiv$ in Step~(A-a).
 While $\Psi_j $  is chosen in a deterministic fashion for ASIS,  
  typically involving a specific factor loading $\load_{n_j,j}$ in each column,
 $\Psi_j $ is  sampled from a working prior for MDA.
 This  leads  to
different priors  $p(\Psiv)$  and $ p(\tilde{\facload}_\nfactrue  | \Psiv)$ in (\ref{fac6px}).

\paragraph*{ASIS for fractional priors.}
For boosting based on  ASIS, a nonzero factor loading $\load_{n_j,j}$  is chosen in   column $j$
 to define
the current value of  $\Psi_j$ as $\sqrt{\Psi_j}=\load_{n_j,j}$. This creates a factor loading matrix $\tilde{\facload}_\nfactrue$ in the expanded model (\ref{fac1px}) where
 $\tilde{\load}_{n_j,j}=1$  whereas $\tilde{\load}_{i,j}= \load_{ij}/\load_{n_j,j}$ for $i\neq n_j$.  $\load_{n_j,j}$ can be chosen as the pivot element in each column, i.e. $n_j=l_j$, or
 such that $|\load_{n_j,j}|$ is maximized for all loadings in column $j$.  Apart from this choice, ASIS requires no further tuning.
 The  implied prior of $\Psi_j=  \load_{n_j,j}^2$ is given by  $p(\Psi_j) \propto  \Psi_j^{-1/2}$, whereas
$  p(\tilde{\facload}_\nfactrue | \Psiv ) \propto \prod_{j=1}^\nfactrue \Psi_j ^{(d_j-1)/2}$,
since  $p(\tilde{\load_{ij}} |\idiov_i, \Psi_j,\delta_{ij}=1) \propto  \Psi_j^{1/2}$
for all $i\neq n_j$ with $\delta_{ij}=1$. Hence,
 $ p(\Psiv | \tilde{\facm}_\nfactrue, \tilde{\facload}_\nfactrue )$ defined in (\ref{fac6px}) reduces to a product of inverted Gamma distribution
  where for all $j = 1, \ldots,\nfactrue$:
 \begin{eqnarray*}
 \Psi_j \new |  \facm _\nfactrue, \facload _\nfactrue  \sim \Gammainv{\frac{T-d_j}{2}, \frac{\load_{n_j,j}^2}{2}  \sum_{t=1}^T \fac_{jt}^2 }, %
\end{eqnarray*}
where $d_j=\sum_{i=1} ^\dimy \delta_{ij}$ counts the  nonzero elements in column $j$ (which is
at least equal to 2 by definition of $\deltav_\nfactrue$).

\paragraph*{MDA for fractional priors.}
In MDA,  the current value $\Psiv$  is drawn from a working prior $p(\Psiv)$ which is independent  both  of $\facload_\nfactrue$ and $\Vare_\nfactrue$. This  guarantees that the prior distribution of  $\facload_\nfactrue$  remains unchanged, despite moving between the two models. Below, an inverted Gamma working prior $\Psi_j  \sim \Gammainv{\nu_j,q_j}$ is  applied.
The prior in the expanded model %
reads
$  p(\tilde{\facload} _\nfactrue | \Psiv ) \propto
 \prod_{j=1}^\nfactrue \Psi_j ^{d_j/2} $,
since  $p(\tilde{\load_{ij}} |\idiov_i, \Psi_j,\delta_{ij}=1) \propto  \Psi_j^{1/2}$
for all $i$ with $\delta_{ij}=1$.   The two last terms  in (\ref{fac6px}) factor  into a product of   independent inverted Gamma distributions  which leads to an inverted Gamma posterior for each $\Psi_j $:
  \begin{eqnarray*}
 \Psi_j \new | \Psi_j , \facm _\nfactrue   \sim \Gammainv{\nu_j - d_j/2 + \frac{T}{2}, q_j +  \Psi_j /2   \sum_{t=1}^T \fac_{tj}^2}. %
\end{eqnarray*}

\paragraph*{Boosting for hierarchical Gaussian priors.}
  Column boosting is based on interweaving $\taucol_j$
  into the state equation by choosing $\Psi_j= \taucol_j$.
  This leads to a prior for the loadings $\tilde \load_{ij} $ which is independent of $\taucol_j$,
  while $\taucol_j$ acts as  variance of the $j$th factors in the expanded model,
  $\tilde{\fac}_{jt} | \taucol_j \sim \Normal{0,\taucol_j}$.
Hence, in Step~(A), under the inverse gamma prior $\taucol_j \sim \Gammainv{\ccol,\bcol_j}$, $\taucol_j$
  is updated in the expanded model,
  \begin{eqnarray*}
 \taucol_j \new | \taucol_j , \facm _\nfactrue   \sim \Gammainv{\ccol  + \frac{T}{2},B_j},
  \quad B_j = \bcol_j +  1/2 \sum_{t=1}^T \tilde{\fac}_{jt}^2 =
   \bcol_j +  \taucol_j /2   \sum_{t=1}^T \fac_{tj}^2.
\end{eqnarray*}
 Moving back,  all factors and factor loadings in the CFA model are updated:
\begin{eqnarray*} %
 \load_{ij} \new = \sqrt{\taucol_j \new} \tilde \load_{ij}=
 \frac{\sqrt{\taucol_j \new}}{\sqrt{\taucol_j}}  \load_{ij},
 \quad
 \fac_{jt} \new = \tilde{\fac}_{jt} /\sqrt{\taucol_j \new}= \frac{\sqrt{\taucol_j}}{\sqrt{\taucol_j \new}} \fac_{jt} .
 \end{eqnarray*}
A boosting step that we found to be useful for hierarchical Gaussian priors in addition to column boosting is interweaving the global shrinkage parameter $\tauglob$ into the prior  $ \taucol_j \sim \Gammainv{\ccol, \bcol_j}$ of the shrinkage parameter
$\taucol_j$
through the transformation $\tilde \taucol_j= \tauglob  \taucol_j$:
 \begin{eqnarray*} %
&& \load_{ij}| \delta_{ij}=1, \tauloc_{ij}, \tilde \taucol_j, \idiov_i   \sim \Normal{0, \tilde \taucol_j \tauloc_{ij}\idiov_i}, \\ %
&& \tilde \taucol_j| \tauglob \sim \Gammainv{\ccol, \tauglob  \bcol_j}, \quad \tauglob \sim \Gammainv{\cglob,\bglob}.
\end{eqnarray*}
The global shrinkage parameter $\tauglob$ is then updated in Step~(A) under this representation, combining the likelihood
\begin{eqnarray*}
p( \tilde \taucol_1, \ldots, \tilde \taucol_\nfactrue | \tauglob) \propto
\tauglob^{\nfactrue \ccol} \exp \left(- \tauglob \sum_{j=1}^\nfactrue \frac{\bcol_j}{ \tilde \taucol_j} \right),
\end{eqnarray*}
 which is the kernel of a gamma density in $\tauglob$ with the inverse gamma prior $p(\tauglob)$. This yields a generalized inverse Gaussian distribution\footnote{The generalized inverse Gaussian distribution, $\rvY \sim \GIG{p,a,b}$, is a three-parameter family of probability distribution with support  $y \in \Real^+$. The  density is given by
\begin{eqnarray*}
\displaystyle  f(y) =
 \frac{(a/b)^{p/2}}{2 K_p(\sqrt{ab})} y^{p-1} e^{-(a/2)y} e^{ -b/(2y)},
\end{eqnarray*}
where $K_p(z)$ is the modified Bessel function of the second kind, $a > 0$, $b > 0$ and $p$ is a real parameter.}
to update $\tauglob \new$:
 \begin{eqnarray*}
\tauglob  \new | \taucol_1, \ldots, \taucol_\nfactrue,  \tauglob   \sim \GIG{  \nfactrue \ccol - \cglob ,  \frac{2}{\tauglob} \sum_{j=1}^\nfactrue \frac{\bcol_j}{ \taucol_j} ,2 \bglob}.
 \end{eqnarray*}
Given $\tauglob  \new $, the column specific parameters are updated as
\begin{eqnarray*}
\taucol_j \new =  \tilde \taucol_j /\tauglob  \new = \frac{\tauglob}{\tauglob  \new} \taucol_j,
\qquad j=1, \ldots,  \nfactrue.
 \end{eqnarray*}  %

\section{Initialising  Algorithm~\ref{Algo3}}    \label{init}

To check the mixing of the MCMC chain,   two (or more) independent runs with different initial dimensions  $\nfactrue$ of the CFA model and a positive number $\nfacsp$ of spurious columns are performed.
 On the one hand,  $\nfactrue$  is chosen
 to be  smaller than the expected number of factors,
 on the other hand a large  value close to $\nfac - \nfacsp$ is chosen.

Initial  values for the leading indices   $\lm_{\nfactrue}=(l_1, \ldots,  l_{\nfactrue})$
of the sparsity matrix $\deltav_{\nfactrue}$
are chosen  by first sampling  $ l_1$ from $\{1, \ldots, u_1 \}$, where $u_1$ is a small number, e.g. 5. Then  for $j=2, \ldots, \nfactrue$, we  sample  $ l_j  $  uniformly from the set $\leadset{\lm_{-j}}$ with  $\lm_{-j}=\{l_1, \ldots,  l_{j-1}\}$.
Factor loadings below the leading elements are allowed to be zero with positive probability $p_0$, e.g. $p_0=0.5$.
 We draw at most 100  initial values $\deltav_\nfactrue$ (including the leading indices) in this way, until a matrix  $\deltav_\nfactrue$ is obtained which satisfies the \CountAR\ counting rule. If no such indicator matrix is found, then we add enough nonzero elements in each  column of $\deltav_\nfactrue$ (e.g., by setting  $\delta_{l_j+1,j}=1,
 \ldots, \delta_{l_j+3,j}=1$) to ensure  variance identification for the initial value.

To obtain starting values for the factors $\facm_\nfactrue$,
  we perform a few (say 100) MCMC iterations in the confirmatory factor model corresponding to   $\deltav_\nfactrue$, which is initialized  by sampling the factors $\facm_\nfactrue=(\facmk{\nfactrue}{1},\ldots,\facmk{\nfactrue}{T})$ from the prior:  $\fac_{jt} \sim \Normal{0,1}$, $j=1, \ldots, \nfactrue$, $t=1, \ldots,T$.
While holding   $\deltav_\nfactrue$ fixed, we iterate between sampling  the model parameters  $\facload_\nfactrue$ and $\idiov_1,\ldots,\idiov_{\dimy}$
as in Step~(P)  and  sampling  the factors $\facm_\nfactrue$  as in Step~(F) of  Algorithm~\ref{Algo3}.
The resulting factors $\facm_\nfactrue$  serve as starting values for the full-blown MCMC scheme described in
Algorithm~\ref{Algo3}.

\section{More details for the exchange rate data}    \label{Dataex}

\paragraph*{Data.} The 22 exchange rates analysed in Section~\ref{applicEx22} are listed in Table~\ref{abbrev}.

\begin{Tabelle}{Exchange rate data; currency abbreviations.}{abbrev}
{  \small \begin{tabular}{cccc}
\begin{tabular}{rllrlrl}
  \hline
1 &   AUD & Australia dollar  \\
2 &   CAD & Canada dollar \\
3&  CHF & Switzerland franc \\
4 &  CZK & Czech R.\ koruna \\
5  &DKK & Denmark krone \\
 6 & GBP & UK pound \\
7  & HKD & Hong Kong dollar \\
8 &IDR & Indonesia rupiah \\
9 &JPY & Japan yen \\
10 &KRW & South Korea won \\
11  & MXN& Mexican Peso\\  \hline
\end{tabular} &&
\begin{tabular}{rrlrlrl} \hline
12  &MYR & Malaysia ringgit \\
13  &NOK & Norway krone \\
 14 &NZD & New Zealand dollar \\
 15 &PHP & Philippines peso \\
 16 &PLN & Poland zloty \\
 17 &RON & Romania fourth leu \\
 18  &RUB & Russian ruble \\
 19 &SEK & Sweden krona \\
 20 &SGD & Singapore dollar \\
 21 &THB & Thailand baht \\
22  &USD & US dollar \\  \hline
\end{tabular} \end{tabular}
}
\end{Tabelle}

\paragraph*{Running MCMC.} 
All computations are based on 
Algorithm~\ref{Algo3} and \ref{AlgoRJMCMC}. For tuning,  we choose $p_s=0.5$ in Step~(R) and $\pshift = p_{\mbox{\rm \footnotesize switch}}=1/3$ and $p_a=0.5$ in Step~(L). Boosting in Step~(A) relies on ASIS with  $\sqrt{\Psi_j}$  being the largest loading (in absolute value) in each non-zero column, see  Appendix~\ref{boost_frac}.

\begin{Tabelle}{Exchange rate data; inclusion probabilities for the sparsity matrix $\deltav_4$ for the 2PB prior,  averaged over all ordered GLT draws with pivots equal to $\lm ^\star= (1,2,5,7) $.}{ken_mpm_tab2}
{  \small \begin{tabular}{ccccc}  \hline
 Currency & Factor~1 & Factor~2 & Factor~3 & Factor~4   \\ \hline
      AUD &   1 &     0 &     0 &     0 \\ 
    CAD   &   1 &     1 &     0 &     0 \\ 
    CHF & 0.01 &  0.12 &     0 &     0 \\ 
    CZK &   0 &   0.2 &     0 &     0 \\ 
    DKK & 0.01 &     1 &     1 &     0 \\ 
    GBP & 0.04 &     1 &  0.05 &     0 \\ 
    HKD  &  0.01 &     1 &  0.97 &     1 \\ 
    IDR  & 0.02 &     1 &  0.03 &     1 \\ 
    JPY & 0.09 &     1 &  0.01 &  0.02 \\ 
    KRW  &    0 &     1 &  0.07 &  0.02 \\ 
    MXN &  0 &  0.16 &  0.01 &  0.02 \\ 
    MYR  &    1 &  0.06 &  0.01 &  0.01 \\ 
    NOK  & 0 &     1 &  0.01 &  0.03 \\ 
    NZD  &   0.06 &  0.45 &  0.04 &  0.02 \\ 
    PHP &   0 &     1 &  0.94 &  0.04 \\ 
     PLN &    0 &     1 &  0.02 &  0.75 \\ 
    RON & 0.1 &  0.07 &  0.25 &  0.01 \\ 
    RUB &   0.21 &   0.1 &   0.1 &  0.16 \\ 
    SEK &  0 &     1 &  0.01 &     1 \\ 
    SGD  &  0.02 &     1 &  0.03 &  0.99 \\ 
   THB  &  0 &     1 &  0.01 &  0.02 \\ 
  USD &  0.01 &     1 &     1 &  0.01 \\ 
     \hline
     \end{tabular}}
\end{Tabelle}

\paragraph*{Further results.}
Table~\ref{ken_mpm_tab2} reports   $\Prob{\delta_{ij}=1|\ym, \lm ^\star}$ for the 2PB prior,
averaged over all ordered GLT draws with pivots equal to $\lm ^\star= (1,2,5,7) $. This table is used to derive 
the MPM.
 Table~\ref{ken_fac4} shows the posterior mean of the factor loading matrix,
the idiosyncratic variances and the communalities, obtained by averaging over all draws
where the pivots of $\deltav_4$  coincide with $\lm ^\star$. 
  As expected,  nonzero factor loadings have
       relatively high  communalities for the different currencies, whereas for zero rows the communalities are practically equal to zero percent.

\begin{Tabelle}{Exchange rate data; posterior mean of the factor loadings  $\loadtrue_{ij}$,  the communalities  $R^2_{ij}$ (in percent)
and the idiosyncratic variances  $\sigma_i^2$   (2PB prior)
 for a 4-factor model with the GLT constraint
$\lm ^\star= (1,3,5,7)$. }{ken_fac4}
{\small \begin{tabular}{lrrrrrrrrr} \hline
 &   \multicolumn{4}{c}{Factor loadings} &  \multicolumn{4}{c}{Communalities}&  \\ %
  Currency & $\loadtrue_{i1}$ & $\loadtrue_{i2}$ & $\loadtrue_{i3}$  & $\loadtrue_{i4}$  & $R^2_{i1} $ & $R^2_{i2}$ & $R^2_{i3}$  & $R^2_{i4}$ %
  & $\sigma_i^2$ \\ \hline
  AUD & 0.962 &      0 &          0 &      0 &      88.4 &       0 &        0 &         0 &    0.12 \\
  CAD  &  0.391 &      0.601 &          0 &      0 &      16.9 &    38.6 &        0 &     0 &    0.42 \\
   CHF &  $\approx 0$ &    -0.014 &  0 &  0 &    0.01 &   0.3 &        0 &         0 &    0.98 \\
    CZK &  $\approx 0$ &      0.034 &     0 &   0 &    0.01 &   0.8 &        0 &         0 &    0.98 \\
     DKK & $\approx 0$ &    1.06 &    0.22 &    0 & $\approx 0$ &    95.3 &      4.1 &   0 &  0.01 \\
     GBP  & 0.01 &      0.569 &    -0.01 &    0 &     0.26 &    31.4 &    0.27 &         0 &    0.71 \\ HKD &   $\approx 0$ &  0.502 &   0.398 &  0.762 &  $\approx 0$ &    21.8 &     14.2 &  49.4 &  0.17 \\
IDR   &  $\approx 0$ &  0.8 &   $\approx 0$ &   0.427 &   0.05&   57.2 &   0.06 &   16.9 &    0.28 \\
  JPY &0.013 &   0.924 &$\approx 0$&$\approx 0$&  0.23 &    75.6 &  $\approx 0$ &    0.01 &    0.27 \\
  KRW  & $\approx 0$ & 1.06 &$\approx 0$& $\approx 0$ & $\approx 0$ & 95.9 & 0.04 & $\approx 0$ & 0.05\\
MXN &  $\approx 0$ & 0.025 & $\approx 0$& $\approx 0$ & $\approx 0$&  0.58 &   0.04 &  0.04 &   0.98\\
MYR &0.793 &$\approx 0$&  $\approx 0$ & $\approx 0$ &  61.2 &  0.04 &  0.01 &    0.01 &  0.40 \\ NOK &$\approx 0$ & 0.88 & $\approx 0$& $\approx 0$ & $\approx 0$ &  69.7 &  0.01 &  0.05 &  0.33 \\
 NZD  &    0.016 &  0.108 &  -0.01 &$\approx 0$& 0.50 & 3.17 & 0.29 &    0.04 &  0.95 \\
 PHP  & $\approx 0$&  0.549 &     -0.415 & $\approx 0$ &    0.01 &    28.7 &  17.9 &  0.17 &    0.56 \\ PLN  & $\approx 0$&  0.995 & $\approx 0$ &  0.128 & $\approx 0$ & 85.6 &  0.01 &      2.02 &    0.14 \\
 RON &  0.027 & $\approx 0$ &    -0.08 & $\approx 0$ &   0.87 &  0.09 &     3.08 &  0.02 &    0.95 \\
 RUB &  -0.067 &     0.012 &     0.029 & 0.043 & 2.37 &   0.28 &    0.95 &   1.37 &  0.94 \\
 SEK &  $\approx 0$ &  0.974 & $\approx 0$ & 0.31 & $\approx 0$ &    81.8 &  0.01 & 8.7 &    0.11 \\
  SGD &  $\approx 0$ &  0.748 & $\approx 0$ & 0.392 &    0.05 &    50.9 &    0.05 &   14.6 &    0.38 \\
THB   &  $\approx 0$ &  0.583 & $\approx 0$ & $\approx 0$ &   0.01 &    32.7 &   0.01 &    0.05 &  0.70 \\
USD & $\approx 0$&  1.06 &      0.215 &  $\approx 0$ & $\approx 0$ &    95.3 & 4.1 & $\approx 0$&  0.01\\
  \hline
\end{tabular}

\footnotesize{Note: entries  with
$|\loadtrue_{ij}|  <0.01$  and entries with
$ R^2_{ij} < 0.1$  are  indicated by $\approx 0$.}}
\end{Tabelle}

\end{document}